\documentclass[11pt,draft]{IEEEtran}

\usepackage{cite}
\usepackage{color}      
\usepackage{epsf,psfrag,amssymb,amsfonts,latexsym,verbatim}
\usepackage[mathscr]{eucal}

\def\psfancypar#1#2{\begingroup\def\par{\endgraf\endgroup\lineskiplimit=0pt}
               \setbox2=\hbox{\large\sc #2}
               \newdimen\tmpht \tmpht \ht2 \advance\tmpht by \baselineskip
               \font\hhuge=Times-Bold at \tmpht
               \setbox1=\hbox{{\hhuge #1}}
               \count7=\tmpht \count8=\ht1
               \divide\count8 by 1000 \divide\count7 by \count8 
               \tmpht=.001\tmpht\multiply\tmpht by \count7 
               \font\hhuge=Times-Bold at \tmpht
               \setbox1=\hbox{{\hhuge #1}}
               \noindent
                \hangindent1.05\wd1
               \hangafter=-2 {\hskip-\hangindent
               \lower1\ht1\hbox{\raise1.0\ht2\copy1}%
                \kern-0\wd1}\copy2\lineskiplimit=-1000pt}

\newcommand{\E}{\mbox{{\rm E}}}

\def\boxit#1{\vbox{\hrule\hbox{\vrule\kern3pt
        \vbox{\kern3pt#1\kern3pt}\kern3pt\vrule}\hrule}}

\def\reals{ { {\rm  I \kern-0.15em R }  } }
\def\complex{ {\,{{\rm C} \kern-0.50em \raise0.20ex {  |}}\, }}

\def\gammabf{\hbox{\boldmath$\gamma$\unboldmath}}

\def\mubf{\hbox{\boldmath$\mu$\unboldmath}}

\def\omegabf{\hbox{\boldmath$\omega$\unboldmath}}
\def\Sigmabf{\hbox{$\bf \Sigma$}}

\def\Lambdabf{\mbox{$ \bf \Lambda $}}

\def\hbf{{\bf h}}
\def\ibf{{\bf i}}
\def\jbf{{\bf j}}
\def\kbf{{\bf k}}

\def\ubf{{\bf u}}
\def\vbf{{\bf v}}

\def\xbf{{\bf x}}
\def\ybf{{\bf y}}

\def\xbf{{\bf x}}
\def\ybf{{\bf y}}
\def\Abf{{\bf A}}

\def\Cbf{{\bf C}}

\def\Ibf{{\bf I}}

\def\Qbf{{\bf Q}}
\def\Rbf{{\bf R}}
\def\Sbf{{\bf S}}

\def\Ubf{{\bf U}}
\def\Vbf{{\bf V}}
\def\Wbf{{\bf W}}

\def\Ac{{\cal A}}

\def\Dc{{\cal D}}
\def\Ec{{\cal E}}

\def\Gc{{\cal G}}

\def\Ic{{\cal I}}

\def\Nc{{\cal N}}

\def\Xc{{\cal X}}

\def\be{\vskip .3cm \begin{equation}}
\def\ee{\end{equation} \vskip .4cm \noindent}
\def\defeq{{\stackrel{\Delta}{=}}}
\newcommand{\R}{\mbox{$\hat {\bf R}_{N}$}}

\def\Rxx{\Rbf_{\ssstyle X\kern-.1em X}}

\let\ssstyle=\scriptscriptstyle

\def\Kout{\setbox1=\hbox{\Huge\bf K}\hbox to
1.05\wd1{\hspace{.05\wd1}
}}
\def\Sout{\setbox1=\hbox{\Huge\bf S}\hbox to 1.05\wd1{\hspace{.05\wd1}
}}

  \ifx\LabelFigloaded\MYundefined\relax
  \else
   \fi

  \def\LabelFigloaded{\relax}

  \chardef\LabelFigCatAt\the\catcode`\@
  \catcode`\@=11

 \let\LabelFigwlog@ld\wlog
 \def\wlog#1{\relax}

 \ifx\\\MYundefined@
    \let\\\relax
 \fi

  \def\ms@g{\immediate\write16}

 \def\N@wif{\csname newif\endcsname }
 \def\Temp@ {\N@wif\ifIN@}
 \ifx\INN@\MYundefined@
    \else \let\Temp@\relax
 \fi
 \Temp@

  \def\IN@{\expandafter\INN@\expandafter}
  \long\def\INN@0#1@#2@{\long\def\NI@##1#1##2##3\ENDNI@
    {\ifx\m@rker##2\IN@false\else\IN@true\fi}%
     \expandafter\NI@#2@@#1\m@rker\ENDNI@}
  \def\m@rker{\m@@rker}
 
  \newtoks\Initialtoks@  \newtoks\Terminaltoks@
  \def\SPLIT@{\expandafter\SPLITT@\expandafter}
  \def\SPLITT@0#1@#2@{\def\TTILPS@##1#1##2@{%
     \Initialtoks@{##1}\Terminaltoks@{##2}}\expandafter\TTILPS@#2@}

 \def\Shifted@@#1#2#3{\setbox0=\hbox{#3}%
   \raise -\dp0\vbox {\kern-#2%
       \hbox {\kern#1\unhbox0\kern-#1}%
           \kern#2}}

 \newcount\gridcount
 \newbox\auxGridbox@ \newbox\hGridbox@ \newbox\vGridbox@
 \newbox\Labelbox@ \newbox\auxLabelbox@
 \newbox\Coordinatebox@
 \newtoks\Labeltoks@
 \newdimen\Wdd@ \newdimen\Htt@
 \newdimen\Wddd@ \newdimen\Httt@
 
 \def\Wr@{\immediate\write16}

 \newdimen\GL@wd
 \def\GridLineWidth#1{\GL@wd=#1}

 \def\gobble#1{}
 \def\EdgeErr@{\Wr@{}%
      \Wr@{\string\Edges\space argument
      1, 10, 100 or 1000 please\string!}%
      }

 \newcount\Edgect@

 \def\Sweepup#1\endSweepup{}

 \def\SetEdges@{%
    \edef\Zr@@s{\expandafter\gobble\number\Edgect@\empty}%
        \count255=0\Zr@@s\relax
        \ifnum\count255=\z@\else\EdgeErr@\show\tailtest\fi
        \count255=1\Zr@@s\relax
        \ifnum\count255=\Edgect@\relax\else\EdgeErr@\show\leadtest\fi
    \EdgGl@b\edef\Zr@s{\expandafter\gobble\Zr@@s\empty}
    \ifnum\Edgect@>\@ne\relax\EdgGl@b\let\L@Dc\empty
        \else\EdgGl@b\edef\L@Dc{\string.}\fi
    \ifnum\Edgect@>\@ne\relax
        \EdgGl@b\edef\Edgescale@##1{\divide##1 by \Edgect@}%
        \else\EdgGl@b\edef\Edgescale@##1{}\fi
    }

 \def\Edges#1{\Edgect@=#1\relax
     \let\EdgGl@b\global \SetEdges@}

 \Edges{1}

 \def\hhrule{\hrule height \GL@wd\vskip-.\GL@wd}

 \def\hRule@{%
   \advance\gridcount -2%
   \vfil\hhrule\vfil
   \llap{\smash{\raise -2.5pt
     \hbox{\L@Dc\number\gridcount\Zr@s\kern2pt}}}%
   \hhrule
   }

\def\vvrule{\vrule width \GL@wd \kern-\GL@wd}

 \def\vRule@{\advance\gridcount 2%
   \hfil\vvrule\hfil
   \setbox\auxGridbox@=\vbox to 0pt
      {\vskip \Htt@\vskip 2pt
        \hbox to 0pt{\hss\L@Dc\number\gridcount\Zr@s\hss}\vss}%
      \wd\auxGridbox@=0pt \box\auxGridbox@
   \vvrule
   }

 \def\PlaceGrid@@{\gridcount=10 
  \setbox\hGridbox@=\hbox{%
        \hbox{%
             \hskip-.4pt\vrule
             \vbox to \Htt@{%
               \offinterlineskip\parindent=\z@\relax
               \hbox to \Wdd@{\hfil}
               \hRule@\hRule@\hRule@\hRule@
               \vfil\hhrule\vfil}%
             \vrule\hskip-.4pt}
    }%
  \gridcount=0%
  \setbox\vGridbox@=\hbox{%
      \vbox{\offinterlineskip\parindent=0pt\hsize=0pt
         \vskip-.4pt\hrule%
         \hbox to \Wdd@{%
                 \vtop to \Htt@{\vfil}%
                 \vRule@\vRule@\vRule@\vRule@
                 \hfil\vvrule\hfil}%
         \hrule\vskip-.4pt}}%
  \wd\hGridbox@=0pt\ht\hGridbox@=0pt
  \wd\vGridbox@=0pt\ht\vGridbox@=0pt
  \hbox{\box\hGridbox@\box\vGridbox@}%
  }

 \def\LabelsGlobal{\def\LabGl@b{\global}}
 \def\LabelsLocal{\def\LabGl@b{}}
 \LabelsGlobal 

 \def\SetLabels#1\endSetLabels{%
   \LabGl@b\Labeltoks@={#1()\\}%
   }

 \LabGl@b\Labeltoks@={()\\}

 \def\ShowGrid{\LabGl@b\let\PlaceGrid@\PlaceGrid@@}
 \def\HideGrid{\LabGl@b\let\PlaceGrid@\relax}
 \def\Grids{\ShowGrid\LabGl@b\let\GridSwitch@\ShowGrid}
 \def\noGrids{\HideGrid\LabGl@b\let\GridSwitch@\HideGrid}

 \noGrids

 \def\bAdjust@@{%
     \setbox\auxLabelbox@=\hbox{\raise \dp\auxLabelbox@
            \box\auxLabelbox@}}
 \def\bAdjust@{\let\vAdjust@\bAdjust@@}

 \def\eAdjust@@{\dimen0=-.5\ht\auxLabelbox@
     \advance\dimen0 by .5\dp\auxLabelbox@
     \setbox\auxLabelbox@=
            \hbox{\raise\dimen0\box\auxLabelbox@}}
 \def\eAdjust@{\let\vAdjust@\eAdjust@@}

 \def\tAdjust@@{%
     \setbox\auxLabelbox@=\hbox{\raise-\ht\auxLabelbox@
            \box\auxLabelbox@}}
 \def\tAdjust@{\let\vAdjust@\tAdjust@@}

 \let\vAdjust@\relax

 \def\lAdjust@{\let\hAdjust@\rlap}
 \def\rAdjust@{\let\hAdjust@\llap}

 \let\hAdjust@\relax\let\vAdjust@\relax

 \def\FetchLabel@#1(#2)#3\\{%
     \IN@0#2@@\ifIN@
        \setbox0=\hbox{\ignorespaces#1#3\unskip}%
        \ifdim\wd0>0pt
           \ms@g{}%
           \ms@g{ !!! Bad label(s)? !!!}%
           \message{ #1(#2)#3}%
        \fi
        \def\LabelMole@##1\endFetchLabel@{%
            \IN@0()\\@##1@%
            \ifIN@\def\Temp@{\FetchLabel@##1\endFetchLabel@}%
            \else\def\Temp@{}%
            \fi
            \Temp@
           }%
     \else
       \ignorespaces#1\unskip
       \setbox\auxLabelbox@=%
         \hbox to 0pt{\hss\ignorespaces\hAdjust@
          {\ignorespaces#3\unskip}\hss}%
       \vAdjust@
       \let\hAdjust@\relax\let\vAdjust@\relax
       \AugmentLabelBox@@{#2}%
       \ht\Labelbox@=0pt\dp\Labelbox@=0pt
       \let\LabelMole@\FetchLabel@%
     \fi\LabelMole@}

 \newtoks\XYSep@ 
 \def\SetXYSeparator#1{%
     \IN@0#1@@\ifIN@\XYSep@{*}%
     \else
     \XYSep@{#1}%
     \fi
     }

 \SetXYSeparator*

 \def\AugmentLabelBox@@#1{%
     \IN@0\the\XYSep@ @#1@\ifIN@
       \SPLIT@0\the\XYSep@ @#1@%
       \setbox\Labelbox@=\hbox to 0pt{%
         \unhbox\Labelbox@
         \Shifted@@{\the\Initialtoks@\Wddd@}%
         {\the\Terminaltoks@\Httt@}%
         {\box\auxLabelbox@}}%
     \else
         \ms@g{}%
         \ms@g{ !!! Bad insertion point. !!!}%
         \message{ (#1\ this point was rejected.)}%
     \fi
    }

 \def\FetchOption@#1[#2]#3\endFetchOption@{%
    \def\temp{#1}
    \ifx\temp\empty
       \Edgect@=#2\relax
       \let\EdgGl@b\relax
       \SetEdges@
       \Cleaner@#3%
    \fi}

 \def\Cleaner@#1[@]{\Labeltoks@{#1}}
     
 \def\PlaceLabels@@{\mathsurround=0pt
     \def\Cr@{\\}%
     \let\L\lAdjust@\let\R\rAdjust@
     \let\B\bAdjust@\let\E\eAdjust@\let\T\tAdjust@
     \expandafter\FetchOption@\the\Labeltoks@[@]\endFetchOption@
     \Wddd@=\Wdd@ \Edgescale@\Wddd@ 
     \Httt@=\Htt@ \Edgescale@\Httt@
     \expandafter\FetchLabel@\the\Labeltoks@\endFetchLabel@
     \box\Labelbox@
     }%

 \let \PlaceLabels@\PlaceLabels@@

 \def\AffixLabels#1{\setbox\Coordinatebox@=\hbox{#1}%
      \Wdd@=\wd\Coordinatebox@ \Htt@=\ht\Coordinatebox@
      \advance\Htt@ \dp\Coordinatebox@
      \hbox{\copy\Coordinatebox@\kern-\Wdd@ 
           \Shifted@@{0pt}{-\dp\Coordinatebox@}%
           {\PlaceLabels@\PlaceGrid@}%
           \kern\Wdd@}%
      \GridSwitch@ 
      \LabGl@b\Labeltoks@{()\\}%
      }
 
   \let\wlog\LabelFigwlog@ld   
   \catcode`\@=\LabelFigCatAt  

                                By

              Raymond S\'eroul <A18645@FRCCSC21.BITNET>
                                and 
              Laurent Siebenmann <lcs@topo.math.u-psud.fr>
    
              VERSIONS: July 1991, Oct 1991, Jan 1992, July 1992

INTRODUCTION

      This labelling package is intended for TeX users who
rely on non-TeX sources for for their graphics inserts.  It
provides means for adding TeX labels to such inserts with a
minimum of fuss. 

       For most labels, TeX users have in the past found it
reasonably convenient to rely on non-TeX sources. Typical
occasions when an inescapable need for TeX labels seemed to
arise are

 (a) when the graphics program lacks certain exotic or complex
mathematical symbols

 (b) when the very highest typographical quality is wanted for the
labels

 (c) when labels included with the graphics fail to print, 
 and you cannot figure out why (cf. boxedeps.doc).  The labels
 provided by labelfig.tex are 100

       Since this package first appeared, many users, who in the
past scarcely dreamed of using TeX labels, have come to use
nothing but.  So it is now appropriate to add

Intoxication Warning:  TeX labels may be addictive and expensive. 

     If you have a fast preview you may disagree, and even find
that this package provides an agreeable paste-up environment; see
extra applications at end.

     Note to publishers: It is possible and convenient to ultimately
export the TeX labels produced by labelfig.tex to become an integral
part of the EPS file. This is often desired by a publisher who typically
uses an "upmarket" graphics or page layout program, with which the
staff is skilled in perfecting figures.  See Appendix I for
a recipe.

     The authors are grateful to Patrick Ion of Math Reviews for
helpful comments and encouragement.

BASIC INSTRUCTIONS

    After reading in the macro file using

preview or proof your figure with a coordinate grid printed on
top, by typing the following:

    \ShowGrid  
    \AffixLabels{<the graphics insertion>}

Here <the graphics insertion> is what you would type to insert
the graphics object alone without the grid.  This must provide
for the space around it. For example <the graphics insertion>
might well be \BoxedEPSF{MyFigure scaled 700} using the
boxedeps.tex macro package (from same source); this provides a
TeX box containing the encapsulated PostScript insert specified by
the file MyFigure. \AffixLabels{...} provides the grid (supposing
\ShowGrid is present) and later, once you have specified labels
using the grid, it will "tack on" the labels.

     The grid is a sort of (usually elongated) checkerboard of
ten rows and ten columns and its (internal) partitions are by
default numbered  .1, ... ,.9  both horizontally (X-coordinate
running left to right) and vertically (Y-coordinate running bottom
to top).  Thus the points enclosed by the grid correspond to the
points of the unit square in the cartesian "X-Y" plane, the lower
left corner corresponding to the origin (0,0).  By extrapolation,
the full page corresponds to a larger rectangle in the plane.

     These coordinates serve to position labels as follows.
Before the \AffixLabels{...} command type label specifications:

  \SetLabels
   (<X-coordinate>*<Y-coordinate>) <first label> \\
   .
   .
   .
   (<X-coordinate>*<Y-coordinate>)  <last label> \\
  \endSetLabels

Each row specifies one label and is terminated by \\.  In each
row, the position indicator comes first; it is written as a
standard cartesian point except that the X- and Y- coordinates
are separated by * rather than a comma because TeX allows a
comma as decimal point. There are no dimension units to specify
as the unit is the grid itself.

     By default, this cartesian point specifies where the middle
of the baseline of the label will be located.  However if you precede
the point by \L [or \R] the left [or right] edge of the baseline will
be located there. Similarly you may also precede the point by \T, \E,
or \B to vertically align the top equator or bottom of the label box
at the specified point.  This gives nine standard positions of
the label with respect to the insertion point --- corresponding to
the eight principle points of the compas and the center

                     \L\T     \T      \R\T

                     \L\E     \E      \R\E

                     \L\B     \B      \R\B

But this neglects the default "baseline" level of TeX,
giving potentially three more positions

                     \L    <no tag>   \R

For text, the baseline level is often the preferred. Its relation to
the others is variable. It will often coincide with the bottom level,
as happens for "X".  But it is often distinct, as for "g", in which
case you have in all 12 distinct positions rather than 9.

     It is convenient to think of this specification of label
position as attaching the label by a thumb-tack to the coordinate
grid. There are up to twelve positions of the thumb-tack on the
label, while the position of the thumb-tack on the coordinate grid is
arbitrary.  Normally, one choses the position of the thumb-tack on
the label to be the one that is the closest to the item being
labeled.  There are good reasons for this "rule of thumb":

   (a)  It facilitates correct positioning at first try.

   (b)  If the scale of the figure must be altered after labels
have been affixed, the labels have a good chance of remaining well
positioned.

   (c)  The visible grid need not extend beyond the "bounding box"
for the figure, because the best preferred position is always
(at least almost) within the bounding box .

The second reason is particularly important. Indeed it often
happens that scale has to be altered after labelling begins, in
order to either provide space for the labels, or to adjust
proportions between the labels and the figure.  (The size of labels
is unaffected by scaling.)

     Here is an artificial but self-contained test which uses
TeX rules to make a graphics object.

TEST

    Do not skip this!

 \def\FrameIt#1{\hbox{\vrule$\vcenter {\hrule\kern3pt%
             \hbox {\kern3pt #1\kern3pt}%
               \kern3pt\hrule}$\relax\vrule}}

 \def\Caption#1#2{\FrameIt{%
       \vtop {\hsize=#1\relax \parindent=0pt
         \leftskip=0pt \rightskip=0pt plus15pt
         \parfillskip=0pt
         \lineskip=1pt\baselineskip=0pt
         #2}}}

 \def\FirstQuadrant{\hbox to 100pt{\vrule\vbox to 100pt{%
        \hbox to 100pt{\hfil}\vfil\hrule}\hss}}

  \SetLabels
    \R(.5*.2) $\zeta\,\cdot$\\
    (.9*-.10) $\xi$\\
    \R(-.03*.9) $\eta$\\
    \T(.5*.9) \Caption{70pt}{%
          \it The norm of
          $g(\xi+i\eta)$ is indicated on
          contours of this invisible surface.}\\
  \endSetLabels

  \AffixLabels{\FirstQuadrant}

  \end

  Note that the coordinates to use for labels are indicated on the
edges of the grid (when visible) corresponding to the conventional
x- and y- axes of the Cartesian plane. By default the grid is
1-by-1. However, by the command \Edges{100}, you can change this
to 100-by-100 and many users find this alternative most
convenient. Place the command \Edges{...} in your style file (or
header) since its effect is is global. Other possible edge values
are 10 and 1000.

  If you use the command \Edges{...} at all, do so with care.  For
if you accidentally delete an \Edges{...} command your labels will
abruptly be badly misplaced and may logically but mysteriously
generate "dimension too big" errors under TeX and "off page" errors
under your driver.  

  You can dictate the edgescale for an individual figure by giving
the scale in brackets immediately after \SetLabels.  Thus, to
import into an article using say \Edge{100} a figure labelled using
another edgescale, say the original 1-by-1 default, you can use
\SetLabels[1]...\endSetLabels.

GETTING IT DOWN PAT

     Complicated labeling deserves the same respect as
complicated mathematics.  Do not expect it to come out perfect the
first time!  What is needed in either case is a mechanism to
repeatedly typeset troublesome pieces.

     One mechanism is always available.  One does complicated
labelling in a separate "test" file involving just the figure being
labelled;  a texpert will know how to \dump TeX's current state as
a temporary format that restarts rapidly at each retry.  Usually,
one then pastes the completed labelled figure back into the main
TeX file, but, of course, one can also \input it as an auxiliary
file.

     If you do not have a TeXpert at handy, here is a first
approximation to an efficient setup. By deletions reduce a copy
of your article to just a few lines before and after the figure.
Now label the figure, and finally, copy and paste the labelled
figure to the original article. Then copy the next figure to label
into this testbed and repeat. The TeXpert can improve the  speed
at which TeX starts up, by compiling a format specifically for
your article; just one caution: best NOT include in the format
ephemeral details of setup like \Set<mydriver>ArtSpecials (from
boxedeps.tex because this reads  figure dimensions which you may
change during your work session.

     An improved mechanism to repeatedly typeset troublesome
pieces is now available on the Macintosh; it is called LinoTeX;
see the same ftp sources.  It could be set up on many types
of computer.

     Before using labelfig.tex to attach labels to a graphics
object inserted using boxedeps.tex or BoxedArt.tex, make it a
firm rule to carefully adjust the bounding box using the trimming
commands of these packages, and also at least tentatively scale
and position the object. Beware of changing the grid inadvertently
after the labels have been positioned.  For example, correcting
the bounding box of a PostScript graphics object can foul up the
labels by changing the coordinate grid to which the labels are
attached. This is particularly true for the trimming  commands of
boxedeps.tex and BoxedArt.tex. However, as noted already, change
of scale is much less disruptive, and modest adjustments should be
well tolerated.

     Sometimes the labels protrude so far from the bounding box
of a figure that the figure has to be repositioned.  Best do this
by ad hoc spacing, say using \hglue and \vglue; altering the
bounding box would create a vicious circle.

     Remember that you are responsible for preventing labels
from overlapping. You are responsible for all label typography
including size and style. A label is really just about anything
that can be put in a TeX box. Note that spaces at the beginning
and end of labels will normally be suppressed; if you really want
them you must protect them with TeX braces.

     This package temporarily sets the \mathsurround parameter
of TeX to zero  while the labels are being affixed. This is done
because nonzero \mathsurround space would influence the position
of left and right aligned labels; then, when a texpert or printer
modifies mathsurround, diagram labeling might be disastrously
altered. There is a small price to pay involving labels that are
formatted as caption boxes including mathematics: you  may want or
need to specify an explicit mathsurround space within the caption
box; it will not influence anything outside.

     Those hostile to the use of * as separator between
the X and Y coordinates of label insertion points, are free to
impose another using \SetXYSeparator{<the new separator>}.  
Americans may prefer "," to "*" since they never use a 
comma as a decimal point; on the other hand, * may be more visible.

APPENDIX (I)  MERGING labelfig.tex LABELS INTO AN EPSF GRAPHICS OBJECT.

     As promised in the introduction, here is a recipe useful for
publishers. It works at least on Macintosh and at least for vectorized
graphics and Adobe type1 fonts.  (There is surely a similar recipe for
PCs under MSWindows.)

 (a)  Use boxedeps.tex utility to integrate the figure given by the eps
file, "x.eps" say, with a visible frame around it.  See
\ShowDisplacementBoxes command in boxedeps.tex.  To get precise results
automatically it is important to use the \Trim... commands of
boxedeps.tex making the "DisplacementBox" neatly fit the figure.

 (b)  Use the TeX printer driver and LaserWriter (versions >= 8.1.1) to
export to an EPSF the DVI page containing the integrated, labelled
figure. You now have an EPS file  "xx.eps"  that contains too much, and at
the wrong scale, and at wrong position.

 (c)  Convert the EPSF to an Adode Illustrator format EPSF using
the shareware utility called epsConvert by Sam Weiss
1993-- (currently $25).

 (d)  In Illustrator (or a compatible program), group the labels and the
"DisplacementBox"; copy them to the clipboard and paste them into "x.ps".
This step requires that all the label fonts be "visible to the Macintosh.

 (e)  Translate and scale the pasted group consisting of the labels plus
the "DisplacementBox" so as to make the "DisplacementBox" the bounding
box of (labelless) figure represented by "x.eps".  At this point the
labels will be correctly placed on the figure "x.eps".

 (f)  Ungroup and delete the "DisplacementBox".  The result is the
desired single EPS file, "x+.eps" say, It contains the original figure
plus its labels.  

     Using grouping and ungrouping appropriately in "x+.eps", a
publisher's staff can very efficiently improve label positions etc.

APPENDIX II)  SOME EXOTIC APPLICATIONS

     The grid of labelfig.tex is analogous to a light-table in
classical page makeup with wax or latex glue.  In principle, you
can use it to compose any page from its indivisible parts.  This
even has some of the artisanal charm of classical paste-up
provided you have a fast screen preview to make the process
"interactive".

     In practice labelfig.tex is a tool for nonstandard jobs.
Here are a few going beyond the labelling already discussed.

(I)  GRAPHICS INTEGRATION.

     This is accomplished by treating the imported graphics
objects as labels.  The underlying graphics object is then
typically an empty  \vbox to <dimension>{\vfill} in a TeX
\midinsert...\endinsert construction.  A label line
might be of the form

   (.1*.1) \\

The exact form of the special command varies from driver to
driver.  However, in the case of encapsulated PostScript graphics
(EPSF norm), by relying on boxedeps.tex, one can have the
following standard syntax (independant of driver  (see
boxedeps.doc for details.
  
  (.1*.1) \BoxedEPSF{MyFigure scaled <scale in mils>}\\

This may be slow since it requires TeX to read the PostScript
file to read bounding box using many complex macros.  So you
may want to try

  (.1*.1) \EPSFSpecial{MyFigure}{<scale in mils>}\\

which is fast and driver independant, but it squashes the
bounding box, normally to its lower left corner.

     Similarly for graphics of the Macintosh PICT norm ---
using BoxedArt.tex (same sources) in place of boxedeps.tex.

     This approach to integration is to be recommended when
one is assembling a composite graphics object.

 (II)  COMMUTATIVE DIAGRAM ENHANCEMENT

     Commutative diagrams or arrays of mathematical objects
connected by arrows of various sorts are common in mathematics.
The mathematical objects require the use of TeX.  Recently TeX
acquired a good collection of arrows of all slopes --- that of
LamSTeX --- plus pwerful macros to build the diagrams.

     However, even the LamSTeX collection is often
inadequate; it lacks for example double shafted arrows, dotted
arrows and curved arrows. Fortunately it is possible to produce
such arrows on an individual basis using sophisticated graphics
programs such as Illustrator and AldusFreehand (both serving
the EPSF norm) or using Metafont (with its public domain norm).
Since the creation of each new arrow is a work of love, you
probably want to limit the number of arrows by using LamSTeX
for most arrows. The 40K commutative diagram module of LamSTeX
has been adapted to work with AmSTeX and a copy may be posted
with LabelFig and related files. Unfortunately no one has yet
offered a version that works with Plain TeX or LaTeX.

       Suffice it here to say that when the exotic arrow has
been somehow imported into TeX, labelfig.tex treats it as a
label that one affixes to the commutative diagram.  Two other
steps will be treated in separate notes, namely the matter of
extracting the dimension specifications for the arrow and the
construction of the arrow --- for these steps are far from
unique and often depend intimately on your computer environment. 
Notes for the Macintosh-Textures-Illustrator combination are
found in the file ExoticArrows.doc.

 (III) NESTING 

Ingenuity pays off in exploiting labelfig.tex. One can
mix graphics and typography quite freely.  labelfig.tex is good
for freeform or overlapping arrangements, while boxedeps.tex (or
BoxedArt.tex) is best for regimented non-overlapping
arrangements --- and the two can be combined.

     The default behavior of labelfig.tex is not ideal 
for nesting objects, because to prevent trouble for beginners
the register for labels is globally cleared when \AffixLabels
concludes.  But there are switches available

      \LabelsGlobal      \LabelsLocal

which change this.  To understand this, extend the above test 
by something like:

 \LabelsLocal

 \SetLabels
    (.5*.5) AAA\\
 \endSetLabels

 {
 \SetLabels
    (.5*.5) ZZZ\\
 \endSetLabels
   \AffixLabels{\FirstQuadrant}
 }

   \AffixLabels{\FirstQuadrant}

     There are however potential pitfalls.  Neither
labelfig.tex nor boxedeps.tex has been tested under extreme
conditions. Problems may occur if their procedures are
indiscriminately nested. For boxedeps.tex (not labelfig.tex)
there is a precise cause for worry, namely many of its
variables are "global", which means that TeX braces will not
provide the protection one might expect.

COMMAND SUMMARY FOR labelfig.tex

  Here [...] means optional (one or zero)
       [...]* means any number of such constructs

  \SetLabels
    [[<P>](<X><Sep><Y>) <label> \\]*
  \endSetLabels
  \ShowGrid  
  \AffixLabels{<the figure>}

   --- <P> is tack position, one of eleven or empty
              order irrelevant

                   \L\T      \T      \R\T

                   \L\E      \E      \R\E

                     \L               \R

                   \L\B      \B      \R\B

   --- (<X><Sep><Y>) insertion point;
  <Sep> is separator, = * by default;
  \SetXYSeparator{<Sep>} changes it.
   <X> and <Y> are real numbers

  --- <label> a label to attach 

  --- <the figure> the figure to label 

  \GlobalLabels (default)     
  \LocalLabels  setting for nested constructs.

 \Grids makes ALL grids appear; \HideGrid then makes just next disappear.
 \noGrids returns to default.  The commands are always global.

 \GridLineWidth{<dimension>} adjusts width of grid lines. Default is very
small, to give "hairline" effect. If your grid lines are missing try
setting \GridLineWidth{1pt}.

 \Edges#1 globally changes the edge size of all grids to the numerical 
value #1, which must be 1, 10, 100, or 1000.  The default is 1.

VERSION HISTORY.
 --- Jan 1993: \Edges#1 and [??] option after \SetLabels
 --- July 1992: \Grids, \noGrids, \HideGrid;
       Gridlines become hairlines; \GridLineWidth{<dimension>}.
 --- Oct 1991, Jan 1992: \SetXYSeparator{<Sep>},  \LabelsGlobal,
       \LabelsLocal.
 --- July 1991: first release

Address for bugs and other feedback:

        Raymond S\'eroul
        IREM and Lab. de Typographie Informatise
        Univ. Rene Descartes
        Strasbourg

    Tel 33-88-41-63-45
    Email:  A18645@FRCCSC21.BITNET

        Laurent Siebenmann
        Mathematique, Bat. 425,
        Univ de Paris-Sud,
        91405-Orsay,
        France

    Tel 33-1-6941-7949; 
    Email: lcs@topo.math.u-psud.fr

\def\scalefig#1{\epsfxsize #1\textwidth}
\def\defeq{\stackrel{\Delta}{=}}

\newcommand{\SNR}{\mbox{SNR}}
\newcommand {\Ebb}{{\mathbb{E}}}

\newcommand {\Zbb}{{\mathbb{Z}}}

\newcommand{\Kmsc}{{\mathscr{K}}}

\newcommand{\Imsc}{{\mathscr{I}}}

\newcommand{\beq}{\begin{equation}}
\newcommand{\eeq}{\end{equation}}

\newtheorem{theorem}{Theorem}
\newtheorem{lemma}{Lemma}
\newtheorem{definition}{Definition}

\newtheorem{corollary}{Corollary}

\newtheorem{remark}{Remark}

\newtheorem{problem}{Problem}

\title{{\Large\bf How Much Information can One Get from a Wireless {\em Ad Hoc} Sensor Network over a Correlated Random Field?}}

\author{Youngchul Sung$^\dagger$\thanks{$^\dagger$Corresponding author}, H. Vincent
Poor and Heejung Yu
\thanks{Youngchul Sung and Heejung Yu are
with the Dept. of Electrical Engineering, KAIST, Daejeon 305-701,
South Korea. Email: ysung@ee.kaist.ac.kr and
hjyu@stein.kaist.ac.kr. H. V. Poor is with the Dept. of Electrical
Engineering, Princeton University, Princeton, NJ 08544. Email:
poor@princeton.edu. The work of Y. Sung  was supported
  by the IT R\&D program of MKE/IITA. [2008-F-004-01 ``5G mobile communication systems based on
  beam-division multiple access and relays with group cooperation''.] The work of H. V. Poor was supported in part by
the U. S. National Science Foundation under Grants ANI-03-38807
and CNS-06-25637. } } 
\begin{document}
\maketitle

\def\asadd#1{{\textbf{\textsl #1}}}
\def\ignore#1{}

\begin{abstract}
New large deviations results that characterize the asymptotic
information rates for general $d$-dimensional ($d$-D) stationary
Gaussian fields are obtained. By applying the general results to
sensor nodes on a two-dimensional (2-D) lattice, the asymptotic
behavior of {\em ad hoc} sensor networks deployed over correlated
random fields for statistical inference is investigated. Under a
2-D hidden Gauss-Markov random field model with symmetric first
order conditional autoregression and the assumption of no
in-network data fusion, the behavior of the total obtainable
information [nats] and energy efficiency [nats/J] defined as the
ratio of total gathered information to the required energy is
obtained as the coverage area, node density and energy vary. When
the sensor node density is fixed, the energy efficiency decreases
to zero with rate $\Theta\left( \mbox{area}^{-1/2} \right)$ and
the per-node information under fixed per-node energy also
diminishes to zero with rate $O(N_t^{-1/3})$ as the number $N_t$
of network nodes increases by increasing the coverage area. As the
sensor spacing $d_n$ increases, the per-node information converges
to its limit $D$ with rate $D-\sqrt{d_n}e^{-\alpha d_n}$ for a
given diffusion rate $\alpha$. When the coverage area is fixed and
the node density increases, the per-node information is inversely
proportional to the node density. As the total energy $E_t$
consumed in the network increases, the total information
obtainable from the network is given by $O\left(\log E_t \right)$
for the fixed node density and fixed coverage case and by $\Theta
\left(E_t^{2/3} \right)$  for the fixed  per-node sensing energy
and fixed density and increasing coverage case.

\end{abstract}

\section{Introduction}  \label{sec:intro}

Sensor networks have drawn much attention in recent years because
of their promising applications such as scientific research,
environmental monitoring, and
surveillance\cite{Estrin&etal:02PerComp}. In the design of sensor
networks, there are several distinctive features. First, sensor
networks are designed to sense and monitor various physical
  phenomena such as temperature, humidity, density of a certain gas
or stress level of different locations in a structure. Many of
these physical processes  can be modelled as two-dimensional (2-D)
random fields over a certain area, where the uncertainty of the
underlying signal is captured as the randomness of samples and the
proximity of samples close in location is modelled by the
correlation among the samples. Second, sensors in different
locations should be able to deliver the measured data to a control
center (or fusion center) where the decision is made, and thus the
communication capability is required  as in {\em ad hoc}
communication networks. Such communication functionality can be
provided by networking sensor nodes, for example, using multi-hop
routing. Third, energy is one of the critical issues in sensor
network design since both sensing and communication require energy
and it is difficult to recharge batteries in already deployed
sensor nodes. Hence, it is of interest to design energy efficient
sensor networks.
\begin{figure}[htbp]
\centerline{
    \begin{psfrags}
    \psfrag{info}[c]{\textsf{Information}}  %
    \psfrag{sn}[l]{\textsf{Sensor network}} %
    \psfrag{phy}[c]{\textsf{Physical process}} %
    \psfrag{un}[c]{\textsf{(Uncertainty)}} %
    \scalefig{0.65}\epsfbox{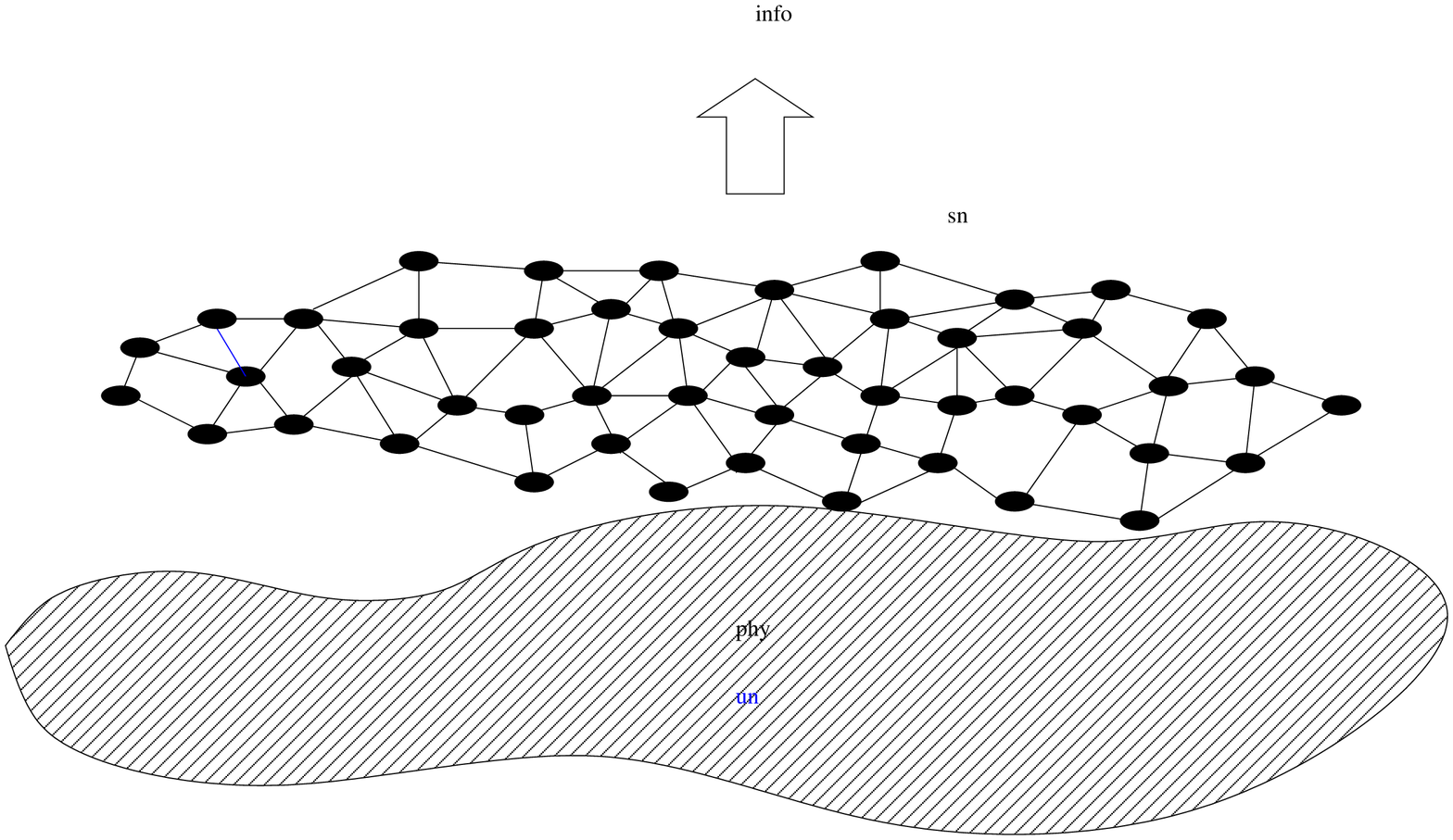}
    \end{psfrags}
} \caption{{\em Ad hoc} sensor network over physical process}
\label{fig:introduction}
\end{figure}

In this paper, we consider the design of such sensor networks, and
investigate the behavior and efficiency of these networks  from an
information-theoretic perspective.  From the information-theoretic
viewpoint, the process of
 sensing and communication mentioned above can be viewed as
extracting information (about the underlying 2-D physical process)
using imperfect sensor nodes by expending energy for statistical
inference such as detection or reconstruction of the sensed signal
field \cite{Chamberland&Veeravalli:06IT, Dong&Tong&Sadler:06SP},
as shown in Fig. \ref{fig:introduction}. Relevant questions
regarding the network design are as follows. How much information
can one obtain from the network for given coverage and node
density? How does the amount of gathered information change as we
increase the coverage area or node density?  How do the field
correlation and measurement signal-to-noise (SNR) affect the
amount of information obtainable from the network?  What is the
optimal node density? What are the information and energy
trade-offs in such a sensor network with {\em ad hoc} routing?
Answering these questions is difficult, especially, because of the
2-D spatial correlation structure of the signal process inherent
to the two dimensionality of network deployment. To circumvent
this problem, several studies based on one-dimensional (1-D)
spatial signal models have been conducted (see, e.g.,
\cite{Chamberland&Veeravalli:06IT},
\cite{Sung&Zhang&Tong&Poor:08SP},
\cite{Pillutal&Krishnamurthy:08ICC}). However, there is an
important difference between 1-D signal models and actual spatial
signals. Suppose that we take observations from sensors located
equidistantly along a line transect laid over an area. The
observations may then be viewed as samples generated by a 1-D
process along the line transect and results from time series
analysis could be applied to examine their statistical properties.
In the 2-D case, however, there is no natural notion of signal
flow or dependence direction along the transect as there is in a
more traditionally obtained time series. For samples from sensors
placed over a 2-D area, it is necessary to consider the signal
dependence in all direction in the plane.

\subsection{The Approach and Summary of Results}

In this paper, we consider {\em ad hoc} sensor networks deployed
for making statistical inferences about underlying 2-D random
fields, and address the above questions  in a general 2-D setting.
In particular, we investigate the amount of information obtainable
 from the network and related trade-offs among information, coverage, density and energy in various asymptotic
settings, and reveal the fundamental behavior of large scale
planar {\em ad hoc} sensor networks. We model the signal field as
a 2-D Gauss-Markov random field (GMRF), which is suitable for many
physical processes, and consider the Kullback-Leibler information
(KLI) and mutual information (MI) as our information measures
\cite{Liese&Vajda:06IT,Kullback:book}. Our approach for
calculating the total obtainable information is based on the large
deviations principle (LDP). Under a stationarity assumption, the
amount of information from a sensor node becomes independent of
sensor location as the network size grows, and the total amount of
information is approximately given by the product of the number of
sensor nodes and the asymptotic information rate or asymptotic
per-node information. (Thus, the units of these quantities is
nats/node.) To quantify the information content, we first derive
closed-form expressions for the asymptotic per-node KLI and MI for
stationary Gaussian fields in a general $d$-dimensional ($d$-D)
lattice in the spectral domain, and then apply these results to
the 2-D case. We do so by exploiting the spectral structure of
$d$-D stationary Gaussian signals and the relationship between the
eigenvalues of the block circulant approximation to a block
Toeplitz matrix describing the $d$-D correlation structure.
However, the general expressions obtained in this way render the
investigation of the field correlation and SNR difficult. To
address this problem, we adopt the {\em conditional autoregression
(CAR) model}, which is a generalization of the autoregressive (AR)
model of classical time series analysis. We further investigate
the properties of the asymptotic per-node KLI and MI as functions
of the field correlation and the measurement SNR under the
symmetric first order conditional autoregression (SFCAR) model,
which captures the 2-D correlation on the plane effectively. In
this case, the asymptotic per-node KLI and MI are given explicitly
in terms of the SNR and the field correlation.
 The behavior of the asymptotic per-node KLI and MI as functions of
 correlation strength is seen to divide into two regions depending on
the value of the SNR. At high SNR, uncorrelated observations
maximize the per-node information  for a given SNR, whereas there
is non-zero optimal correlation at low SNR. Interestingly, it is
 seen that there is a discontinuity in the optimal correlation
 strength as a function of SNR.  In the
 perfectly correlated case, the asymptotic per-node KLI and MI are zero as expected.
 As a function of SNR, the asymptotic per-node information increases as
 $\log \mbox{SNR}$ for a given correlation strength at high SNR.
 At low SNR, the two information measures show different rates of
 convergence to zero.

Based on the derived expressions for asymptotic per-node
information and their properties under the SFCAR and corresponding
correlation function, we then investigate the fundamental behavior
of large scale {\em ad hoc} sensor networks deployed over
correlated random fields for statistical inference. Specifically,
we examine the total information [nats] (about the underlying
physical process) obtainable from the network and the energy
efficiency [nats/J]
 defined as the ratio of total gathered
information to the required energy as the coverage, density and
energy vary.  We assume that sensors are located on a 2-D lattice
and all sensor nodes in the network deliver the measured data to a
fusion center in the center of the 2-D lattice via  minimum hop
routing {\em without in-network data fusion}. Under these
assumptions, we have the following results on the trade-offs among
the information, coverage, density and energy, and the results
provide guidelines for the design of sensor networks for
statistical inference about many interesting physical processes
that can be modelled as 2-D correlated random fields:

\begin{itemize}

\item[(1)] When the sensor node density is fixed, the amount of
total information increases linearly with respect to (w.r.t.) the
coverage area, and the energy efficiency decreases to zero with
rate $\Theta\left( \mbox{area}^{-1/2} \right)$ as the coverage
area increases. Further, in this case the amount of information
per sensor node diminishes to zero as the network size grows with
fixed energy per node.

\item[(2)]  As the sensor spacing $d_n$ increases, the per-node
information converges to its limit $D$ with rate
$D-\sqrt{d_n}e^{-\alpha d_n}$ for a given diffusion rate $\alpha$.
Hence, the per-node information saturates almost exponentially as
we increase the sensor spacing.

\item[(3)]  When the coverage area is fixed and the node density
increases, the per-node information is inversely proportional to
the node density for any nontrivial diffusion rate. Hence, the
total amount of information from a given area is upper bounded
unless the random field is spatially white.

\item[(4)] As the total energy $E_t$ consumed in the network
increases, the total information obtainable from the network is
given by $\Theta \left(E_t^{2/3} \right)$ for fixed node density
and increasing coverage, whereas the total information increases
only with rate of $O\left(\log E_t \right)$ for fixed node density
and fixed coverage.

\end{itemize}

\subsection{Related Work}

 Large deviations analysis of Gaussian
processes in Gaussian noise has been considered previously, e.g.,
\cite{Donsker&Varadhan:85CMP,Benitz&Bucklew:90IT,Bryc&Dembo:97JTP,Bercu&Gamboa&Rouault:97SPA,Sung&Tong&Poor:06IT,Leong&Dey&Evans:07SP}.
However, most work in this area considers only 1-D signals or time
series. A closed-form expression for the asymptotic KLI rate was
obtained and its properties were investigated for 1-D hidden
Gauss-Markov random processes in \cite{Sung&Tong&Poor:06IT}. Large
deviations analyses were used to examine the issues of optimal
sensor density and optimal sampling in a 1-D signal model in
\cite{Chamberland&Veeravalli:06IT} and
\cite{Sung&Zhang&Tong&Poor:08SP}. For a 2-D setting, an error
exponent was obtained for the detection of 2-D GMRFs in
\cite{Anandkumar&Tong&Swami:07ICASSP}, where the sensors are
located randomly and the Markov graph is based on the nearest
neighbor  dependency enabling a loop-free graph.   Our work here
focuses on the analysis of the  fundamental behavior of 2-D sensor
networks deployed for statistical inference via new large
deviations results for general $d$-D and 2-D stationary Gaussian
random fields and their application to 2-D SFCAR GMRFs, which
enable us to investigate the impact of field correlation and
measurement SNR on the information and the fundamental behavior of
{\em ad hoc} sensor networks for statistical inference with
preliminary presentation of the work in
\cite{Sung&Poor&Yu:08ISIT}.

\subsection{Notation and Organization}   \label{sec:notation}

We will make use of standard notational conventions.
    Vectors and matrices are written
    in boldface with matrices in capitals. All vectors are column vectors.
 For a matrix $\Abf$, $\Abf^T$ indicates the transpose and $\Abf(i,j)$ denotes the $(i,j)$-th element of $\Abf$.
 We reserve $\Ibf_m$ for
    the identity matrix of size $m$ (the subscript is included only
    when necessary).
  For a random vector $\xbf$, $\Ebb_{j}\{\xbf \}$ is
  the expectation of $\xbf$ under probability density $p_{j}, ~j=0,1$.
    The notation $\xbf\sim \Nc(\mubf,\Sigmabf)$ means that $\xbf$ is
   Gaussian distributed with mean vector $\mubf$ and
    covariance matrix $\Sigmabf$.  For a set $\Ac$,
    $|\Ac|$ denotes the cardinality of $\Ac$.

The paper is organized as follows. The background and signal model
are described in Section \ref{sec:systemmodel}. In Section
\ref{sec:InfoAnal}, the closed-form expressions for the asymptotic
KLI and MI rates are obtained in the spectral domain, and their
properties are investigated as functions of the correlation and
the SNR under the symmetric first order CAR model. The trade-offs
related to {\em ad hoc} sensor networks deployed for statistical
inference are presented in Section \ref{sec:adhocsennet},
 followed by conclusions in Section \ref{sec:conclusion}.

\section{Background and Signal Model}\label{sec:systemmodel}

\begin{figure}[htbp]
\centerline{
    \begin{psfrags}
    \psfrag{ij}[l]{{$(i,j)$}}  %
    \psfrag{xij}[c]{{$X_{ij}$}} %
    \psfrag{wij}[l]{{$W_{ij}$}} %
    \psfrag{nt}[c]{{measurement noise}} %
    \psfrag{yij}[c]{{$Y_{ij}$}} %
    \psfrag{Nij}[c]{{Sensor $ij$}} %
    \psfrag{r}[c]{{ $d_n$}}
    \psfrag{xyzxyz}[r]{{$xyzkkkk$}} %
    \scalefig{0.65}\epsfbox{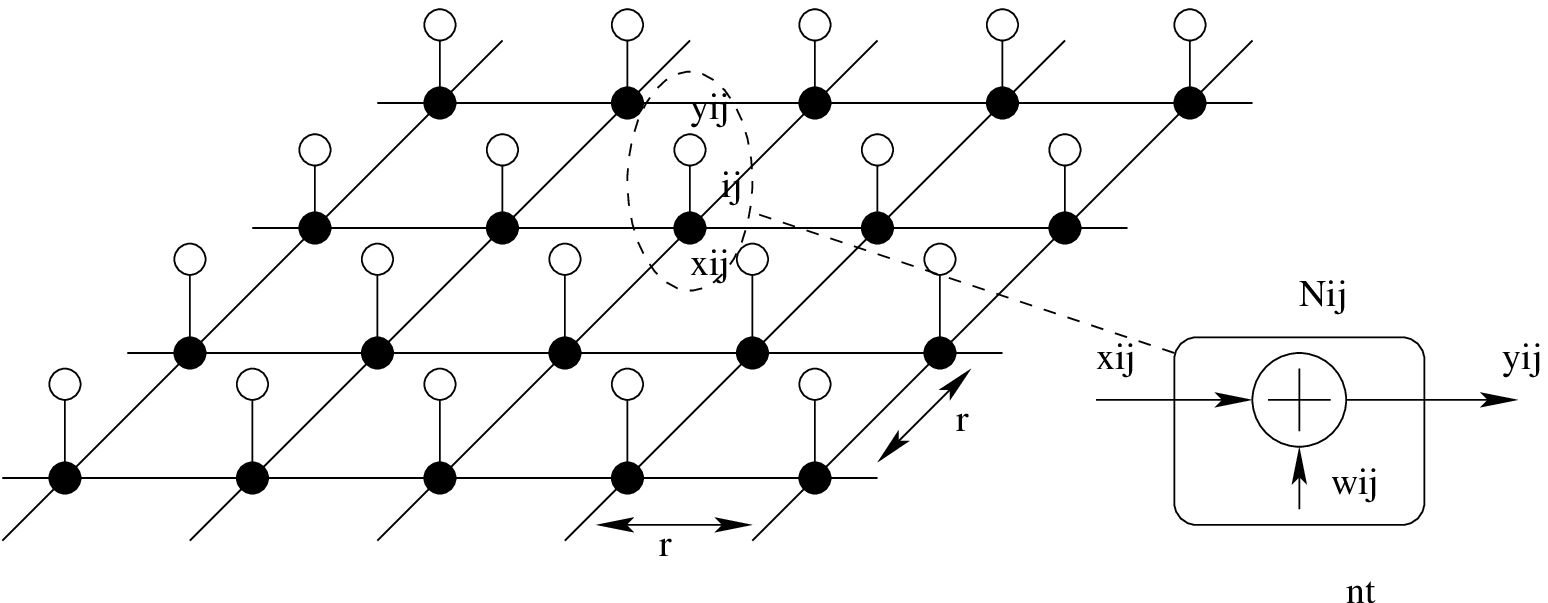}
    \end{psfrags}
} \caption{Sensors on a 2-D lattice: hidden Markov structure}
\label{fig:2dHGMRF}
\end{figure}

We assume that sensors are distributed over a 2-D area and each
sensor measures the underlying signal field at its location. To
simplify the problem and gain insights into behavior in 2-D, we
assume that sensors are located on a 2-D square lattice
\begin{equation}  \label{eq:finiteLattice}
\Ic_n \defeq \{(i,j), ~i= 0,  1,  \cdots, n-1, ~\mbox{and}~ j= 0,
 1, \cdots, n-1\},
\end{equation}
where the distance between two adjacent nodes $(i,j)$ and
$(i+1,j)$ is $d_n$, as shown in Fig. \ref{fig:2dHGMRF}. (We will
use $ij$ to denote  $(i,j)$ when there is no ambiguity of
notation.) We model the 2-D signal field $\{X_{ij}, ij \in
\Ic_n\}$ (or simply $\{X_{ij}\}$) sampled by sensors as a
GMRF\footnote{The Markov dependence structure may be restrictive.
However, it is a meaningful model capturing  2-D spatial
correlation structure and allowing further analysis.} w.r.t. an
undirected graph in which a node corresponds to a sensor node or
its signal sample.  We assume that each sensor has Gaussian
measurement noise. The noisy measurement $Y_{ij}$ of Sensor $ij$
on the 2-D lattice ${\mathcal I}_n$ is then given by
\begin{equation} \label{eq:hypothesis2d}
 Y_{ij} = X_{ij}+ W_{ij}, ~~~ij \in {\cal I}_n,
\end{equation}
where  $\{W_{ij}\}$ represents independent and identically
distributed (i.i.d.) $\Nc(0,\sigma^2)$ noise with a known variance
$\sigma^2$, and the GMRF $\{X_{ij}\}$ is assumed to be independent
of the measurement noise $\{W_{ij}\}$. Thus, the observation
samples form a 2-D hidden GMRF.\footnote{In this paper, we
 focus primarily on the spatial correlation structure of 2-D sensor
fields, and the signal evolution over time is not considered.} In
the following, we briefly review results on GMRFs relevant to our
further development.

\vspace{0.5em}
\begin{definition}[Undirected graph] \label{def:undirectGraph}
An undirected labelled graph $\Gc$ is a collection
$({\Nc},{\mathcal E})$ of nodes and edges, where
${\Nc}=\{1,2,\cdots, N\}$ is the set of nodes in the graph, and
$\Ec$ is the set of edges $\{(l,m): l, m \in {\Nc} ~\mbox{and}~ l
\ne m\}$. There exists an undirected edge between two nodes $l$
and $m$ if and only if $(l,m) \in \Ec$.
\end{definition}
\vspace{0.5em}

\noindent We will use the terms node, sample and sensor
interchangeably hereafter.

\vspace{0.5em}
\begin{definition}[GMRF]\label{def:GMRF}
A Gaussian random vector $\xbf=[X_1,X_2,\cdots,X_N]^T$ $\in
{\mathbb R}^N$ with mean vector $\mubf$ and covariance matrix
$\Sigmabf
> 0$ is a GMRF w.r.t. a labelled graph ${\mathcal G}=({\mathcal
N},{\mathcal E})$ if  $X_l$ and $X_m$ are independent given
$\Xc_{-lm}$ if and only if there exists no edge between nodes $l$
and $m$,
 where $\Xc_{-lm} \defeq \{X_k, k \in \Nc ~\mbox{and}~ k \ne l,
m\}$.
\end{definition}
\vspace{0.5em}

\noindent Note that a GMRF is defined using conditional
independence on a graph. However, its distribution is easily
characterized by the mean $\mubf$ and the precision matrix $\Qbf$
($\defeq \Sigmabf^{-1}$), and is given by
\begin{equation}
p(\xbf) = (2\pi)^{-N/2}|\Qbf|^{1/2}\exp\left( - \frac{1}{2}
(\xbf-\mubf)^T \Qbf (\xbf-\mubf) \right),
\end{equation}
and $Q_{lm} \ne 0$ if and only if $(l,m) \in {\mathcal
E}~\mbox{for all}~ l \ne m$, i.e.,
\begin{equation}
Q_{lm} = 0 \Longleftrightarrow X_l \perp X_m|\Xc_{-lm}.
\label{eq:MarkovStructQ}
\end{equation}
Note that the covariance matrix $\Sigmabf$ is completely dense in
general while the precision matrix $\Qbf$ has nonzero elements
$Q_{lm}$ only when there is an edge between nodes $l$ and $m$ in
the Markov random field. Hence, when the graph is not fully
connected, the precision matrix is sparse \cite{Rue&Held:book}.
The 2-D indexing scheme $(i,j)$ in (\ref{eq:finiteLattice}) and
(\ref{eq:hypothesis2d}) can properly be converted to a 1-D scheme
to apply Definitions \ref{def:undirectGraph} and \ref{def:GMRF}.
From here on, we again use the 2-D indexing scheme for
convenience.

\vspace{0.5em}
\begin{definition}[Stationarity]
A GMRF $\{X_{ij}\}$  on a 2-D infinite lattice ${\mathcal
I}_\infty$ is said to be (second order) {stationary} if the mean
vector is constant and the covariance between samples $X_{ij}$ and
$X_{i^\prime j^\prime}$ depends only on the difference of the node
index, i.e.,
\[
\mbox{Cov}( X_{ij}, X_{i^\prime j^\prime}) =  \Ebb \{(X_{ij}-\mu)
(X_{i^\prime j^\prime}-\mu)\}= c(i-i^\prime, j-j^\prime)
\]
for some function $c(\cdot, \cdot)$, where $\mu$ is the mean of
the stationary field.
\end{definition}
\vspace{0.5em}

\noindent Without loss of generality, we assume that the signal
GMRF $\{X_{ij}\}$ is zero-mean.\footnote{Of course, if a
stationary GMRF has a known and non-zero mean,  the known mean can
be subtracted to yield a zero-mean field.}
  For a 2-D zero-mean and stationary GMRF
$\{X_{ij}\}$, the covariance $\{\gamma_{ij}\}$ is defined as
\begin{equation}  \label{eq:2Dautocovariancefunction}
\gamma_{ij} \defeq  \Ebb \{ X_{i^\prime j^\prime} X_{i^\prime+i,
j^\prime +j}\} =\Ebb \{ X_{00} X_{ij}\},
\end{equation}
which does not depend on $i^\prime$ or $j^\prime$ due to the
stationarity. The spectral density function of a stationary GMRF
$\{X_{ij}\}$ on $\Ic_\infty$ with covariance $\gamma_{ij}$ is
defined as
\begin{equation}  \label{eq:2DDTFT}
f(\omega_1,\omega_2) =  \frac{1}{(2\pi)^2}\sum_{ij \in \Ic_\infty}
\gamma_{ij} e^{-\iota(i\omega_1 + j\omega_2) },
\end{equation}
where $\iota = \sqrt{-1}$ and $(\omega_1,\omega_2) \in
[-\pi,\pi)^2$. Note that (\ref{eq:2DDTFT}) is a 2-D extension of
the conventional 1-D Fourier transform.  We can express
$\{\gamma_{ij}\}$ from the spectral density function via the
inverse transform
\begin{equation}  \label{eq:2DIDTFT}
\gamma_{ij} = \int_{-\pi}^\pi  \int_{-\pi}^\pi
f(\omega_1,\omega_2)e^{\iota(i\omega_1 + j\omega_2) } d\omega_1
d\omega_2.
\end{equation}

A stationary GMRF can be implicitly specified by a conditional
autoregressive (CAR) model, which is a natural generalization of
the autoregressive (AR) model arising in 1-D time series and which
provides an efficient tool for capturing the spatial correlation
structure of the sensor field considered here.

\vspace{0.5em}
\begin{definition}[The conditional autoregression \cite{Rue&Held:book}]
A zero-mean CAR GMRF is defined by a set of full conditional
normal distributions with mean and precision:
\begin{equation}
\Ebb \{ X_{ij}|\Xc_{-ij}\} =  -\frac{1}{\theta_{00}}
\sum_{i^\prime j^\prime \in \Ic_\infty \backslash \{00\}}
\theta_{i^\prime j^\prime} X_{i+i^\prime,j+j^\prime},
\label{eq:condMean2DInf}
\end{equation}
and
\begin{equation}
 \Ebb^{-1}\{X_{ij}^2|\Xc_{-ij}\} = \theta_{00}
 > 0, \label{eq:condPrec2DInf}
\end{equation}
where $\Xc_{-ij}$ denotes the set of all variables except
$X_{ij}$.
\end{definition}
\vspace{0.5em}

\noindent  Note in (\ref{eq:condMean2DInf}) that the the
conditional mean of $X_{ij}$ given all other node variables
depends on nodes $(i+i^\prime,j+j^\prime)$ such that
$\theta_{i^\prime j^\prime} \ne 0$, and  the relationship between
the CAR model of (\ref{eq:condMean2DInf}) and
(\ref{eq:condPrec2DInf}) and the precision matrix is given by
\begin{equation}
\Qbf_{(i,j),(i+i^\prime,j+j^\prime)} = \theta_{i^\prime j^\prime}.
\end{equation}
Hence, the Markov dependence structure on the graph is easily
captured by the CAR model through  (\ref{eq:MarkovStructQ}), and
$\{\theta_{i^\prime j^\prime}\}$ directly represent the
connectivity of the Markov graph.

\vspace{0.5em}
\begin{theorem}[Spectrum of a CAR model \cite{Rue&Held:book}]
\label{theo:CARspectrum} The GMRF defined by the CAR model of
(\ref{eq:condMean2DInf}) and (\ref{eq:condPrec2DInf}) is a
zero-mean stationary Gaussian process on ${\mathcal I}_\infty$
with the spectral density function
\begin{equation}  \label{eq:CARspectrum}
f(\omega_1,\omega_2) =  \frac{1}{(2\pi)^2} \frac{1}{\sum_{ij \in
{\mathcal I}_\infty} \theta_{ij} \exp(-\iota (i\omega_1 +
j\omega_2))},
\end{equation}
 if
\begin{equation}
|\{\theta_{ij} \ne 0\}| < \infty, ~~~~ \theta_{ij} =
\theta_{-i,-j}, ~~~~ \theta_{00} >0, \label{eq:CARcond1}
\end{equation}
and
\begin{equation}
 \{\theta_{ij}\} ~\mbox{is such that}~
f(\omega_1,\omega_2)>0, ~~~ \forall (\omega_1,\omega_2) \in
[-\pi,\pi)^2. \label{eq:CARcond4}
\end{equation}
\end{theorem}
\vspace{0.5em}

\noindent Henceforth, we assume that the 2-D stochastic signal
$\{X_{ij}\}$ in (\ref{eq:hypothesis2d}) is given by a stationary
GMRF defined by the CAR model of  (\ref{eq:condMean2DInf}) and
(\ref{eq:condPrec2DInf}) satisfying  (\ref{eq:CARcond1}) and
(\ref{eq:CARcond4}) as $n\rightarrow \infty$.

The SNR of the observation $Y_{ij}$ in (\ref{eq:hypothesis2d}) is
well defined due to the stationarity as $n\rightarrow\infty$, and
is given by
\begin{equation} \label{eq:SNRorg}
 \mbox{SNR} = \frac{\Ebb\{X_{ij}^2\}}{\Ebb\{W_{ij}^2\}}=
 \frac{P}{\sigma^2}, ~~~\forall~ ij,
\end{equation}
where the signal power is constant  over $(i,j)\in \Ic_\infty$ and
is given, using the inverse Fourier transform of
(\ref{eq:2DDTFT}), by
\begin{eqnarray}
P &=& \gamma_{00} = \int_{-\pi}^\pi \int_{-\pi}^\pi
f(\omega_1,\omega_2) d\omega_1 d\omega_2.
\end{eqnarray}

\section{Asymptotic Information Rates: Closed-Form Expressions and Impact of Correlation and Signal-to-Noise
Ratio} \label{sec:InfoAnal}

In this section, we derive closed-form expressions for the
asymptotic KLI and MI rates under the 2-D CAR GMRF model discussed
in the previous section. We further investigate the properties of
the asymptotic information rates under a symmetric correlation
assumption. For the MI, the signal model (\ref{eq:hypothesis2d})
is directly applicable, whereas for the KLI the probability
density functions of the null (noise-only) and alternative
(signal-plus-noise) distributions are given  by
\begin{eqnarray}
p_0 (Y_{ij}) &:&  Y_{ij} = W_{ij} , ~~ij \in {\cal I}_n, ~~~~~~~\mbox{and}\label{eq:hypothesis2da1}\\
p_1(Y_{ij})  &:& Y_{ij} = X_{ij}+ W_{ij}, ~~ij \in {\cal I}_n,
\label{eq:hypothesis2da2}
\end{eqnarray}
respectively.  The asymptotic KLI rate $\Kmsc$ is defined as
\begin{equation}  \label{eq:defaKLI}
\Kmsc = \lim_{n \rightarrow \infty} \frac{1}{|\Ic_n|} \log
\frac{p_{0}}{p_{1}}(\{Y_{ij}, ij \in \Ic_n\}) ~~~\mbox{almost
surely (a.s.) under}~p_{0},
\end{equation}
where $p_0$ and $p_1$ are given by (\ref{eq:hypothesis2da1}) and
(\ref{eq:hypothesis2da2}), respectively.  Under a Neyman-Pearson
detection formulation, the miss probability $P_M$ decays
exponentially in many cases, including (\ref{eq:hypothesis2da1})
and (\ref{eq:hypothesis2da2}), and the error exponent is defined
as the exponential decay rate
\begin{equation}
 \lim_{|\Ic_n| \rightarrow \infty} - \frac{1}{|\Ic_n|} \log P_M,
\end{equation}
where $|\Ic_n|$ is the total number of samples in $\Ic_n$. It is
known that the error exponent is given by the asymptotic KLI rate
$\Kmsc$  defined in (\ref{eq:defaKLI}) in this case
\cite{Bahadur&Zabell&Gupta:80book}. Hence, a larger KLI rate (or
per-node KLI) implies better detection performance with a given
network size,  or a smaller network size required for a given
level of performance.

While the asymptotic KLI rate determines the error exponent for
Neyman-Pearson detection,  the asymptotic MI rate is interpreted
as the amount of uncertainty reduction about the hidden signal
field resulting from one observation sample, in the large sample
size regime. The asymptotic MI rate $\Imsc$ is given by
\begin{eqnarray}
\Imsc &=& \lim_{n \rightarrow \infty} \frac{1}{|\Ic_n|}
I(\{X_{ij},
ij \in \Ic_n\};\{Y_{ij}, ij\in \Ic_n\}), \nonumber\\
&=& \lim_{n \rightarrow \infty} \frac{1}{|\Ic_n|} [ H(\{X_{ij}, ij
\in \Ic_n\}) - H( \{X_{ij}, ij \in \Ic_n\}|\{ Y_{ij}, ij \in
\Ic_n\} )].
\end{eqnarray}
It is shown in the sequel that the asymptotic KLI rate is smaller
than the asymptotic MI rate and that the two information measures
converge when SNR increases. Thus, at high SNR the two information
measures are equivalent.

\subsection{Asymptotic Information Rates in General $d$-Dimension}

While the 2-D results are relevant to our analysis of fundamental
trade-offs in planar sensor networks, it is of theoretical
interest to investigate the statistical properties of stationary
Gaussian random fields in general higher dimension. In this
section, we first derive closed-form expressions for the
asymptotic KLI and MI rates for stationary Gaussian random fields
in $d$-D, and then apply the results to the 2-D case. For a
stationary $d$-D Gaussian random field $\{Y_\ibf, ~\ibf \in
\Zbb^d\}$, where  ${\mathbb Z}$ is the set of all integers, the
autocovariance function under $p_1$ is given by
\begin{equation}  \label{eq:AutoCordD}
\gamma_\hbf = \Ebb_1 \{ Y_{\ibf} Y_{\ibf + \hbf} \}, ~~~\hbf =
(h_1,h_2,\cdots, h_d) \in {\mathbb Z}^d,
\end{equation}
and the corresponding Fourier transform (i.e., the power spectral
density) and its inverse
 are given by
\begin{equation}  \label{eq:SpecdD}
f_1(\omegabf) = \frac{1}{(2\pi)^d} \sum_{\hbf \in \Zbb^d}
\gammabf_\hbf e^{- \iota \hbf \cdot \omegabf}, ~~~~\omegabf =
(\omega_1, \omega_2,\cdots, \omega_d) \in [-\pi,\pi)^d,
\end{equation}
and
\begin{equation}  \label{eq:IDTFTdD}
\gammabf_\hbf = \int e^{\iota \hbf \cdot \omegabf} f_1(\omegabf)
d\omegabf,
\end{equation}
respectively, where the integration is over $\omegabf \in [-\pi,
\pi)^d$, and  $\hbf \cdot \omegabf$ denotes the inner product
between $\hbf$ and $\omegabf$.  Note that (\ref{eq:AutoCordD}),
(\ref{eq:SpecdD}) and (\ref{eq:IDTFTdD}) are the extensions of
(\ref{eq:2Dautocovariancefunction}), (\ref{eq:2DDTFT}) and
(\ref{eq:2DIDTFT}), respectively, to $d$-D. The null and
alternative distributions arising in the  KLI in $d$-D are given
by
\begin{equation}  \label{eq:dDnullalter}
\left\{ \begin{array}{lcl}
p_0(Y_{\ibf}) &:& Y_{\ibf} = W_{\ibf} , ~~\ibf \in {\cal D}_n,\\
p_1(Y_{\ibf}) &:& Y_{\ibf} = {Y}_{\ibf}^{(1)}, ~~\ibf \in {\cal
D}_n,
\end{array}
\right.
\end{equation}
where  $\{W_\ibf\}$ are i.i.d. Gaussian from $\Nc(0,\sigma^2)$,
$\{{Y}_{\ibf}^{(1)}\}$ is a stationary $d$-D Gaussian random field
with spectrum $f_1(\omegabf)$\footnote{Note that
$\{{Y}_{\ibf}^{(1)}\}$ need not be a hidden Markov field.}, and
\begin{equation}
\Dc_n \defeq [0,1,\cdots, n-1]^d.
\end{equation}

Based on the previous work \cite{Kent&Mardia:96JSPI}, we further
exploit the relationship between the eigenvalues of block
circulant and block Toeplitz matrices representing correlation
 structure in $d$-D and the i.i.d. null distribution, and  obtain the KLI for (\ref{eq:dDnullalter}) given by the following
 theorem.

\vspace{0.5em}
\begin{theorem}[Asymptotic KLI rate in $d$-D]  \label{theo:KLI}  Suppose that
\begin{itemize}
\item[A.1] the alternative spectrum $f_1(\omegabf)$ has a positive
lower bound, and %
\item[A.2]  $\exists ~M < \infty ~~\mbox{such that} ~~\forall ~ k
= 1,2,\cdots, d, ~~ \sum_{\hbf \in {\mathbb Z}^d}
(1+|h_k|)|\gamma_\hbf| < M.$
\end{itemize}
 Then, the
asymptotic KLI rate $\Kmsc$ for (\ref{eq:dDnullalter}) is given by
\begin{eqnarray}
\Kmsc &=& \frac{1}{(2\pi)^d} \int_{[-\pi,\pi)^d} \biggl[
\frac{1}{2}\log \frac{(2\pi)^d f_1(\omegabf)}{\sigma^2}
 -\frac{1}{2}\left(1 -
\frac{\sigma^2}{(2\pi)^d f_1(\omegabf)} \right)
\biggr]d\omegabf \label{eq:errorexponentspectraldD}\\
&=&\frac{1}{(2\pi)^d} \int_{[-\pi,\pi)^d}
D(\Nc(0,\sigma^2)||\Nc(0,(2\pi)^df_1(\omegabf))
  )d\omegabf, \label{eq:errorexponentspectraldD2}
\end{eqnarray}
 where   $D(\cdot||\cdot)$ denotes the Kullback-Leibler distance.
\end{theorem}

\vspace{0.5em}  {\em Proof:} See Appendix I. \vspace{0.5em}

\noindent Theorem \ref{theo:KLI} is an extension to general $d$-D
of the asymptotic KLI rate in 1-D obtained in
\cite{Sung&Tong&Poor:06IT}, and  shows that the frequency binning
interpretation of (\ref{eq:errorexponentspectraldD2}) holds in the
general $d$-D case under some regularity conditions on the
alternative spectrum. Note that the integrand in
(\ref{eq:errorexponentspectraldD2}) is the Kullback-Leibler
information between two zero-mean Gaussian distributions with
variances $\sigma^2$ and $(2\pi)^d f_1(\omegabf)$, respectively.
For each $d$-D frequency segment $d\omegabf$, the spectra can be
thought of as being flat, i.e., the signals are independent, and
Stein's lemma \cite{Bahadur:71SIAM} can be applied for the
segment. The overall KLI is the sum of contributions from each
bin. The smoothness of the spectrum $f_1(\omegabf)$ is a
sufficient condition for Assumption {\textit{A.2}} for
second-order stationary fields, and thus the frequency binning in
Theorem \ref{theo:KLI} is valid for a wide class of spectra.
Theorem \ref{theo:KLI} follows from the fact that $\Kmsc$ is given
by the almost-sure limit of the normalized log-likelihood ratio in
(\ref{eq:defaKLI}) and that we have Gaussian distributions for
$p_0$ and $p_1$. That is, $\Kmsc$ is given by the almost sure
limit
\begin{equation} \label{eq:LLRdecomp}
\Kmsc = \lim_{n\rightarrow \infty} \frac{1}{|\Dc_n|} \left(
\frac{1}{2}\log
\frac{\det(\Sigmabf_{1,|\Dc_n|})}{\det(\Sigmabf_{0,|\Dc_n|})} +
\frac{1}{2} \ybf_{|\Dc_n|}^T (\Sigmabf_{1,|\Dc_n|}^{-1} -
\Sigmabf_{0,|\Dc_n|}^{-1})\ybf_{|\Dc_n|}\right) ~~\mbox{under}~
p_{0},
\end{equation}
where $\ybf_{|\Dc_n|}$ is a vector consisting of $|\Dc_n|$
observation samples $\{Y_{\ibf}, \ibf \in \Dc_n\}$ with elements
arranged in lexicographic order; for example, in  2-D
\begin{equation}  \label{eq:2Dto1DconversionVector}
\ybf_{|\Ic_n|} = [y_1,\cdots,y_{|\Ic_n|}]^T \defeq [
Y_{00},Y_{10}, \cdots, Y_{n-1,0}, Y_{01},\cdots, Y_{n-1,n-1} ]^T,
\end{equation}
and $\Sigmabf_{0,|\Dc_n|}$ and $\Sigmabf_{1,|\Dc_n|}$ are the
covariance matrices of $\ybf_{|\Dc_n|}$ under $p_{0}$ and $p_{1}$,
respectively. Note that the log-likelihood ratio in
(\ref{eq:LLRdecomp}) consists of two terms: one is a deterministic
term and the other is a quadratic random term. The overall
convergence follows from the convergence of each of the two terms.
Note that the deterministic term in (\ref{eq:LLRdecomp}) is simply
the mutual information between $\{X_{\ibf}, \ibf \in \Dc_n\}$ and
$\{Y_{\ibf}, \ibf \in \Dc_n\}$ for the model
\begin{equation}
Y_\ibf = X_\ibf + W_\ibf, ~~ \ibf \in \Dc_n.
\end{equation}
Using the convergence of the first term in the right-handed side
(RHS) of (\ref{eq:LLRdecomp}), the asymptotic MI rate $\Imsc$ for
$d$-D is given by
\begin{equation}
\Imsc = \frac{1}{(2\pi)^d} \int_{[-\pi, \pi)^d} \frac{1}{2}\log
\frac{\sigma^2+(2\pi)^df(\omegabf)}{\sigma^2}
  d\omegabf,\label{eq:persensormutualinformationdD}\\
\end{equation}
where $f(\omegabf)$ is the spectrum of the signal $\{X_\ibf\}$.
This is simply a $d$-D extension of the 1-D MI rate in spectral
form \cite{Cover&Thomas:book}, and shows the validity of the log
(1+ SNR) formula and frequency binning approach in general $d$-D
under some regularity conditions on the spectrum; a sufficient
condition is provided in Theorem \ref{theo:KLI}.

Applying the $d$-D results  to the 2-D hidden GMRF model of
(\ref{eq:hypothesis2da1}) and (\ref{eq:hypothesis2da2}), we have
the following corollary for 2-D.

\vspace{0.5em}
\begin{corollary}[Asymptotic information rates in 2-D]  \label{cor:KLI2D}  Assuming that the conditions
 (\ref{eq:CARcond1}) and (\ref{eq:CARcond4}) hold, the asymptotic KLI and MI rates for the hidden CAR GMRF model
 with (\ref{eq:hypothesis2da1}) and (\ref{eq:hypothesis2da2})  are given by
\begin{eqnarray}
\Kmsc &=& \frac{1}{4\pi^2} \int_{-\pi}^{\pi} \int_{-\pi}^{\pi}
\biggl[ \frac{1}{2}\log
\frac{\sigma^2+4\pi^2f(\omega_1,\omega_2)}{\sigma^2}
 -\frac{1}{2}\left(1 -
\frac{\sigma^2}{\sigma^2+4\pi^2f(\omega_1,\omega_2)} \right)
\biggr]d\omega_1d\omega_2,\label{eq:errorexponentspectral}
\end{eqnarray}
and
\begin{equation}
\Imsc = \frac{1}{4\pi^2} \int_{-\pi}^{\pi} \int_{-\pi}^{\pi}
\frac{1}{2}\log \frac{\sigma^2+ 4\pi^2
f(\omega_1,\omega_2)}{\sigma^2}
  d\omega_1 d\omega_2,\label{eq:persensormutualinformation}
\end{equation}
where   $f(\omega_1,\omega_2)$ is the 2-D spectrum of the signal
GMRF $\{X_{ij}, ij \in \Ic_\infty\}$ defined in
(\ref{eq:CARspectrum}).
\end{corollary}

\vspace{0.5em}  {\em Proof:} See Appendix I. \vspace{0.5em}

Comparing (\ref{eq:errorexponentspectral}) and
(\ref{eq:persensormutualinformation}), we note that the asymptotic
KLI rate is strictly less than the asymptotic MI rate for any
positive signal spectrum, and that the two information measures
converge with a fixed offset of -1/2 as the SNR increases without
bound since $\frac{\sigma^2}{\sigma^2+4\pi^2 f(\omega_1,\omega_2)}
\rightarrow 0$ in (\ref{eq:errorexponentspectral}) as $\SNR
\rightarrow \infty$. Hence, the two information measures can be
equivalently used at high SNR.

\subsection{Symmetric First-Order Conditional Autoregression}

In the previous section, we have derived closed-form expressions
for the asymptotic KLI and MI rates for  hidden CAR GMRFs with
general 2-D spectra defined in (\ref{eq:CARspectrum}) in the
spectral domain. However, these general spectral expressions
render further analysis infeasible. To investigate the impact of
the field correlation and the SNR on the information rates, we
further adopt the symmetric first order conditional autoregression
(SFCAR) model, described by the conditions
\begin{equation}
\Ebb \{ X_{ij}|\Xc_{-ij}\} =  \frac{\lambda}{\kappa}
(X_{i+1,j}+X_{i-1,j}+X_{i,j+1}+X_{i,j-1}), \label{eq:SFCARmean}
\end{equation}
and
\begin{equation}
 \Ebb^{-1}\{X_{ij}^2|\Xc_{-ij}\} = \kappa >
0, \label{eq:SFCARprec}
\end{equation}
where $0 \le \lambda \le \frac{\kappa}{4}$.\footnote{This is a
sufficient condition to satisfy (\ref{eq:CARcond1}) and
(\ref{eq:CARcond4}).}  Note that the parameters in
(\ref{eq:condMean2DInf}) and (\ref{eq:condPrec2DInf}) for this
model are given by $\theta_{00}=\kappa$, $\theta_{1,0} =
\theta_{-1,0} = \theta_{0,1} = \theta_{0,-1} = -\lambda$ and all
other $\theta_{ij} = 0$.
\begin{figure}[htbp]
\centerline{
    \begin{psfrags}
    \psfrag{m1}[c]{{ $-\lambda$}} %
    \psfrag{4d}[c]{{ $\kappa$}} %
    \psfrag{1}[l]{{ $(i,j)$}} %
    \psfrag{2}[c]{{ $(i,j+1)$}} %
    \psfrag{3}[r]{{ $(i-1,j)$}} %
    \psfrag{4}[l]{{ $(i+1,j)$}} %
    \psfrag{5}[c]{{ $(i,j-1)$}} %
    \scalefig{0.30}\epsfbox{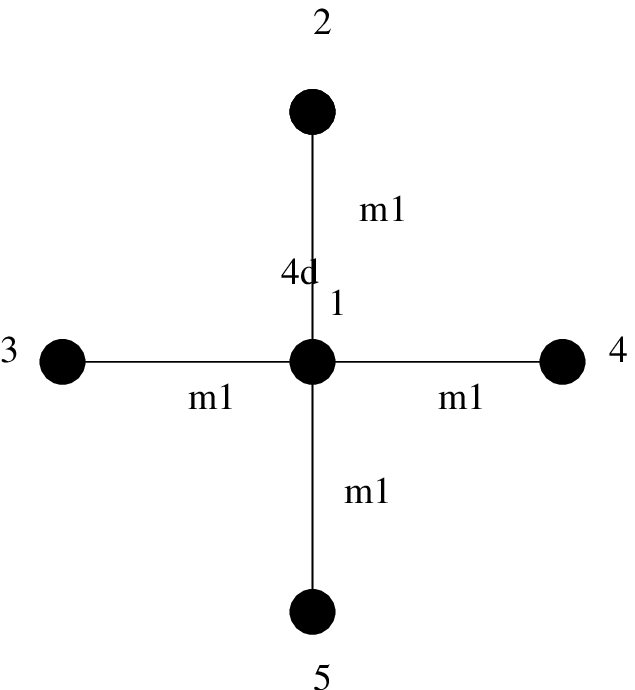}
    \end{psfrags}
}\caption{Symmetric first order conditional autoregression model}
\label{fig:2DSFCAR}
\end{figure}
In this model, the correlation is symmetric for each set of four
neighboring nodes, as seen in Fig. \ref{fig:2DSFCAR}.
 The SFCAR model is a simple yet
meaningful extension of the 1-D first order autoregression (AR)
model which has the conditional causal dependency only on the
previous sample.  Here in the 2-D SFCAR  we have the conditional
dependency on four neighboring nodes in the four (planar)
directions. By Theorem \ref{theo:CARspectrum} the spectrum of the
SFCAR is given by
\begin{equation} \label{eq:SFCARspectrum}
f(\omega_1,\omega_2) = \frac{1}{4\pi^2 \kappa (1 - 2 \zeta
\cos\omega_1 - 2 \zeta \cos\omega_2)},
\end{equation}
where we define the {\em edge dependence factor} $\zeta$ as
\begin{equation} \zeta
\defeq \frac{\lambda}{\kappa}, ~~~~ 0 \le \zeta \le 1/4.
\end{equation}
Note that for the range of $0 \le \zeta \le 1/4$ the 2-D spectrum
(\ref{eq:SFCARspectrum}) is always non-negative and the conditions
(\ref{eq:CARcond1}) and (\ref{eq:CARcond4}) are satisfied. Note
also that $\zeta =0$ corresponds to the i.i.d. case whereas $\zeta
=1/4$ corresponds to the perfectly correlated case, i.e., $X_{ij}
= X_{i^\prime j^\prime}$ for all $i,j,i^\prime, j^\prime$.  Hence,
the correlation strength can be captured in this single quantity
$\zeta$ for
 2-D SFCAR signals: larger $\zeta$ implies stronger correlation.
  The power of the SFCAR signal is obtained using the
inverse Fourier transform via the relation (\ref{eq:2DDTFT}), and
is given by \cite{Besag:81JRSS}
\begin{equation}  \label{eq:gamma00}
P = \gamma_{00} =  \frac{2K(4\zeta)}{\pi \kappa}, ~~~\left(0 \le
\zeta \le \frac{1}{4} \right),
\end{equation}
where $K(\cdot)$ is the complete elliptic integral of the first
kind. The SNR is given by
\begin{equation} \label{eq:SNR}
 \mbox{SNR} = \frac{P}{\sigma^2} =
\frac{2K(4\zeta)}{\pi \kappa \sigma^2}.
\end{equation}
Using  (\ref{eq:errorexponentspectral}), (\ref{eq:SFCARspectrum})
and (\ref{eq:SNR}), we now obtain the asymptotic KLI and MI rates
in the SCFAR signal case, denoted by $\Kmsc_s$ and $\Imsc_s$ and
given in the following corollary to Corollary \ref{cor:KLI2D}.

\vspace{0.5em}
\begin{corollary} \label{corol:eeSFA}
For the hidden 2-D SFCAR signal model the asymptotic per-node KLI
$\Kmsc_s$  is given by
\begin{eqnarray}
\Kmsc_s &=& \frac{1}{4\pi^2} \int_{-\pi}^{\pi} \int_{-\pi}^{\pi}
\biggl[ \frac{1}{2}\log \left(1+\frac{ \mbox{SNR}}{
(2/\pi)K(4\zeta) (1 - 2 \zeta \cos\omega_1 - 2 \zeta
\cos\omega_2)}\right) \nonumber\\
&& ~~~~~~~~~~~~~~~~~~~~~~~~~~~~~-\frac{1}{2} \left( 1-
\frac{1}{1+\frac{ \mbox{SNR}}{ (2/\pi)K(4\zeta) (1 - 2 \zeta
\cos\omega_1 - 2 \zeta \cos\omega_2)}} \right) \biggl]
d\omega_1d\omega_2, \label{eq:errorexponentSFA}
\end{eqnarray}
and the asymptotic per-node MI $\Imsc_s$ is given by
\begin{equation}
\Imsc_s = \frac{1}{4\pi^2} \int_{-\pi}^{\pi} \int_{-\pi}^{\pi}
\frac{1}{2}\log \left(1+\frac{ \mbox{SNR}}{ (2/\pi)K(4\zeta) (1 -
2 \zeta \cos\omega_1 - 2 \zeta \cos\omega_2)}
\right)d\omega_1d\omega_2. \label{eq:SFCARMI}
\end{equation}
\end{corollary}
\vspace{1em}

{\em Proof:} The result follows upon substitution of
(\ref{eq:SFCARspectrum}) and (\ref{eq:SNR})
 into (\ref{eq:errorexponentspectral}) and
(\ref{eq:persensormutualinformation}), respectively.

\hfill{$\blacksquare$}

\vspace{0.5em}

\noindent Note that the SNR for the hidden SFCAR model is
dependent on correlation through $\zeta$ (see (\ref{eq:SNR})).
However, the SNR and correlation are separated in the expressions
(\ref{eq:errorexponentSFA}) and (\ref{eq:SFCARMI}) for the
asymptotic per-node information, which enables us to investigate
the effects of each term on the per-sample information separately.

\subsubsection{Properties of the asymptotic per-node KLI and MI
 for the hidden SFCAR model}\label{subsec:errorexponent2D}

First, it is readily seen from Corollary \ref{corol:eeSFA} that
the asymptotic per-node KLI $\Kmsc_s$ and MI $\Imsc_s$ are
continuously differentiable functions  of the edge dependence
factor $\zeta$ ($0 \le \zeta \le 1/4$) for a given SNR since $f: x
\rightarrow K(x)$ is a continuously differentiable $C^\infty$
function for $0 \le x < 1$ \cite{Erdelyi:53book}. Now we examine
the asymptotic behavior of $\Kmsc_s$ and $\Imsc_s$ as functions of
$\zeta$. The values of $\Kmsc_s$ at the extreme correlations are
given by noting that the values of the complete elliptic integral
at the two extreme correlation points
\[
K(0) = \frac{\pi}{2} ~~ \mbox{and} ~~  K(1)= \infty.
\]
Therefore, in the i.i.d. case (i.e., $\zeta =0$), Corollary
\ref{corol:eeSFA} reduces to Stein's lemma\cite{Bahadur:71SIAM} as
expected, and $\Kmsc_s$ is given by
\begin{eqnarray}
\Kmsc_s(0) &=& \frac{1}{2} \log (1+ \mbox{SNR}) - \frac{1}{2}
\left( 1 - \frac{1}{1+\SNR}
\right)\\
&=&D(\Nc(0,1)||\Nc(0,1 + \SNR )).
\end{eqnarray}
For the perfectly correlated case ($\zeta=1/4$), on the other
hand, $\Kmsc_s =0$. In fact, in this case as well as in the i.i.d.
case, the two-dimensionality is irrelevant. The known result in
the 1-D case \cite{Sung&Tong&Poor:06IT} is applicable. With regard
to $\Imsc_s$, we have similar behavior at the extreme
correlations. In the i.i.d. case, the
 mutual information is given by the well known formula
\begin{equation}
\Imsc_s(0) = \frac{1}{2}\log (1 + \SNR),
\end{equation}
whereas  we have $\Imsc_s =0$ in the perfectly correlated case.
Thus, both information measures are zero at perfect correlation
($\zeta = 1/4$). The limiting behavior of the asymptotic
information rates near the extreme correlation values is given by
Taylor's theorem. Due to the differentiability of $\Kmsc_s$ and
$\Imsc_s$ w.r.t. $\zeta$, we have
\begin{equation}
\Kmsc_s (\zeta) = c_1 \cdot (\zeta - 1/4 ) + o ( |\zeta - 1/4 | ),
\label{eq:KmscQuaterZeta}
\end{equation}
and
\begin{equation}
 \Imsc_s (\zeta) =  c_1^\prime \cdot
(\zeta - 1/4 ) + o ( |\zeta - 1/4 | ),
\end{equation}
in a neighborhood of $\zeta = 1/4$  for some constants $c_1$ and
$c_1^\prime$ as $\zeta \rightarrow 1/4$. Similarly, we also have
the linear limiting behavior for $\Kmsc_s$ and $\Imsc_s$ in a
neighborhood of $\zeta =0$ with non-zero limiting values,
$D(\Nc(0,1)||\Nc(0,1+\SNR))$ and $\frac{1}{2}\log(1+\SNR)$,
respectively, as $\zeta \rightarrow 0$. That is,
\begin{equation}  \label{eq:Kmsc0Zeta}
\Kmsc_s(\zeta) = \Kmsc_s(0) + c_2 \zeta + o(\zeta),
\end{equation}
and
\begin{equation}
\Imsc_s(\zeta) = \Imsc_s(0) + c_2^\prime \zeta + o(\zeta),
\end{equation}
for some $c_2$ and $c_2^\prime$, as $\zeta \rightarrow 0$.

\begin{figure}[htbp]
\centerline{ \SetLabels
\L(0.25*-0.1) (a) \\
\L(0.75*-0.1) (b) \\
\endSetLabels
\leavevmode
\strut\AffixLabels{
\scalefig{0.5}\epsfbox{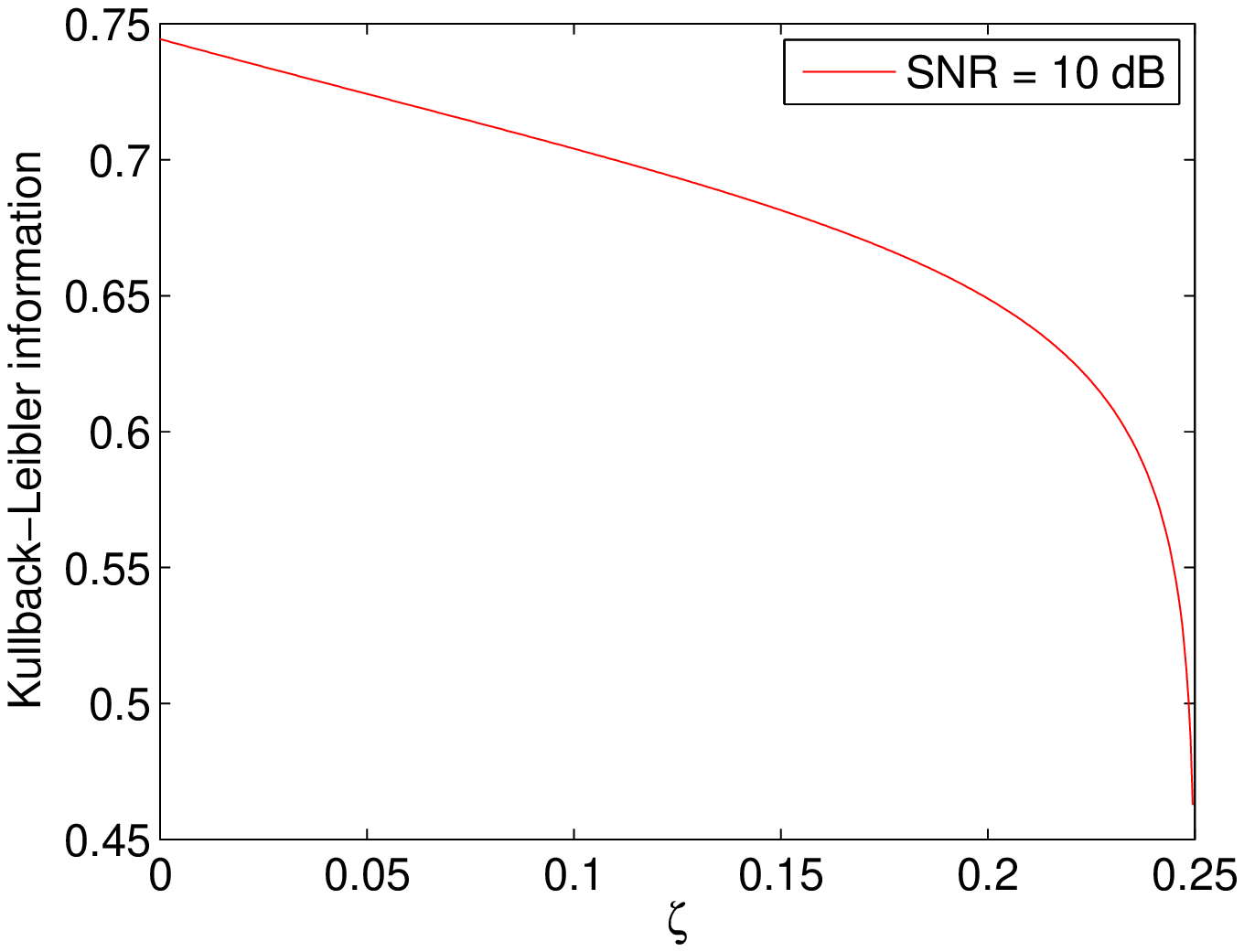}
\scalefig{0.5}\epsfbox{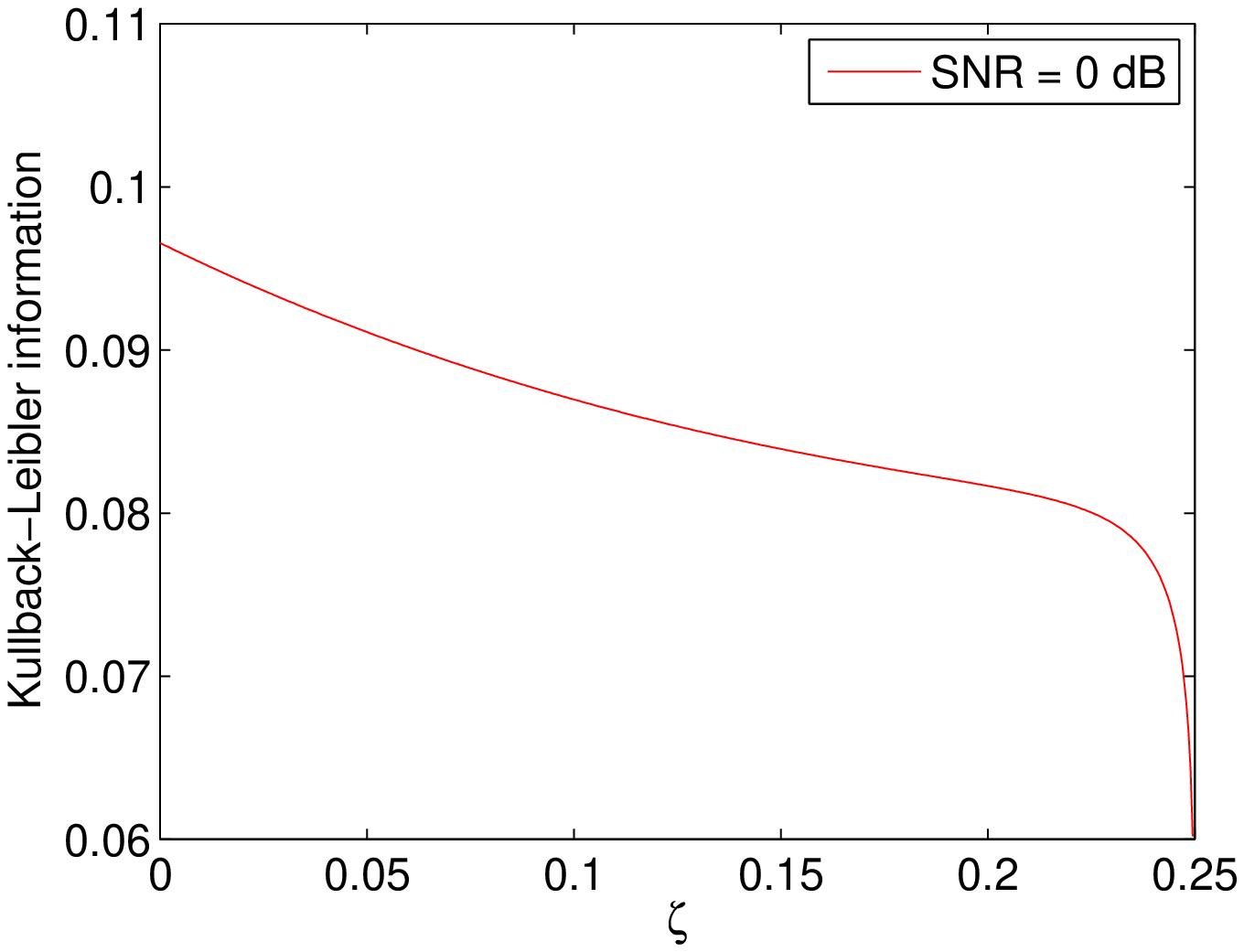} }
} \vspace{0.5cm} \centerline{ \SetLabels
\L(0.25*-0.1) (c) \\
\L(0.75*-0.1) (d) \\
\endSetLabels
\leavevmode
\strut\AffixLabels{
\scalefig{0.5}\epsfbox{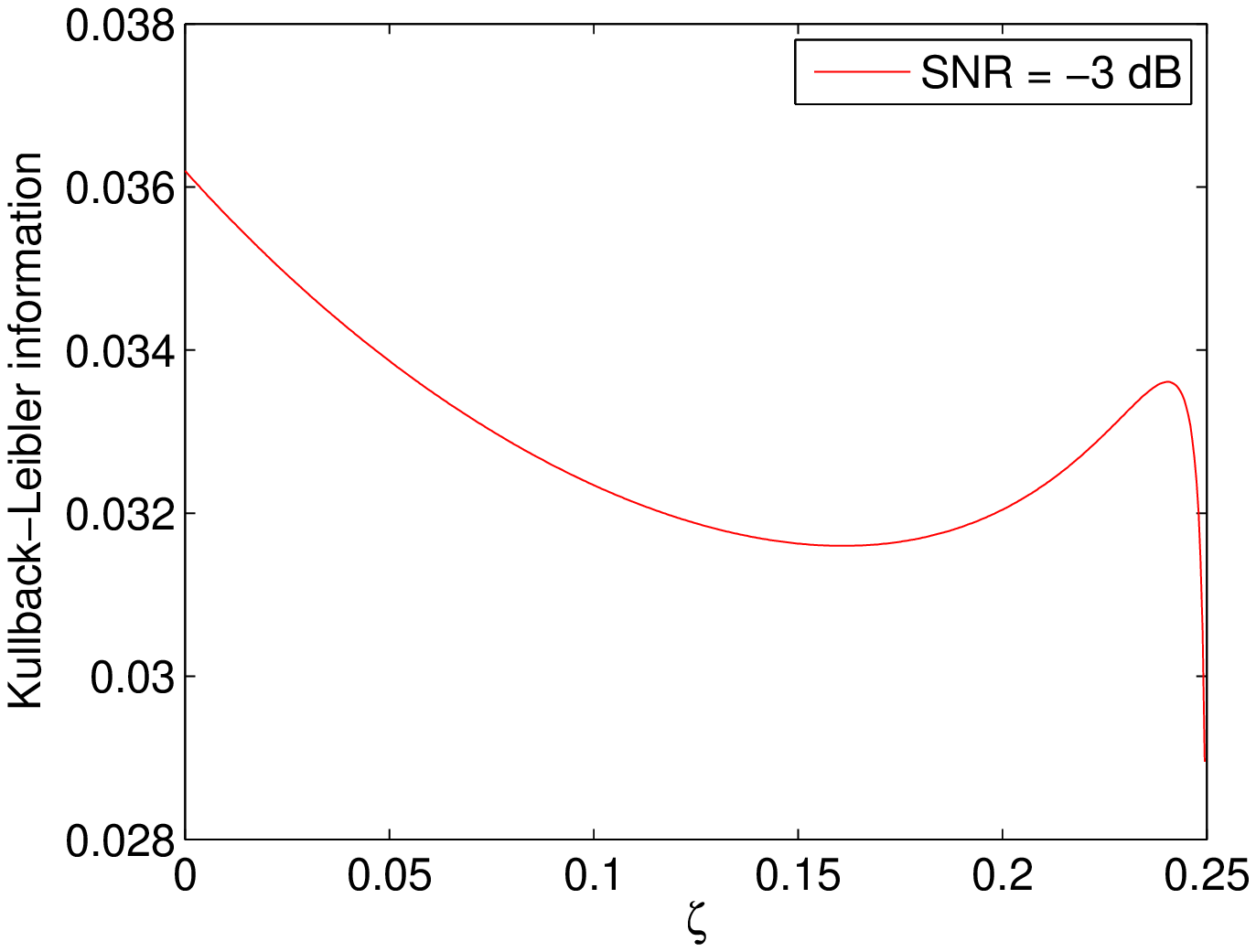}
\scalefig{0.5}\epsfbox{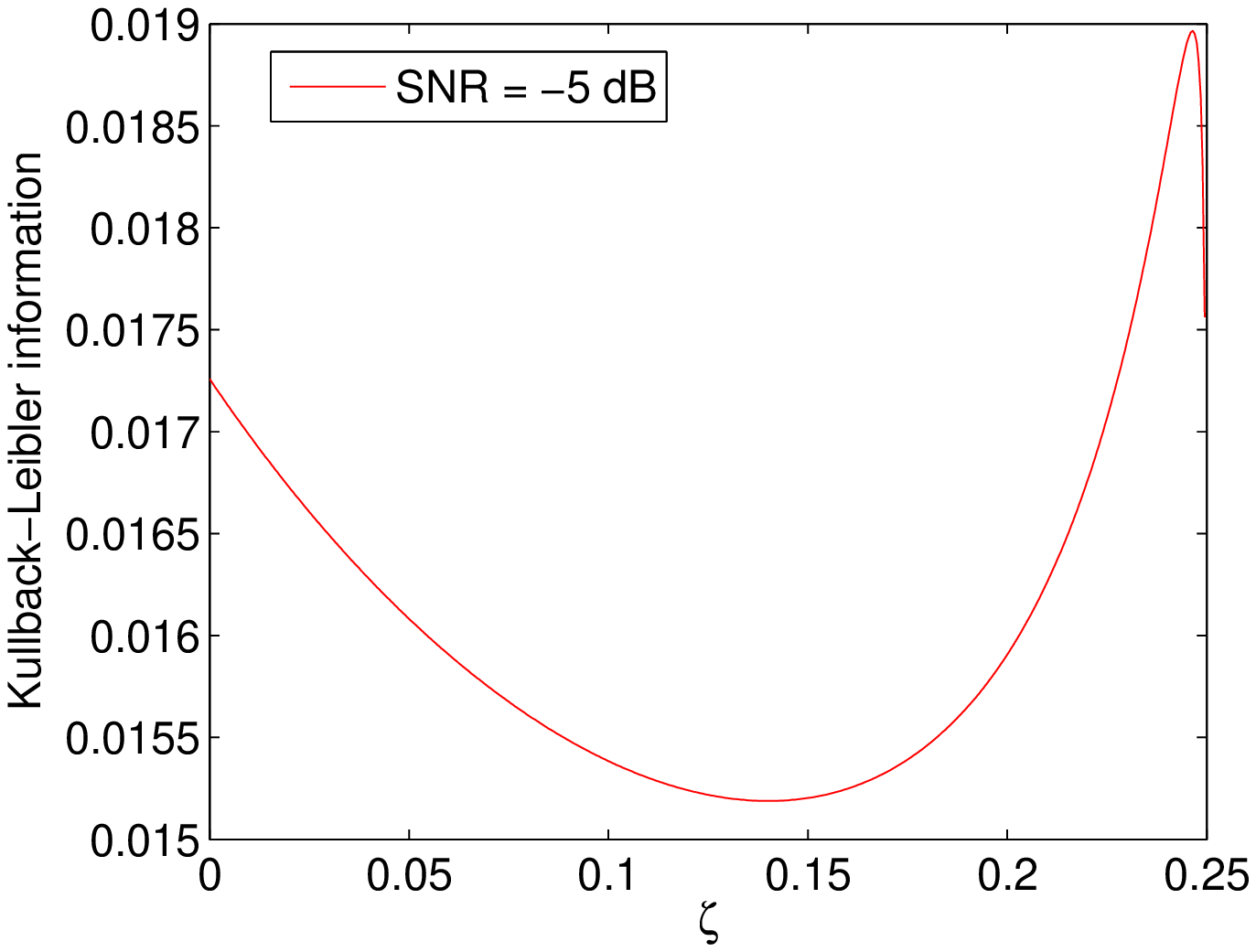} }
} \vspace{0.5cm} \caption{$\Kmsc_s$ as a function of $\zeta$: (a)
SNR = 10 dB, (b) SNR = 0 dB, (c) SNR = -3 dB, (d) SNR = -5 dB}
\label{fig:KcsVsZeta}
\end{figure}
\begin{figure}[htbp]
\centerline{
    \begin{psfrags}
    \psfrag{ij}[l]{{\large $(i,j)$}}
    \psfrag{xij}[c]{{\large $X_{ij}$}}
    \psfrag{wij}[l]{{\large $W_{ij}$}}
    \psfrag{yij}[c]{{\large $Y_{ij}$}}
    \psfrag{Nij}[c]{{\large Sensor $ij$}}
    \scalefig{0.55}\epsfbox{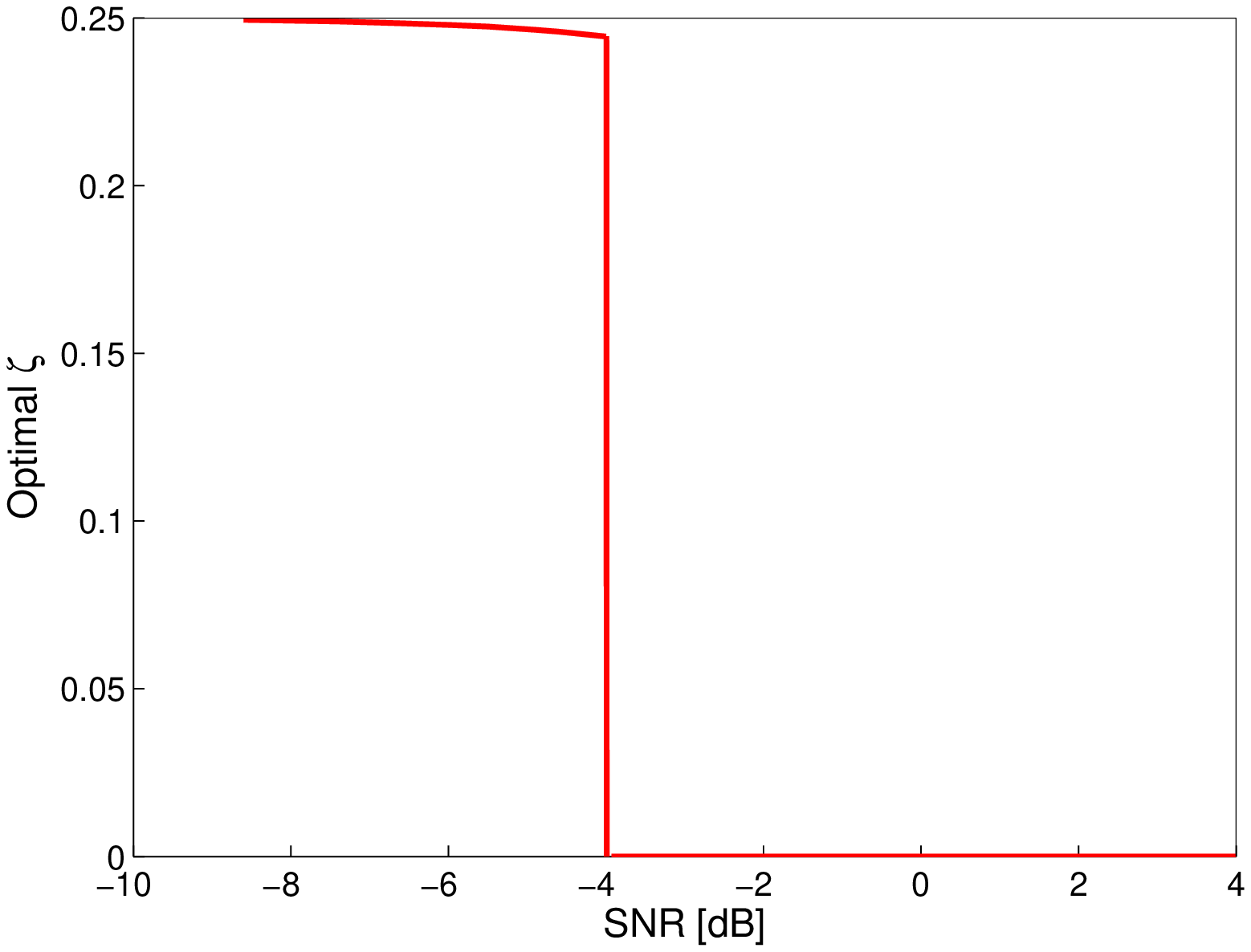}
    \end{psfrags}
}\caption{Optimal $\zeta$ maximizing $\Kmsc_s$ vs. SNR}
\label{fig:KcsVsZeta2}
\end{figure}

For intermediate values of correlation, we evaluate
(\ref{eq:errorexponentSFA}) and (\ref{eq:SFCARMI}) for several
different SNR values, as shown in Fig. \ref{fig:KcsVsZeta}. It is
seen that, at high SNR,  $\Kmsc_s$ decreases monotonically as
$\zeta$ increases.  Hence,  i.i.d. observations yield the largest
per-node information for a given value of SNR when SNR is large,
as in the 1-D case \cite{Sung&Tong&Poor:06IT}. As we decrease the
SNR, it is seen that a second mode grows near $\zeta=1/4$, i.e.,
in the strong correlation region. As we decrease the SNR further,
the value of $\zeta$ of the second mode shifts toward $1/4$, and
the value of the second mode exceeds that of the i.i.d. case.
Hence, there is a discontinuity in the optimal correlation as a
function of SNR in the 2-D case even if the maximal $\Kmsc_s$
itself is continuous, as seen in Fig. \ref{fig:KcsVsZeta2}. That
is, there is a phase transition for optimal correlation w.r.t.
SNR: above a certain SNR value i.i.d. observations yield the best
performance, whereas below that SNR point suddenly strong
correlation is preferred. This is not the case for 1-D
Gauss-Markov time series, where the optimal correlation maximizing
the information rate is continuous w.r.t. SNR.  Although it is not
shown here, the per-node MI $\Imsc_s$ exhibits similar behavior as
a function of the edge dependence factor $\zeta$.

 With regard to $\Kmsc_s$ and $\Imsc_s$  as functions of SNR, it is
straightforward to see from (\ref{eq:errorexponentSFA}) that they
are continuously differentiable functions, and the behavior of
$\Kmsc_s$ and $\Imsc_s$ with respect to SNR is given by the
following theorem.

\vspace{0.5em}
\begin{theorem}[Per-node information vs. SNR] \label{theo:KLIsvsSNR}
The asymptotic per-node KLI $\Kmsc_s$ for the hidden SFCAR model
is continuous and monotonically increasing as SNR increases for a
given edge dependence factor $\zeta \in [0~~1/4]$. Moreover,
$\Kmsc_s$ increases with rate  $\frac{1}{2}\log \mbox{SNR}$ as
$\SNR \rightarrow \infty$. As SNR decreases to zero, on the other
hand, $\Kmsc_s$ converges to zero and the rate of convergence is
given by
\begin{equation}  \label{eq:KLIvsSNRlowSNR}
\Kmsc_s(\SNR) = c_3\cdot \SNR^2  + o(\SNR^2),
\end{equation}
as $\SNR \rightarrow 0$, where $c_3$ is given by
\begin{equation}
c_3= \frac{1}{2^6 K^2(4\zeta)} \int_{-\pi}^{\pi} \int_{-\pi}^{\pi}
 \frac{1}{\left(1 - 2 \zeta \cos\omega_1 - 2 \zeta
 \cos\omega_2\right)^2}
d\omega_1 d\omega_2.
\end{equation}
 The per-node  MI $\Imsc_s$
has similar properties as a function of SNR, i.e., it is a
continuous and monotonically increasing function of SNR. At high
SNR, it increases with rate $\frac{1}{2}\log \SNR$, whereas it
decreases to zero with rate of convergence
\begin{equation}  \label{eq:MIvsSNRlowSNR}
\Imsc_s(\SNR) = c_3^\prime \cdot\SNR + o(\SNR),
\end{equation}
as $\SNR \rightarrow 0$, where $c_3^\prime$ is given by
\begin{equation}
c_3^\prime= \frac{1}{2^3\pi K(4\zeta)} \int_{-\pi}^{\pi}
\int_{-\pi}^{\pi}
 \frac{1}{1 - 2 \zeta \cos\omega_1 - 2 \zeta
 \cos\omega_2}
d\omega_1 d\omega_2.
\end{equation}
\end{theorem}

\vspace{0.5em} {\em Proof:} See the Appendix I. \vspace{0.5em}

\begin{figure}[htbp]
\centerline{
    \begin{psfrags}
    \psfrag{ij}[l]{{\large $(i,j)$}}
    \psfrag{xij}[c]{{\large $X_{ij}$}}
    \psfrag{wij}[l]{{\large $W_{ij}$}}
    \psfrag{yij}[c]{{\large $Y_{ij}$}}
    \psfrag{Nij}[c]{{\large Sensor $ij$}}
    \scalefig{0.6}\epsfbox{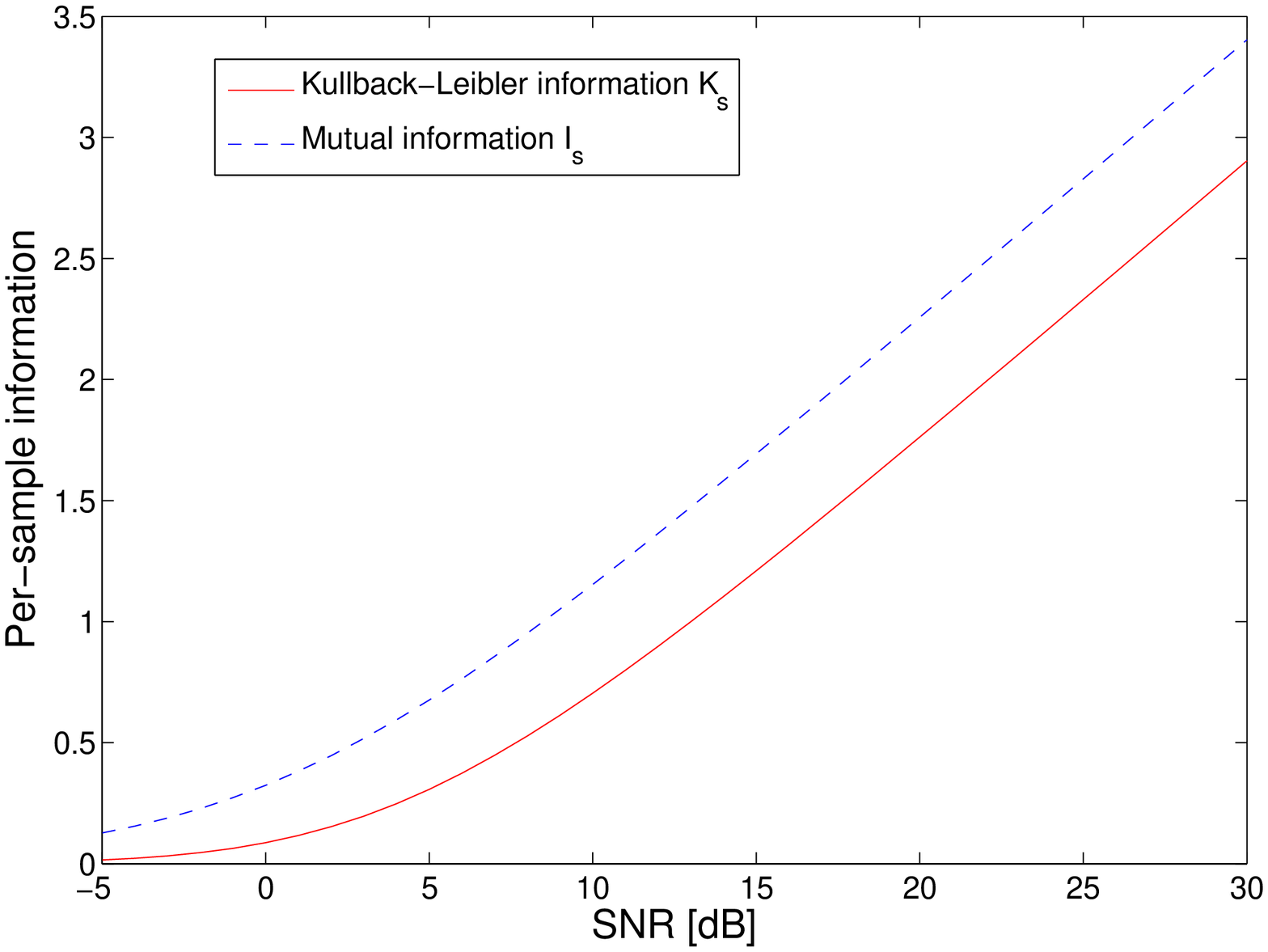}
    \end{psfrags}
}\caption{$\Kmsc_s$ and $\Imsc_s$ as functions of SNR
($\zeta=0.1$)} \label{fig:KLIMIVsSNR}
\end{figure}

\noindent Note that the limiting behavior as $\SNR \rightarrow 0$
is different for $\Kmsc_s$ and $\Imsc_s$; $\Kmsc_s$ decays to zero
quadratically while $\Imsc_s$ decreases linearly.  Fig.
\ref{fig:KLIMIVsSNR} shows $\Kmsc_s$ and $\Imsc_s$ with respect to
SNR for $\zeta =0.1$. The log SNR behavior is evident at high SNR
for both information measures. Note that $\Kmsc_s$ and $\Imsc_s$
increase with the same slope in the logarithmic scale with offset
$1/2$. This is easily seen from (\ref{eq:errorexponentSFA}) and
(\ref{eq:SFCARMI}) because the second term in the integrand of
(\ref{eq:errorexponentSFA}) converges to -1/2, and thus $\Kmsc_s
\rightarrow \Imsc_s -\frac{1}{2}$ as SNR increases. However, the
offset is negligible as SNR increases. It is easy to see from
(\ref{eq:errorexponentSFA}) and (\ref{eq:SFCARMI}) that  for a
given edge dependence factor $\zeta$ the convergence between the
two information measures is characterized by
$\frac{\Kmsc_s}{\Imsc_s} = 1 + O\left( \frac{1}{\log \SNR}\right)$
as $\SNR \rightarrow \infty$.

\section{Ad Hoc Sensor Networks: Fundamental Trade-Offs among Information, Coverage, Density and
Energy} \label{sec:adhocsennet}

Using the results of the previous sections, we now answer the
fundamental questions, raised in Section \ref{sec:intro},
concerning planar {\em ad hoc} sensor networks deployed over
correlated random fields for statistical inference under the 2-D
hidden SFCAR GMRF model. We first derive relevant physical
correlation parameters for the SFCAR from the corresponding
continuous-index stochastic model. Once the physical correlation
parameters for the SFCAR are obtained, the analysis of information
obtainable from an {\em ad hoc} sensor network and related
trade-offs is straightforward.

\subsection{Physical Correlation Model}
\label{subsec:physicalmodel}

\begin{figure}[htbp]
\centerline{ \SetLabels
\L(0.24*-0.1) (a) \\
\L(0.74*-0.1) (b) \\
\endSetLabels
\leavevmode
\strut\AffixLabels{
\scalefig{0.55}\epsfbox{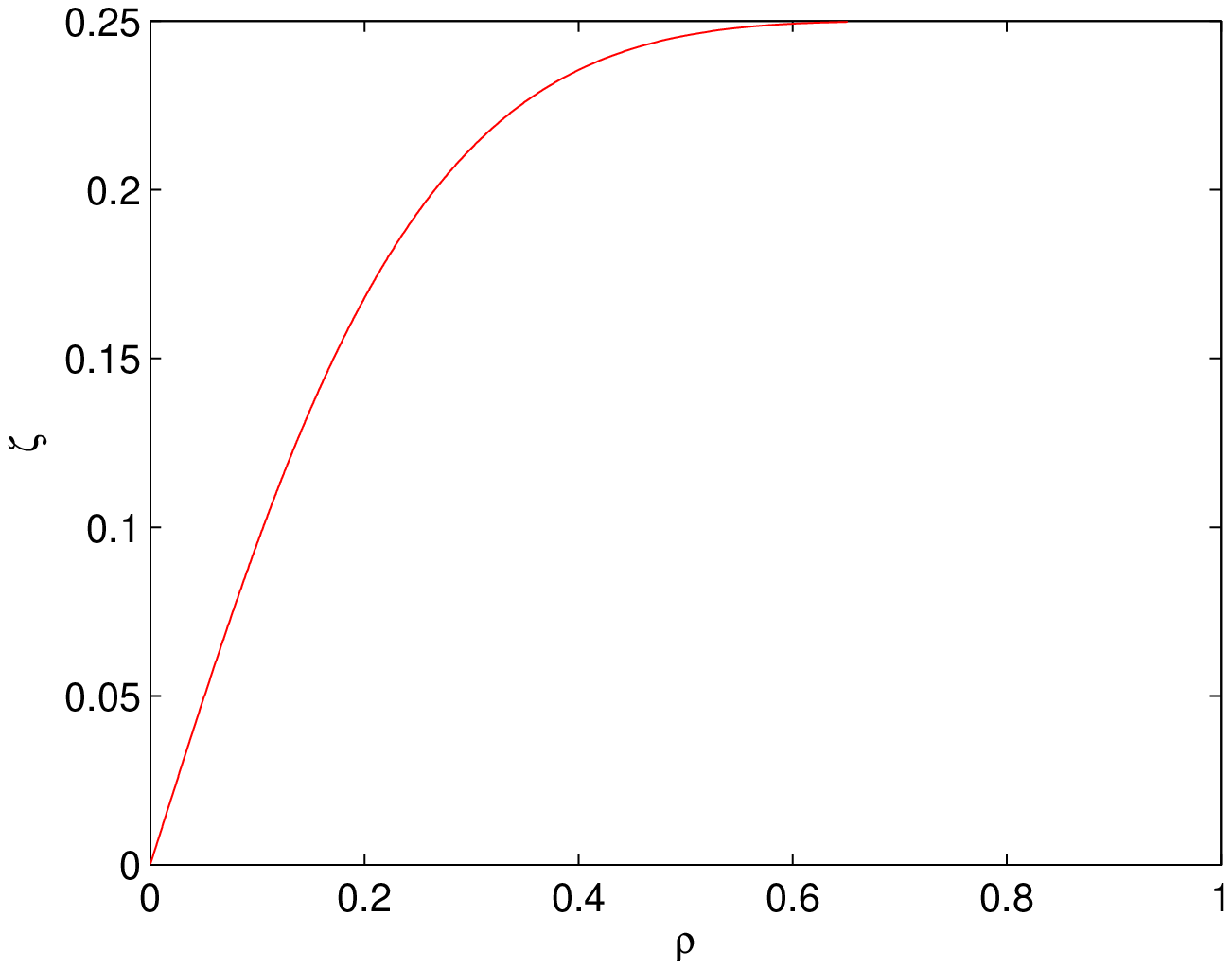}
\scalefig{0.55}\epsfbox{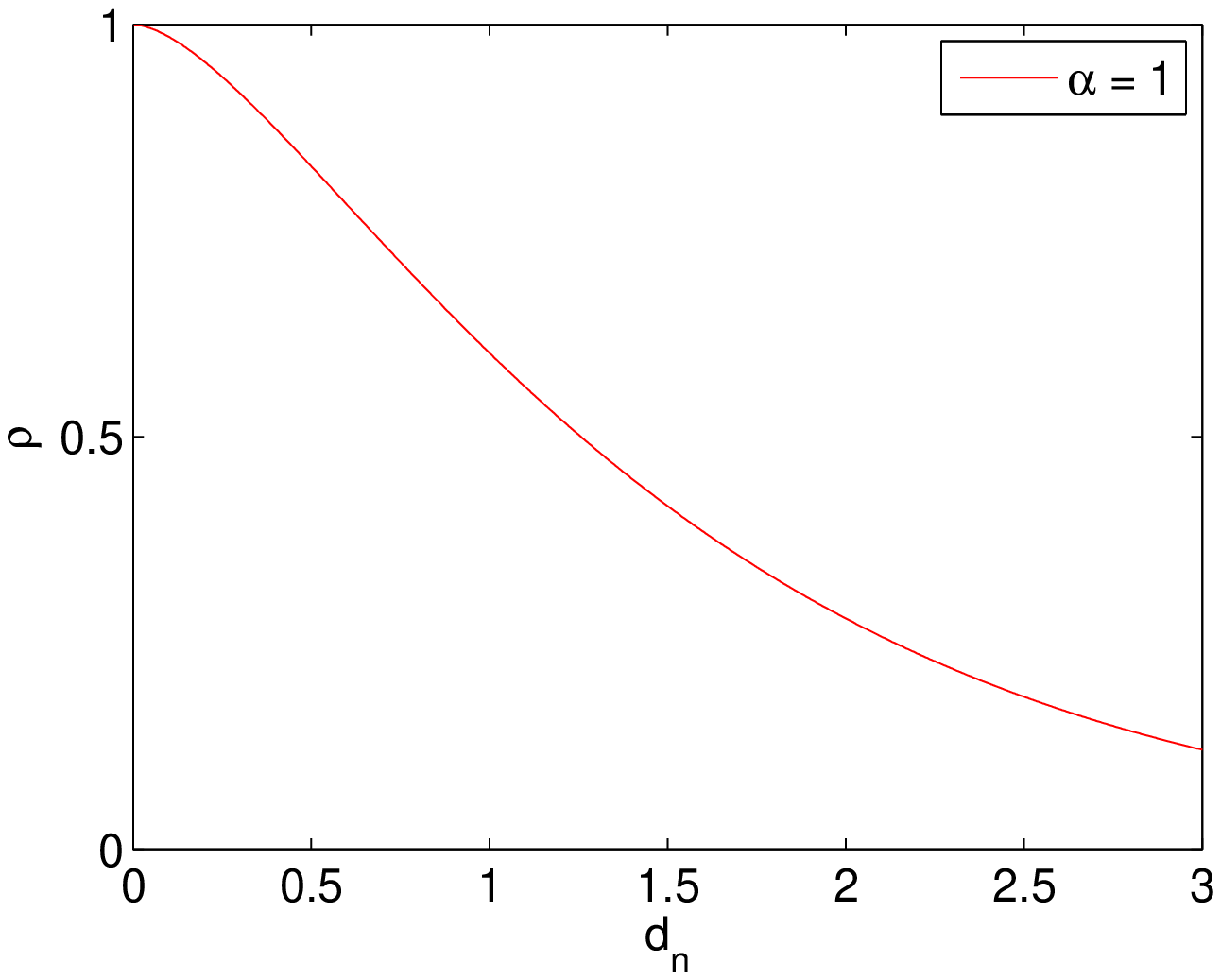} } }
\vspace{0.3cm}
 \caption{(a) edge dependence factor $\zeta$ vs. edge correlation
coefficient $\rho$ and (b) $\rho$ vs. edge length $d_n$}
\label{fig:ZetaVsRho}
\end{figure}

 We first derive how the
physical correlation is related to the edge dependence factor
$\zeta$ in the 2-D SFCAR model.  The edge correlation coefficient
$\rho$ is defined as
\begin{equation}  \label{eq:physicalCorrelation}
\rho \defeq \frac{\gamma_{01}}{\gamma_{00}}=
\frac{\gamma_{10}}{\gamma_{00}}, ~~(0 \le \rho \le 1),
\end{equation}
due to the spatial symmetry, where $\gamma_{ij} = \Ebb \{
X_{00}X_{ij}\}$.   $\rho$ represents the correlation strength
between the signal samples of two adjacent sensor nodes connected
by the Markov dependence graph defined by the SFCAR model. The
edge correlation coefficient $\rho$ is obtained using the
following relationship \cite{Besag:81JRSS}:
\begin{equation}  \label{eq:phyCorRelation}
\kappa \gamma_{00} = 1 + 4\zeta \kappa \gamma_{01}
~~~\Rightarrow~~~ \gamma_{01} = \frac{\kappa \gamma_{00}
-1}{4\kappa \zeta },
\end{equation}
and by substituting (\ref{eq:gamma00}) and
(\ref{eq:phyCorRelation}) into (\ref{eq:physicalCorrelation}), we
have
\begin{equation} \label{eq:ZetaVsRho}
\rho =  \frac{(2/\pi)K(4\zeta)-1}{ (2/\pi)(4\zeta) K(4\zeta)} =:
g^{-1}(\zeta).
\end{equation}
Note that the correlation coefficient $\rho$ is not dependent on
the power factor  $\kappa$ in (\ref{eq:SFCARprec}), as expected,
even though $\gamma_{00}$ and $\gamma_{01}$ are.  Note that
function $g^{-1}: \zeta \rightarrow \rho$ is a  continuous and
differentiable $C^1$ function on the domain $0 \le \zeta \le 1/4$
due to the continuous differentiability of $K(x)$ for $0 \le x <
1$, and $g^{-1}(1)=\lim_{x\rightarrow 1} \frac{(2/\pi)K(x)-1}{
(2/\pi)x K(x)}= 1$ by $K(1)=\infty$. Note also that $g^{-1}(0)=0$
since $K(0)=\pi/2$. Thus, the inverse mapping $g:\rho \rightarrow
\zeta$ from the edge correlation factor $\rho$ to the edge
dependence factor $\zeta$, which maps zero and one to zero and
1/4, respectively, behaves as shown in Fig. \ref{fig:ZetaVsRho}
(a).

Now we consider the correlation coefficient $\rho$ as a function
of the sensor spacing $d_n$. In general, the correlation function
$h: d_n \rightarrow \rho$ is a positive and monotonically
decreasing function of $d_n$ with $h(0) = 1$ and $h(\infty)=0$. It
is well known that for the 1-D first order AR signal a
corresponding underlying (continuous-index) physical model is
given by the Ornstein-Uhlenbeck process
\begin{equation} \label{eq:diffusioneq}
\frac{ds(x)}{dx}= - A s(x)+ B u(x),
\end{equation}
and its discrete-time equivalent is given by
\begin{equation} \label{eq:1Dstatespace}
\left\{
\begin{array}{ccl}
s_{i+1} &=& a s_{i} + u_i,\\
a &=& \Ebb\{s_i s_{i-1}\}/\Ebb\{s_i^2\}=e^{-A d_n},
\end{array}
\right.
 \end{equation}
 where  $A \ge 0$,  ~$B \in {\mathbb R}$, ~$s_i =
s(i
 d_n)$, and the input processes $u(x)$ and $u_i$ are  zero-mean white Gaussian
processes.  Here, $d_n$ is the spacing between two adjacent signal
samples. For the 2-D SFCAR signal, however, the same stochastic
differential equation is not applicable. Note that the dependence
in the signal in (\ref{eq:diffusioneq}) and
(\ref{eq:1Dstatespace}) is only on the past in 1-D space, whereas
the signal (\ref{eq:SFCARmean}) has symmetric dependence in all
four direction in the plane. The SFCAR signal is given by the
solution of a second-order difference equation
\begin{equation}  \label{eq:laplaceSDEdiscrete}
X_{ij} = \zeta (X_{i+1,j} + X_{i-1,j} + X_{i,j+1}+ X_{i,j-1}) +
\epsilon_{ij},
\end{equation}
and the corresponding continuous-index physical model is given by
the {\em stochastic Laplace equation} \cite{Whittle:54Biometrika}.
\begin{equation} \label{eq:laplaceSDE}
\left[ \left( \frac{\partial}{\partial x} \right)^2 + \left(
\frac{\partial}{\partial y} \right)^2 -  \alpha^2 \right] X (x,y)
= \epsilon(x,y),
\end{equation}
where  $\alpha ~(\ge 0)$ is  the physical diffusion rate, and
$\epsilon_{ij}$ and $\epsilon(x,y)$ are  2-D white zero-mean
Gaussian perturbations. Note that the solution of
(\ref{eq:laplaceSDE}) is circularly symmetric, i.e., it depends
only on $r=\sqrt{x^2+y^2}$, and samples of the solution
 $X(x,y)$ of (\ref{eq:laplaceSDE}) on lattice $\Ic_n$ do not
form a discrete-index SFCAR GMRF. However, (\ref{eq:laplaceSDE})
is still the continuous-index counterpart of
(\ref{eq:laplaceSDEdiscrete}), and we use its correlation function
for the SFCAR model.  The correlation function corresponding to
(\ref{eq:laplaceSDE}) is given by \cite{Whittle:54Biometrika}
\begin{equation} \label{eq:2DcorrelationFunc}
\rho = h(d_n) =  \alpha d_n K_1(\alpha d_n),
\end{equation}
where   $K_1(\cdot)$ is the modified Bessel function of the second
kind. Fig. \ref{fig:ZetaVsRho} (b) shows the correlation function
w.r.t. $d_n$ for $\alpha=1$. The asymptotic behavior of $K_1(x)$
is given by  
\begin{equation} \label{eq:modifiedBessel}
\left\{
\begin{array}{cccl}
K_1(x) &\rightarrow& \sqrt{\frac{\pi}{2x}}e^{-x} & \mbox{as}~ x
\rightarrow \infty,\\
K_1(x) &\rightarrow& \frac{1}{x} & \mbox{as}~ x
\rightarrow 0.\\
\end{array}
\right.
\end{equation}
The correlation function (\ref{eq:2DcorrelationFunc}) can be
regarded as the representative  correlation in 2-D, similar to the
exponential correlation function $e^{-Ad_n}$ in 1-D. Both
functions decrease monotonically w.r.t. $d_n$. However, the 2-D
correlation function is flat at $d_n=0$
\cite{Whittle:54Biometrika}, i.e.,
\begin{equation} \label{eq:K1flatTop}
 \left( \frac{d \rho}{d d_n}\right)_{d_n=0}=0,
\end{equation}
and it decays with rate $\sqrt{d_n}e^{-\alpha d_n}$ as $d_n
\rightarrow \infty$. Note that the 2-D correlation function has
$\sqrt{d_n}$ in front of the exponential decay as $d_n \rightarrow
\infty$. However, this polynomial term is not significant and the
exponential decay is dominant for large $d_n$. Thus, we have
$\zeta = g(h (d_n)),$ and for given physical parameters (with a
slight abuse of notation),
\[
\Kmsc_s(\mbox{SNR},\zeta) = \Kmsc_s(\mbox{SNR},g(h(d_n))) =
\Kmsc_s(\mbox{SNR},d_n),
\]
and
\[
\Imsc_s(\mbox{SNR},\zeta) = \Imsc_s(\mbox{SNR},g(h(d_n))) =
\Imsc_s(\mbox{SNR},d_n).
\]
We will use the arguments SNR, $\zeta$ and $d_n$ for $\Kmsc_s$ and
$\Imsc_s$ properly as needed for exposition.

\subsection{Scaling Laws in  Ad Hoc Sensor Networks over Correlated Random Fields}

In this section,  we investigate the fundamental behavior of
wireless flat multi-hop {\em ad hoc} sensor networks deployed for
statistical inference based on the 2-D hidden SFCAR model and the
corresponding correlation functions (\ref{eq:ZetaVsRho}) and
(\ref{eq:2DcorrelationFunc}).  We consider several criteria for
determining the efficiency of the sensor network. Specifically, we
consider the total amount of information [nats] obtainable from
the network and the energy efficiency $\eta$ of a sensor network,
defined as
\begin{equation}
\eta = \frac{{\mbox{total gathered information}~
I_t}}{{\mbox{total required energy}~E_t}} ~~~[\mbox{nats/J}],
\end{equation}
where the gathered information is about the underlying physical
process.

In the following, we summarize the assumptions for the planar {\em
ad hoc} sensor network that we consider.
\begin{itemize}
\item[(A.1)]  $n^2$ sensors are located on the grid $\Ic_n$ with
spacing $d_n$, as shown in Fig. \ref{fig:2dHGMRF}, and a fusion
center is located at the center $(\lfloor n/2 \rfloor, \lfloor n/2
\rfloor)$. The network size is $L \times L$, where $L = n d_n$.
Thus, the node density $\mu_n$ on $\Ic_n$ is given by
\begin{equation}
\mu_n = \frac{n^2}{L^2}= \frac{n^2}{(nd_n)^2}.
\end{equation}

\item[(A.2)]  The observations $\{Y_{ij}\}$ of sensor nodes form a
2-D hidden (discrete-index) SFCAR GMRF on the lattice for each
$d_n
> 0$, and the edge dependence factor is given by  the correlation
functions (\ref{eq:ZetaVsRho}) and (\ref{eq:2DcorrelationFunc}).

\item[(A.3)] The fusion center gathers the measurements from all
nodes using minimum hop routing. Note that the links in Fig.
\ref{fig:2dHGMRF} are not only the Markov dependence edges but
also the routing links.  The minimum hop routing requires a hop
count of $|i-\lfloor n/2 \rfloor|+|j-\lfloor n/2 \rfloor|$ to
deliver $Y_{ij}$ to the fusion center.

\item[(A.4)] The communication  energy per link is given by
$E_{c}(d_n) = E_0 d_n^{\nu}$, where  $\nu \ge 2$ is the
propagation loss factor of the wireless channel.

\item[(A.5)] Sensing requires energy, and the sensing energy per
node is denoted by $E_{s}$.  Moreover, we assume that the {\em
measurement} SNR in (\ref{eq:SNRorg}) is linearly increasing
w.r.t. $E_{s}$, i.e., $\SNR = \beta E_s$  for some constant
$\beta$.
\end{itemize}

\begin{remark} Assumption \textit{(A.2)} facilitates the analysis.
Since discrete samples of a continuous-index GMRF do not form a
discrete-index GMRF almost surely, we assume that for each $d_n$
sensor samples on $\Ic_n$ form a discrete-index SFCAR GMRF, and
match the correlation between two neighboring nodes with the
physically meaningful correlation function
(\ref{eq:2DcorrelationFunc}).
\end{remark}

\begin{remark}
In Assumption \textit{(A.3)} we assume  that there is no data
fusion during the information gathering, i.e., no in-network data
fusion. The fusion center collects the raw measurements from all
sensors.
\end{remark}

\begin{remark}
 We can also consider a routing graph  different from the
Markov dependence graph in Fig. \ref{fig:2dHGMRF}. For example,
sensors not directly connected to the transmitting node via the
Markov dependence edge  can deliver the data to the fusion center.
However, this results in a reduced number of hops with a larger
hop length, and the corresponding routing path consumes  more
energy. Thus, Assumption \textit{(A.3)} of minimum hop routing via
the Markov dependence edge ensures least energy consumption with a
minimum hop routing strategy.
\end{remark}

\begin{remark}
Assumption \textit{(A.5)} does not imply that we can increase the
power of the underlying signal, but it means that we can increase
the SNR of effective sensor samples. Suppose that $E_1$ joules are
required for one sensing to obtain one sample $Y_{ij}(1) =
X_{ij}(1) + W_{ij}(1)$ at location $ij$ and the measurement SNR of
this sample is $\SNR_1$. Now assume that we have $M$ identical
subsensors at location $ij$ and obtain $M$ samples with one sample
per each subsensor, requiring $M\cdot E_1$ joules, and we take an
average of $M$ samples at location $ij$, yielding $Y_{ij} = (1/M)
\sum_{m=1}^M Y_{ij}(m)$ where $Y_{ij}(m)$ denotes the sample at
the  $m$th subsensor at location $ij$. The measurement SNR of the
effective sample $Y_{ij}$ is given by $M\cdot\SNR_1$ assuming that
the measurement noise is i.i.d. across the subsensors. Thus, the
effective measurement SNR at each sensor can be increased linearly
w.r.t. the sensing energy. However, this linear SNR model is an
optimistic assumption since the observation SNR may saturate as
the sensing energy is increased without bound in practical
situation.
\end{remark}

From here on, we consider various asymptotic scenarios and
investigate the fundamental behavior of {\em ad hoc} sensor
networks deployed over correlated random fields for statistical
inference under assumptions \textit{(A.1)}-\textit{(A.5)}.
 Our asymptotic analysis in the previous sections enables us to
calculate the total information $I_t$ for large sensor networks.
The total amount of information is given approximately by the
product of the number of sensor nodes in the network and the
asymptotic per-node information $\Kmsc_s$ or $\Imsc_s$, i.e.,
\begin{equation} \label{eq:adhocTotalInfo}
I_t = n^2 \Kmsc_s(\mbox{SNR},d_n) ~~\mbox{or}~~ I_t = n^2
\Imsc_s(\mbox{SNR},d_n),
\end{equation}
for KLI or MI, respectively.  The total energy $E_t$ required for
data gathering via the minimum hop routing is given by
\begin{eqnarray}
E_t &=& n^2 E_s + E_c(d_n) \sum_{i=0}^{n-1}\sum_{j=0}^{n-1}
(|i-\lfloor n/2 \rfloor|+|j-\lfloor n/2 \rfloor|),\nonumber\\
&=& \left\{ \begin{array}{cc} n^2 E_s + \frac{1}{2}n(n-1)(n+1) E_{c}(d_n) & \mbox{if $n$ odd}, \\
n^2 E_s + \frac{1}{2}n^3 E_{c}(d_n) & \mbox{if $n$ even}.
\end{array}\right.
\label{eq:adhocTotalEnergy}
\end{eqnarray}

First, we consider an infinite area model with fixed density. In
this case, the number of sensor nodes per unit area is fixed and
the total area increases without bound as we increase $n$. The
behavior of the information vs. area and energy in this case is
given in the following theorem.

\vspace{0.5em}
\begin{theorem}[Fixed density and infinite area] \label{theo:infiniteareafixeddensity} For an {\em ad hoc} sensor network
with  a fixed and finite node density and fixed sensing energy per
node, the total amount of information increases linearly w.r.t.
area, but the amount of gathered information per unit energy
decays to zero with rate
\begin{equation} \label{eq:efficiencyIAM}
\eta = \Theta\left(\mbox{area}^{-1/2} \right),
\end{equation}
for any non-trivial diffusion rate $\alpha$, i.e., $0 < \alpha <
\infty$, as we increase the area. Further, in this case the total
amount of information obtainable from the network as a function of
total consumed energy increases  with rate of
\begin{equation} \label{eq:energyAsymptotic2}
\mbox{Total information}~ I_t = \Theta \left( E_t^{2/3} \right),
\end{equation}
for any propagation loss factor $\nu > 0$, as the total energy
$E_t$ consumed by the network increases without bound, i.e., $E_t
\rightarrow \infty$.
\end{theorem}

\vspace{0.5em} {\em Proof:} See Appendix I. \vspace{0.5em}

Theorem \ref{theo:infiniteareafixeddensity} enables us to
investigate the asymptotic behavior of ad hoc sensor networks with
fixed available energy per node. From the detection perspective
the error probability is given by
\begin{equation}  \label{eq:infiniteAreaPM}
P_M \sim e^{-I_t(E_t(N_t(A)))},
\end{equation}
for large networks, where $N_t(A)$ represents the total number of
sensor nodes in the network with coverage area $A$.  Now consider
that each node has a fixed amount of energy denoted by $\bar{E} ~
(< \infty)$. Then, the total energy in the network is given by
\begin{equation}  \label{eq:infiniteAreaEt}
E_t = N_t(A)  \bar{E}.
\end{equation}
Note in this case that the total energy available in the network
increases linearly w.r.t. the number of sensor nodes. The
asymptotic behavior of ad hoc networks with fixed per-node energy
is given by the following corollary to Theorem
\ref{theo:infiniteareafixeddensity}.

\vspace{0.5em}
\begin{corollary} \label{cor:infiniteAreaZeroInformation} For an {\em ad hoc} sensor network
with  a fixed and finite node density and fixed per-node sensing
energy, the information amount per sensor node diminishes to zero
as the network size grows, i.e.,
\begin{eqnarray}
&&\lim_{N_t(A)\rightarrow \infty} - \frac{1}{N_t(A)} \log
P_M(E_t(N_t(A))) \nonumber \\
&=& \lim_{N_t(A)\rightarrow \infty}
O(N_t(A)^{-1/3})= 0,
\end{eqnarray}
if each sensor has a finite amount of available energy.
\end{corollary}

\vspace{1em} {\em Proof:}  Substitute
(\ref{eq:energyAsymptotic2}), (\ref{eq:infiniteAreaPM}) and
(\ref{eq:infiniteAreaEt}) into $I_t$, $P_M$ and $E_t$,
respectively. \vspace{0.5em}

\noindent  Corollary \ref{cor:infiniteAreaZeroInformation} states
that a non-zero per-node information is not achievable as the
coverage increases without in-network data fusion in the case that
each node has only a fixed amount of energy, which is the case in
most network design with fixed amount of battery. In this case,
the per-node information scales with $O(N_t^{-1/3})$ as the
network size grows. This result is by the communication energy
required for ad hoc routing without in-network data fusion. Note
from (\ref{eq:adhocTotalEnergy}) that for the fixed density and
increasing area model the sensing energy increases quadratically
with $n$ while the communication energy without in-network data
fusion increases cubically with $n$
 since $d_n$ is fixed w.r.t. $n$.
Hence, for {\em ad hoc} sensor networks with large coverage areas
the communication energy dominates the sensing energy, and both
the energy efficiency for information and the per-node information
under fixed per-node energy constraint diminish to zero because of
the slower increasing  rate of the total information amount than
that of the communication energy required for {\em ad hoc} routing
without in-network data fusion.

This diminishing energy efficiency and per-node information under
fixed per-node energy constraint can be fixed with {\em in-network
data fusion}. Suppose that in-network data fusion is performed so
that each node needs to deliver (aggregated) data only to the
neighboring node along the minimum hop route to the fusion center
in Fig. \ref{fig:2dHGMRF}. In this case  the number of
transmission associated with one node is just one and the total
number of transmission in the network is given by $\Theta(n^2)$.
So, the communication energy as well as the sensing energy
increases quadratically with $n$. Since the total amount of
information also increases quadratically with $n$, the total
amount of information as a function of total energy is given,
under this aggregation scenario, by
\begin{equation}
I_t = \Theta(E_t),
\end{equation}
as we increase the area, and a non-zero energy efficiency and a
non-zero per-node information  under fixed per-node energy
constraint are achieved. Thus, in-network data fusion is essential
for energy-efficiency in large sensor networks.

Next, we consider the case in which the node density diminishes,
i.e., $d_n \rightarrow \infty$.  Especially, this case is of
interest at high SNR since at high SNR less correlated samples
yield larger per-node information, as seen in Section
\ref{subsec:errorexponent2D}. However, the per-node information is
upper bounded as $d_n \rightarrow \infty$, and the asymptotic
behavior is given by the following theorem.

\begin{theorem} \label{theo:informationvsdninfty}
As $d_n \rightarrow \infty$, the per-node information $\Kmsc_s$
and $\Imsc_s$ converge to $D(\Nc(0,1)||\Nc(0,1+\SNR))$ and
$\frac{1}{2}\log \SNR$, respectively, and the convergence rates
are given by
\begin{equation} \label{eq:theoremdninfty}
\Kmsc_s(d_n) =   D(\Nc(0,1)||\Nc(0,1+\SNR)) - c_4\sqrt{d_n}e^{-
\alpha d_n} + o\left(\sqrt{d_n}e^{-\alpha d_n}\right)
\end{equation}
and
\begin{equation}
 \Imsc_s(d_n) = \frac{1}{2} \log ( 1 + \SNR)  - c_4^\prime\sqrt{d_n}e^{- \alpha
d_n} + o\left(\sqrt{d_n}e^{-\alpha d_n}\right),
\end{equation}
with positive constants $c_4$ and $c_4^\prime$.
\end{theorem}

\vspace{0.5em} {\em Proof:} See Appendix I. \vspace{0.5em}

\begin{figure}[htbp]
\centerline{
    \begin{psfrags}
    \psfrag{ER}[r]{{ $E(R)$}}  %
    \psfrag{R}[c]{{ $R$}} %
    \psfrag{R0}[c]{{ $R_0$}} %
    \psfrag{Rc}[c]{{ $R_c$}} %
    \psfrag{I1}[c]{ $I$} %
    \psfrag{A}[c]{ $A$}
    \psfrag{I}[c]{{Mutual information}} %
     \scalefig{0.65}\epsfbox{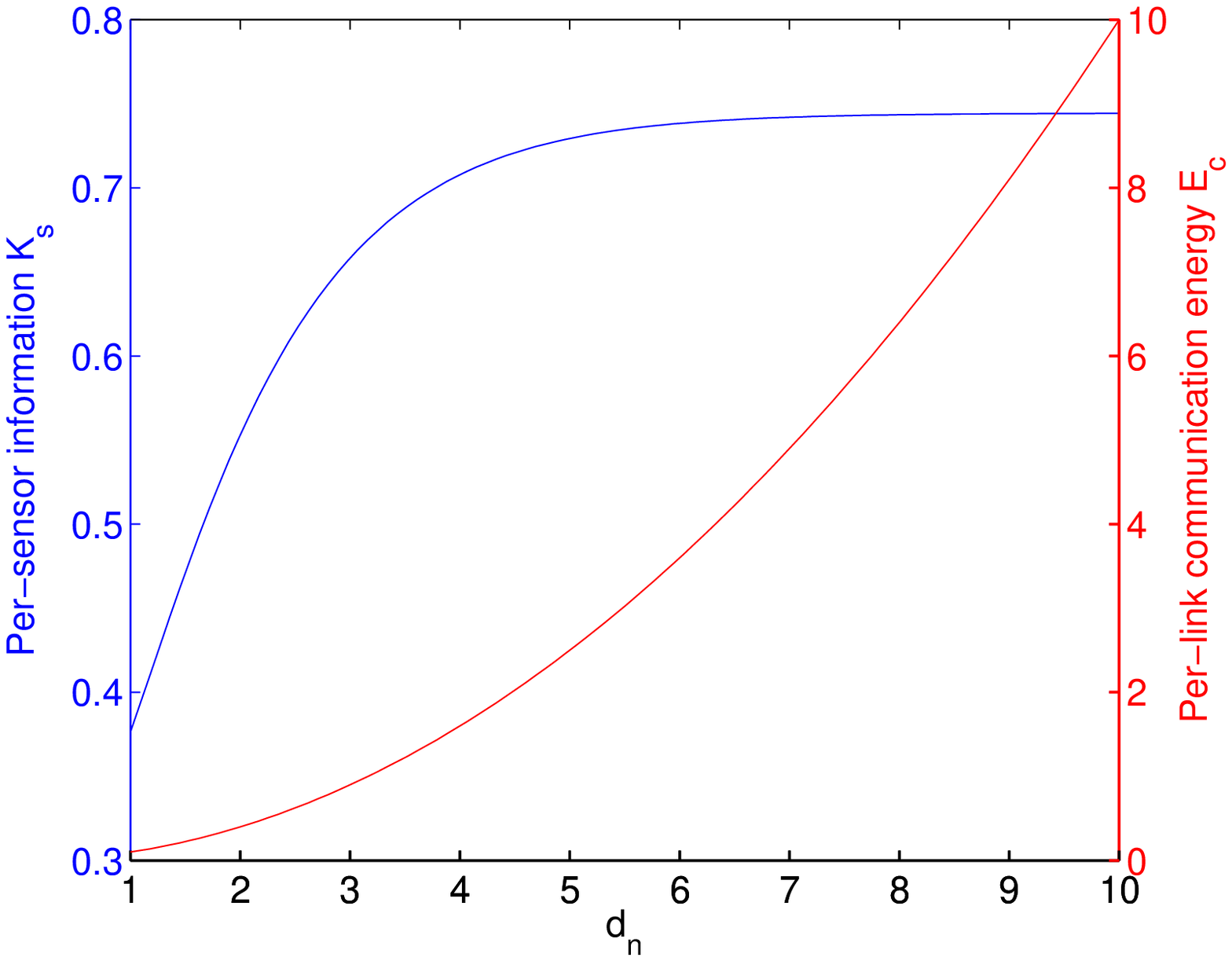}
    \end{psfrags}
} \caption{Per-node information and per-link communication energy
w.r.t. sensor spacing $d_n$ (SNR = 10 dB, $\alpha=1$, $c_4=1$)}
\label{fig:dngoestoinfty}
\end{figure}

\noindent Theorem \ref{theo:informationvsdninfty} explains how
much gain in information  is obtained from less correlated
observation samples by making the sensor spacing larger.  Fig.
\ref{fig:dngoestoinfty} shows the per-node KLI $\Kmsc_s$ and the
communication energy $E_c$ for each link as functions of $d_n$ for
$\alpha =1$, $c_4=1$ and 10 dB SNR. The gain in information is
given by $\sqrt{d_n}e^{-\alpha d_n}$ for large $d_n$,  whereas the
required per-link communication energy increases without bound,
i.e., $E_c (d_n) = E_0 d_n^\nu$ ~($\nu \ge 2$). Since the
exponential term is dominant in the gain as $d_n$ increases, the
information gained by increasing the sensor spacing $d_n$
decreases almost exponentially fast, and no significant gain is
obtained by increasing the sensor spacing further after some
point.  Hence, it is not effective, in terms of energy efficiency,
to increase the sensor spacing too much to obtain less correlated
samples at high SNR.

From Theorem \ref{theo:informationvsdninfty} we have seen that
increasing the sensor spacing is not so effective in terms of the
information gain per unit of consumed energy since the per-link
communication energy increases without bound. On the other hand,
the per-link communication energy can be made arbitrarily small by
decreasing the sensor spacing. To investigate the effect of
diminishing communication energy $E_c$ as $d_n \rightarrow 0$, we
now consider the asymptotic case in which the node density goes to
infinity for a fixed coverage area.  In this case, the per-node
information decays to zero as $d_n \rightarrow 0$ since $\zeta
\rightarrow 1/4$ as $d_n \rightarrow 0$, and  $\Kmsc_s(\zeta)$ and
$I_s(\zeta)$ converge to zero as $\zeta \rightarrow 1/4$, as shown
in Section \ref{subsec:errorexponent2D}.  The asymptotic behavior
in this case is given by the following theorem.

\vspace{0.5em}
\begin{theorem}[Infinite density model]
\label{theo:infinitedensitymodel} For the infinite density model
with a fixed coverage area $S$ with nontrivial diffusion rate
$\alpha$, the per-node information decays to zero with convergence
rate
\begin{equation} \label{eq:infintedensityKs}
\Kmsc_s =   c_5\mu_n^{-1} + o\left( \mu_n^{-1} \right),
\end{equation}
for some constant $c_5$ as the node density $\mu_n \rightarrow
\infty$. Hence, the amount of total information from the coverage
area converges to the constant $c_5 S$ as $\mu_n \rightarrow
\infty$. Furthermore, in the case of no sensing energy, a non-zero
energy efficiency $\eta$ is achievable if the propagation loss
factor $\nu =3$, and even an infinite energy
efficiency\footnote{Of course, this is under Assumption
\textit{(A.4)} for any $d_n >0$. In reality, Assumption
\textit{(A.4)} is valid for $d_n \ge d_{min}$ for some $d_{min}
> 0$. } is achievable under Assumption \textit{(A.4)} if $\nu
> 3$ as $\mu_n \rightarrow \infty$. $\Imsc_s$ has similar
behavior.
\end{theorem}

\vspace{0.5em} {\em Proof:} See Appendix I. \vspace{0.5em}

\begin{remark}  The finite total information for the infinite
density and fixed area model follows our intuition.  The maximum
information provided by the samples from the continuous-index
random field does not exceed the information between $X(x,y)$ and
$Y(x,y)$ except in the case of spatially white fields.  Here, the
relevance of (\ref{eq:K1flatTop}) in 2-D is evident. From
(\ref{eq:K1flatTop}) we have
\begin{equation}
\Kmsc_{s,2-D}(\zeta(\rho(d_n))) = c_6 \cdot d_n^2 + o(d_n^2),
\end{equation}
as $d_n \rightarrow 0$ since $h: d_n \rightarrow \zeta$ has slope
 zero at $d_n=0$ and $\Kmsc_s$ is a continuous and differentiable
function of $\zeta$. In the 1-D case, it is shown in
\cite{Sung&Tong&Poor:06IT} that $\Kmsc_{s,1-D}$ is also a
continuous and differentiable function of $a=e^{-Ad_n}$ for $0\le
a \le 1$ with $\Kmsc_{s,1-D}|_{a=1} =0$. However,  the exponential
correlation $e^{-Ad_n}$ has a nonzero slope at $d_n=0$, and thus
we have
\begin{equation}
\Kmsc_{s,1-D}(a(d_n)) = c_6^\prime \cdot d_n + o(d_n),
\end{equation}
as $d_n \rightarrow 0$. The number of nodes in the space is given
by $\Theta(n^2)$ and $\Theta(n)$ for 2-D and 1-D, respectively,
and $d_n = L/n$ in both cases. Hence, the total amount of
information from the coverage space (given by the product of the
per-node information and the number of nodes in the space)
converges to a constant both in 1-D and 2-D as the node density
increases.  Thus, any proper 2-D correlation function w.r.t. the
sample distance should have a flat top at a distance of zero.
\end{remark}

\begin{remark} It is common that the propagation loss factor $\nu
> 3$ for near field propagation (i.e., $d_n \rightarrow 0$).
Hence, infinite energy efficiency is theoretically achievable
under Assumption \textit{(A.4)} as we increase the node density
for a fixed area assuming that only communication energy is
required. Note that the total amount of information converges to a
constant as we increase the node density. So, the infinite energy
efficiency is achieved by diminishing communication energy as $d_n
\rightarrow 0$.
\end{remark}

\begin{remark} Considering the sensing energy, infinite energy
efficiency is not feasible even theoretically since we have in
this case
\begin{equation}
E_t = n^2 E_s + \Theta( n^{3-\nu}),
\end{equation}
and
\begin{equation}
\eta = \frac{c_5S + o(1)}{n^2 E_s + \Theta( n^{3-\nu}) }, ~~\nu
\ge 2,
\end{equation}
as $n\rightarrow \infty$ for fixed coverage area. In this case the
sensing energy $n^2 E_s$ is the dominant factor for low energy
efficiency, and the energy efficiency decreases to zero with rate
$O\left(\mu_n^{-1}\right)$. Thus,  it is critical for densely
deployed sensor networks to minimize the sensing energy or
processing energy for each sensor.
\end{remark}

\vspace{0.5em} In the infinite density model, we have observed
that energy is an important factor in efficiency. Now we
investigate the change of total information w.r.t. energy.  There
are many possible ways to invest energy in the network. One simple
way is to fix the node density and coverage area and to increase
the sensing energy.  We assume that the network size is
sufficiently large so that our asymptotic analysis is valid. The
energy-asymptotic behavior in this case is given  in the following
theorem under Assumptions \textit{(A.1)-(A.5)}.

\vspace{0.5em}
\begin{theorem} \label{theo:energyasymptotic}
 As we increase the total energy $E_t$ consumed by a sensor
network (including both sensing and communication) with a fixed
node density and fixed area, the total information increases with
rate
\begin{equation} \label{eq:energyAsymptotic1}
\mbox{Total information} ~ I_t = O\left(  \log E_t \right)
\end{equation}
 as $E_t \rightarrow
\infty$.
\end{theorem}
\vspace{0.5em}

{\em Proof:} See Appendix I.

\vspace{0.5em}

\noindent Theorem \ref{theo:energyasymptotic} suggests a guideline
for investing the excess energy.   It is not efficient in terms of
the total amount of gathered information to invest energy to
improve the quality of sensed samples from a limited area. This
only provides an increase in total information at a logarithmic
rate.  Note in Theorem \ref{theo:infiniteareafixeddensity} that
the information gain  is given by
\begin{equation}
I_t = \Theta (E_t^{2/3})
\end{equation}
as we increase the coverage area with fixed density and sensing
energy even without in-network data fusion. Thus, the energy
should be spent to increase the number of samples by enlarging the
coverage area even if it yields less accurate samples. In this
way, we can achieve the information increase with rate at least
$\Theta(E_t^{2/3})$ which is much faster than the logarithmic
increase obtained by increasing the sensing energy.

\subsection{Optimal Node Density}

In the previous section,  we investigated the asymptotic behavior
of the total information obtainable from the network and the
energy efficiency as the coverage, density or energy change. We
now consider another important problem in sensor network design
for statistical inference about underlying random fields, namely,
the optimal density problem. Here, we are given a fixed coverage
area, and are interested in determining an
 optimal node density. The total amount of information gathered
from the network increases monotonically (even if it has an
upperbound) as we increase the node density, as shown in Theorem
\ref{theo:infinitedensitymodel}. Hence, the problem cannot be
properly formulated without some constraint. We consider a total
energy constraint in which  a fixed amount of energy is available
to the entire network for both sensing and communication. Thus, we
consider the following problem.

\vspace{0.5em}
\begin{problem}[Optimal density] Given a fixed coverage area with size $L \times L$  and
total available energy $E_t$, find the density $\mu_n$ that
maximizes the total information $I_t$ obtainable from the sensor
network.
\end{problem}
\vspace{0.5em}

\noindent The above optimization problem can be solved using our
analysis based on the large deviations principle assuming the
asymptotic result is still valid in the low density case, and the
optimal density for the KLI measure is given by
\begin{eqnarray}
\mu_{n}^* &=& \mathop{\arg \max}_{\mu_n} ~L^2 \mu_n \Kmsc_s
(\mbox{SNR}(E_t,\mu_n), d_n(\mu_n)),\\
&& \mbox{s.t.} ~~~n^2 E_s (\mu_n) + \frac{1}{2}n(n-1)(n+1)
E_c(d_n(\mu_n)) \le E_t, \label{eq:optimaldensityconstraint}
\end{eqnarray}
where the sensing energy $E_s$ as well as $n$ and $d_n$ are
functions of the node density $\mu_n$.  From $\mu_n ~(= n^2/L^2)$,
we first calculate $n$ and then $d_n = L/n$. (Here, the
quantization of $n$ to the nearest integer is not performed.) With
the determined $d_n$, $E_c(d_n)$ is obtained from the propagation
parameters $E_0$ and $\nu$, and then $E_s(\mu_n)$ is obtained from
the constraint (\ref{eq:optimaldensityconstraint}). When
$E_s(\mu_n)$ is determined, the measurement SNR is calculated
using Assumption {\em (A.5)}, i.e., SNR = $\beta E_s$, and finally
we evaluate the per-node information $\Kmsc_s(\SNR,
\zeta(\rho(d_n)))$ and $\Imsc_s(\SNR, \zeta(\rho(d_n)))$ from
Corollary \ref{corol:eeSFA}.

\begin{figure}[htbp]
\centerline{ \SetLabels
\L(0.24*-0.1) (a) \\
\L(0.74*-0.1) (b) \\
\endSetLabels
\leavevmode
\strut\AffixLabels{
\scalefig{0.55}\epsfbox{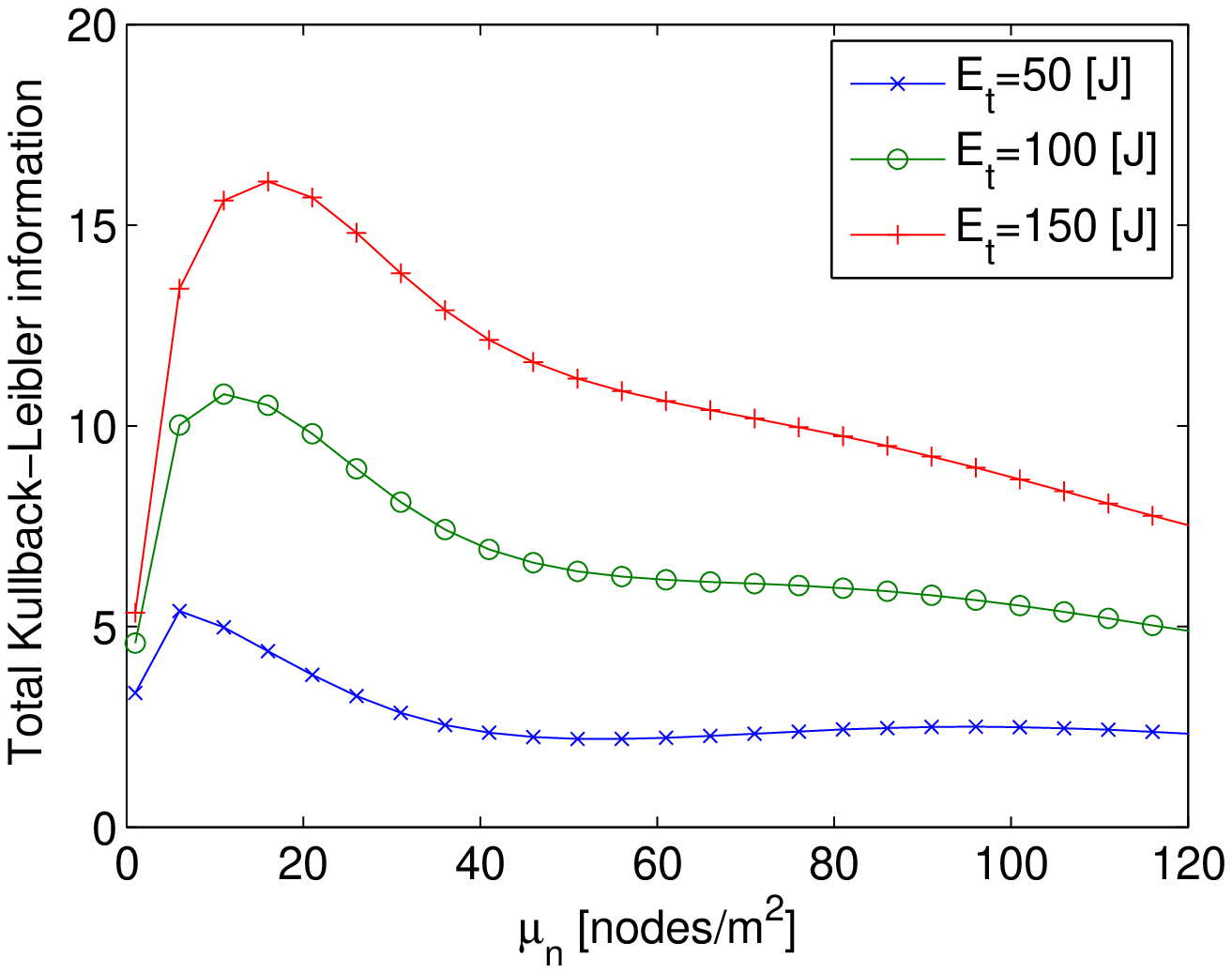}
\scalefig{0.55}\epsfbox{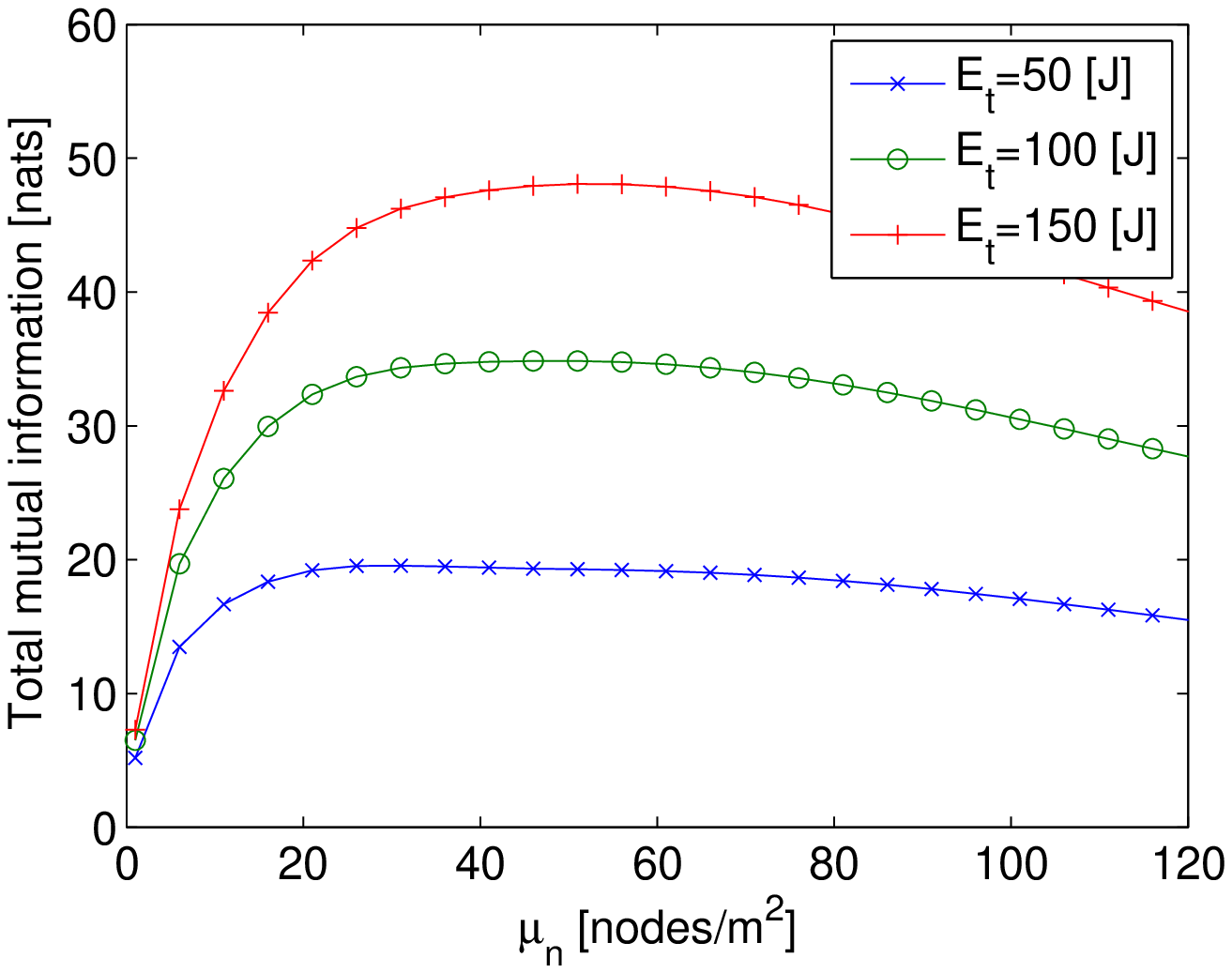} } }
\vspace{0.3cm}
 \caption{(a) total KLI vs. density  and (b) total MI vs. density}
\label{fig:InformationVsDensity}
\end{figure}

Fig. \ref{fig:InformationVsDensity} shows the total information
obtainable from a 2 meter $\times$ 2 meter area as we change the
node density $\mu_n$ with a fixed total energy budget of $E_t$
joules. Other parameters that we use are given by
\[
\alpha = 100, ~\beta=1, ~ E_0 = 0.1~ \mbox{and}~ \nu = 2.
\]
Here, the values of $E_t$, $E_0$ and $\beta$ are selected so that
the minimum and maximum per-node  sensing SNRs are roughly -10 to
10 dB for maximum and minimum densities, respectively.  The
diffusion rate $\alpha=100$ is chosen for the edge correlation
coefficient $\rho$ to range from almost zero to 0.6 as the node
density varies. It is seen in the figure that there is an optimal
density for each value of $E_t$ under either information measure.
It is also seen that the total KLI is sensitive to the density
change whereas the total MI is less sensitive. The existence of
the optimal density is explained as follows. At low densities, we
have only a few sensors in the area. So, the energy for
communication is not large due to the small number of
communicating nodes (see (\ref{eq:infinitedensityproof4}) below)
and most of the energy is allocated to sensing. Here, the per-node
sensing energy is even higher due to the small number of sensors.
However, the per-node information increases only logarithmically
w.r.t. the sensing energy or SNR by Theorem
\ref{theo:energyasymptotic}, and this logarithmic gain cannot
compensate for the loss in the number of sensors. Hence, low
density yields very poor performance, and large gain is obtained
initially as we increase the density from very low values, as seen
in Fig. \ref{fig:InformationVsDensity}. As we further increase the
density, on the other hand, the per-node sensing energy or SNR
decreases due to the increase in the overall communication and the
increase in the number of sensor nodes, and the measurement SNR is
 in the  low SNR regime eventually, where (\ref{eq:KLIvsSNRlowSNR}) and
(\ref{eq:MIvsSNRlowSNR}) hold. From (\ref{eq:adhocTotalEnergy}),
we have
\begin{equation}
E_s (\mu_n) =\beta^{-1}\SNR = O(n^{-2})
\end{equation}
for fixed $E_t$ and $E_c =E_0 (L/n)^2$, as $n \rightarrow \infty$.
By the quadratic decaying behavior of $\Kmsc_s$ at low SNR given
by (\ref{eq:KLIvsSNRlowSNR}), the total Kullback-Leibler
information is given by
\[
\mbox{Total KLI}= L^2 \mu_n \Kmsc_s = O(n^2 n^{-4}) = O(n^{-2}) =
O(\mu_n^{-1}).
\]
By (\ref{eq:MIvsSNRlowSNR}), on the other hand, the mutual
information decays linearly as SNR decreases to zero, and the
total mutual information is given by
\[
\mbox{Total MI} = L^2 \mu_n \Imsc_s = O(n^2 n^{-2}) = O(1).
\]
This explains the initial fast decay after the peak in Fig.
\ref{fig:InformationVsDensity} (a) and flat curve in Fig.
\ref{fig:InformationVsDensity} (b).  In the above equations,
however, the effect of $\zeta$  on $\Kmsc_s$ and $\Imsc_s$ is not
considered. As the node density increases, the sensor spacing
decreases and the edge dependence factor $\zeta$ increases for a
given diffusion rate $\alpha$. The behavior of the per-node
information as a function of $\zeta$ is shown in Fig.
\ref{fig:KcsVsZeta}. Note in Fig. \ref{fig:KcsVsZeta} that the
per-node information  has a second lobe at strong correlation at
low SNR while at high SNR it  decreases monotonically as the
correlation becomes strong. The benefit of sample correlation is
evident in the low energy case ($E_t=50 [\mbox{J}]$) in
\ref{fig:InformationVsDensity} (a); the second peak  around $\mu_n
= 95$ [nodes/$m^2$] is observed. Note that the second peak is not
very significant. Since the per-node  information decays to zero
as $\zeta \rightarrow 1/4$ eventually, the total amount of
information decreases eventually, as seen in the right corner of
the figure, as we increases the node density further.

\section{Conclusion and Discussion}\label{sec:conclusion}

In this paper, we have considered the design of sensor networks
for statistical inference about correlated random fields in a 2-D
setting.   To quantify the information from the sensor network, we
have used a spectral domain approach to derive  closed-form
expressions for asymptotic KLI and MI rates in general $d$-D and
in 2-D in particular, and
 have adopted the 2-D hidden CAR GMRF for our signal model to capture
the spatial correlation and measurement noise for samples in a 2-D
sensor field. Under the first order symmetry assumption, we have
further obtained the asymptotic information rates explicitly
 in terms of the SNR and the edge dependence factor, and have investigated the properties of the asymptotic information rates
 as functions of SNR and correlation. Based on these LDP results, we have then analyzed the asymptotic
behavior of {\em ad hoc} sensor networks deployed over 2-D
correlated random fields for statistical inference. Under the
SFCAR GMRF model, we have obtained fundamental scaling laws for
total information and energy efficiency as the coverage, node
density  and consumed energy change. The results provide
guidelines for sensor network design for statistical inference
about 2-D correlated random fields such as temperature, humidity,
or density of a gas on a certain area.

In closing, we discuss several issues related to some of the
assumptions we have used to simplify our analysis. First, of
course, sensors in a real network may not be located on a 2-D
grid. However, we conjecture that similar scaling behaviors w.r.t.
the coverage, density and energy are valid for randomly and
uniformly deployed sensors. Secondly, the spatial Markov
assumption may be restrictive. However, it is a minimal model that
captures the two dimensionality of the signal correlation
structure in all planar directions and allows analysis to be
tractable. And, finally we have not considered the temporal
evolution of the spatial signal field. In case of i.i.d. temporal
variation, the results here can be applied directly without
modification. When the signal variation over time is correlated,
the modification to spatio-temporal fields is required.

\section*{Appendix I}

{\em Proof of Theorem~\ref{theo:KLI}}

\noindent The asymptotic KLI rate $\Kmsc$ is given by the
almost-sure limit
\begin{equation}
\Kmsc= \lim_{n\rightarrow\infty} \frac{1}{|\Dc_n|} \log
\frac{p_{0}}{p_{1}}(\{Y_{\ibf}, \ibf \in \Dc_n\}),
\end{equation}
evaluated under $p_{0}$\cite{Vajda:book}.   We consider the
following index mapping from $d$-D to 1-D in lexicographic order:
\begin{equation}  \label{eq:indexMapping}
l = f_{id}(\ibf), ~~~(\ibf \in [0,1,\cdots,n-1]^d),
\end{equation}
and the corresponding observation vector $\ybf_{|\Dc_n|}$
generated from $\{Y_{\ibf}, ~\ibf \in \Dc_n\}$. Then,
$\ybf_{|\Dc_n|}$ is a zero-mean Gaussian vector with the
covariance matrices $\Sigmabf_{0,|\Dc_n|}$ and
$\Sigmabf_{1,|\Dc_n|}$ under $p_0$ and $p_1$, respectively. Hence,
the asymptotic KLI rate is given by
\begin{eqnarray} \label{eq:proofKmscRoot}
\Kmsc = \lim_{n\rightarrow \infty} \frac{1}{|\Dc_n|} \left(
\frac{1}{2}\log
\frac{\det(\Sigmabf_{1,|\Dc_n|})}{\det(\Sigmabf_{0,|\Dc_n|})} +
\frac{1}{2} \ybf_{|\Dc_n|}^T (\Sigmabf_{1,|\Dc_n|}^{-1} -
\Sigmabf_{0,|\Dc_n|}^{-1})\ybf_{|\Dc_n|} \right),
\end{eqnarray}
under $p_0$. Now we consider the terms on the RHS of
(\ref{eq:proofKmscRoot}). First, we consider  $\log \det (
\Sigmabf_{0,|\Dc_n|})$. Since $\Sigmabf_{0,|\Dc_n|} = \sigma^2
\Ibf_{n^d}$ under the assumption of an i.i.d. null distribution,
we simply have
\begin{equation}  \label{eq:proofTermOne}
\frac{1}{|\Dc_n|} \log \det \Sigmabf_{0,|\Dc_n|} = \frac{1}{n^d}
\log \det ( \sigma^2 \Ibf_{n^d}) =  \log \sigma^2.
\end{equation}
Next we consider the term $\frac{1}{|\Dc_n|}  \ybf_{|\Dc_n|}^T
\Sigmabf_{0,|\Dc_n|}^{-1}\ybf_{|\Dc_n|}$.  Since $\ybf_{|\Dc_n|}$
is i.i.d. Gaussian, $d$-D is irrelevant in this case,   the known
result  from  \cite[Proposition 10.8.3]{Brockwell&Davis:book}  is
applicable, and we have
\begin{equation}  \label{eq:proofTermTwo}
\frac{1}{|\Dc_n|} \ybf_{|\Dc_n|}^T \Sigmabf_{0,|\Dc_n|}^{-1}
\ybf_{|\Dc_n|} \rightarrow 1 ~~~\mbox{almost surely,}
\end{equation}
assuming that the random vector $\ybf_{|\Dc_n|}$ is generated from
the distribution $p_0$.  Now we consider the term
$\frac{1}{|\Dc_n|} \log \det \Sigmabf_{1,|\Dc_n|}$. This is the
entropy rate of a $d$-D Gaussian process, and the convergence
behavior of this term is studied in \cite{Kent&Mardia:96JSPI}. It
is shown in  \cite[p. 391]{Kent&Mardia:96JSPI} under the
assumption in Theorem~\ref{theo:KLI}  that we have
\[
\left|\log \det \Sigmabf_{1,|\Dc_n|} - \frac{|\Dc_n|}{(2\pi)^d}
\int_{[-\pi,\pi)^d} \log ((2\pi)^d f_1(\omegabf))d\omegabf\right|
= O \left( \frac{|\Dc_n|}{n}\right).
\]
Applying this result, we have
\begin{equation}  \label{eq:proofTermThree}
\frac{1}{|\Dc_n|} \log \det \Sigmabf_{1,|\Dc_n|} \rightarrow
\frac{1}{(2\pi)^d} \int_{[-\pi,\pi)^d} \log((2\pi)^d
f_1(\omegabf))d\omegabf.
\end{equation}
Finally, we consider the random term $\frac{1}{|\Dc_n|}
\ybf_{|\Dc_n|}^T
\Sigmabf_{1,|\Dc_n|}^{-1}\ybf_{|\Dc_n|}$.\footnote{The proof given
in \cite{Brockwell&Davis:book} and \cite{Hannon:73JAP}   for the
convergence of this term for the 1-D index case is not applicable
for general $d$-D, nor is the almost-sure convergence of the term
shown in \cite{Kent&Mardia:96JSPI}, where the convergence  of the
term in probability to an integral involving the periodogram was
shown. Thus, we prove the almost-sure convergence of the term in
Lemma \ref{lem:4thTerm} separately in Appendix II.}
  By
Lemma \ref{lem:4thTerm} in Appendix II,  we have
\begin{equation}   \label{eq:proofTermFour}
\frac{1}{|\Dc_n|} \ybf_{|\Dc_n|}^T \Sigmabf_{1,|\Dc_n|}^{-1}
\ybf_{|\Dc_n|} \rightarrow \frac{1}{(2\pi)^d} \int_{[-\pi,\pi)^d}
\frac{\sigma^2}{(2\pi)^df_1(\omegabf)}d\omegabf,
\end{equation}
almost surely as $n\rightarrow \infty$.

Combining (\ref{eq:proofKmscRoot}) - (\ref{eq:proofTermFour}), we
have
\begin{equation} \label{eq:proofTheo2Last}
\Kmsc = \frac{1}{(2\pi)^d} \int_{[-\pi,\pi)^d}  \biggl[
\frac{1}{2}\log \frac{(2\pi)^df_1(\omegabf)}{\sigma^2}
 -\frac{1}{2}\left(1 -
\frac{\sigma^2}{(2\pi)^df_1(\omegabf)} \right) \biggr]d\omegabf.
\end{equation}
Since
\begin{equation}
D(\Nc(0,\sigma_0^2)||\Nc(0,\sigma_1^2)) = \frac{1}{2} \log
\frac{\sigma_1^2}{\sigma_0^2} -\frac{1}{2}\left(1-
\frac{\sigma_0^2}{\sigma_1^2} \right),
\end{equation}
(\ref{eq:proofTheo2Last}) is given by
\begin{equation}
\Kmsc = \frac{1}{(2\pi)^d} \int_{[-\pi,\pi)^d}
D(\Nc(0,\sigma^2)||\Nc(0,(2\pi)^df_1(\omegabf))) d\omegabf.
\end{equation}
\hfill{$\blacksquare$}

\vspace{1em} {\em Proof of Corollary \ref{cor:KLI2D}}

\noindent  For the 2-D hidden model we have
\begin{equation}  \label{eq:cor1-2D}
f_1(\omega_1,\omega_2) = (2\pi)^{-2} \sigma^2 +
f(\omega_1,\omega_2),
\end{equation}
where $f(\omega_1,\omega_2)$ is the CAR spectrum
(\ref{eq:CARspectrum}) in 2-D satisfying (\ref{eq:CARcond1}) and
(\ref{eq:CARcond4}). First, $f_1(\omega_1,\omega_2)$ has a
positive lower bound, and thus satisfies Assumption {\textit{A.1}}
in Theorem \ref{theo:KLI}. It is also known in \cite{Guyon:book}
that if $\kbf=(k_1,\cdots,k_d) \in {\mathbb{N}}^d$ and if
$f_1(\omegabf)$ is of class $C^\kbf$ (i.e., differentiable up to
the $k_d$-order w.r.t. $\omega_d$), then
\begin{equation}
\mathop{\lim \sup}_{ \hbf \rightarrow \infty}
h_1^{k_1}h_2^{k_2}\cdots h_d^{k_d} |\gamma_\hbf| < \infty,
\end{equation}
where ${\mathbb{N}}$ is the set of all natural numbers, and $\hbf
\rightarrow \infty$ means that at least one coordinate tends to
infinity. Under the condition (\ref{eq:CARcond1}) and
(\ref{eq:CARcond4}), the hidden CAR spectrum
$f_1(\omega_1,\omega_2)$ in (\ref{eq:cor1-2D}) is
$C^{(\infty,\infty)}$, i.e., smooth both in $\omega_1$ and
$\omega_2$. This ensures that Assumption {\textit{A.2}} in Theorem
\ref{theo:KLI} is satisfied, and  the corollary follows by
substituting (\ref{eq:cor1-2D}) and $d=2$ into
(\ref{eq:errorexponentspectraldD}). \hfill{$\blacksquare$}

\vspace{1em} {\em Proof of Theorem~\ref{theo:KLIsvsSNR}}

\noindent The continuity is straightforward. The monotonicity is
shown as follows. Let $s=1 + \SNR g_\zeta(\omegabf)$ where
$g_\zeta(\omegabf) = ((2/\pi)K(4\zeta)(1-2\zeta\cos\omega_1-2\zeta
\cos \omega_2))^{-1}$. Then, the partial derivative of $\Kmsc_s$
w.r.t. $\SNR$ is given by
\begin{equation}
\frac{\partial \Kmsc_s}{\partial \SNR} = \frac{1}{(2\pi)^2}
\int_{\omegabf \in [-\pi,\pi)^2} \frac{\partial }{\partial
s}\left(\frac{1}{2} \log s + \frac{1}{2s} -\frac{1}{2}\right)
\frac{\partial s}{\partial \SNR } d\omegabf,
\end{equation}
where
\begin{equation}
 \frac{\partial }{\partial s}
\left(\frac{1}{2}\log s +\frac{1}{2s} -\frac{1}{2}\right) =
\frac{1}{2}\frac{s-1}{s^2}=\frac{1}{2}\frac{\SNR
g_\zeta(\omegabf)}{s^2} \ge 0,
\end{equation}
and
\begin{equation}
\frac{\partial s}{\partial \SNR } = g_\zeta(\omegabf) \ge 0
\end{equation}
for $0\le \zeta \le 1/4$.  Hence,
\begin{equation}
\frac{\partial \Kmsc_s}{\partial \SNR} \ge 0,
\end{equation}
and  $\Kmsc_s$ increases monotonically as SNR increases for a
given $\zeta$ ($ 0 \le a \le 1/4$).

As SNR $\rightarrow \infty$, we have
\begin{eqnarray*}
\Kmsc_s &\approx& \frac{1}{(2\pi)^2} \int_{\omegabf \in
[-\pi,\pi)^2} \frac{1}{2} \log (\SNR g_\zeta(\omegabf))
d\omegabf,\\
&=& \frac{1}{2} \log \SNR + \frac{1}{(2\pi)^2} \int_{\omegabf \in
[-\pi,\pi)^2} \frac{1}{2} \log (g_\zeta(\omegabf)) d\omegabf.
\end{eqnarray*}
Thus, we have $\frac{1}{2} \log \SNR$ behavior at high SNR.

For (\ref{eq:KLIvsSNRlowSNR}) and (\ref{eq:MIvsSNRlowSNR}), take
the Taylor expansion around $\SNR =0$ to obtain
\begin{eqnarray*}
\log ( 1 + \SNR g_\zeta(\omegabf)) &=&  \SNR g_\zeta(\omegabf) -
\SNR^2 g_\zeta^2(\omegabf)/2+ \cdots,\\
\frac{1}{1+\SNR g_\zeta(\omegabf)} &=& 1 - \SNR g_\zeta(\omegabf)
+ \SNR^2 g_\zeta^2(\omegabf) - \cdots,
\end{eqnarray*}
and then integrate. \hfill{$\blacksquare$}

\vspace{1em} {\em Proof of
Theorem~\ref{theo:infiniteareafixeddensity}}

\noindent In this case, the edge length $d_n=d$ for all $n$, and
thus the asymptotic per-sensor information $\Kmsc_s(d_n)$ or
$\Imsc_s(d_n)$ does not change with $n$. Considering the
Kullback-Leibler information, we have $I_t = n^2 \Kmsc_s(d)$, and
$\mbox{area} = \Theta(n^2)$. Hence, the total information is
linear w.r.t. area.  The total energy $E_t$ required for data
gathering is given by
\begin{eqnarray}
E_t &=& n^2 E_s + E_c(d) \sum_{i=0}^{n-1}\sum_{j=0}^{n-1}
(|i-\lfloor n/2\rfloor|+|j-\lfloor n/2\rfloor|),\nonumber\\
&=& n^2 E_s + \Theta(n^3) E_{c}(d),
\end{eqnarray}
where the first term is the sensing energy and the second term is
the energy consumed for communication. The energy efficiency is
given by
\begin{equation}
\eta = \frac{n^2\Kmsc_s(d)}{n^2 E_s + \Theta(n^3) E_{c}(d)} =
\Theta\left(\frac{1}{n}\right),
\end{equation}
as $n\rightarrow \infty$.  Since $\mbox{area} = \Theta(n^2)$,
(\ref{eq:efficiencyIAM}) follows.

For the second statement we have $E_t = \Theta(n^3)$. The total
information is given by $n^2 \Kmsc_s(\mbox{SNR},d)$. Since
$\Kmsc_s$ is fixed, the total information is $\Theta(n^2)$ as
$n\rightarrow \infty$, and we have (\ref{eq:energyAsymptotic2}).
\hfill{$\blacksquare$}

\vspace{1em} {\em Proof of
Theorem~\ref{theo:informationvsdninfty}}

\noindent The proof is by the asymptotic behavior of the modified
Bessel function  $K_1(\cdot)$ of the second kind and Taylor
expansion of $\Kmsc_s$ (as a function of $\zeta$) and $\zeta$ (as
a function of $\rho$), which is allowed because of their
continuous differentiability. From (\ref{eq:2DcorrelationFunc})
and (\ref{eq:modifiedBessel}) we have
\begin{equation} \label{eq:proofTheoremdninfty}
\rho(d_n) = \sqrt{\frac{\pi}{2}} \alpha d_n e^{-\alpha d_n} +
o\left(\alpha d_n e^{-\alpha d_n}\right)
\end{equation}
as $d_n\rightarrow\infty$. From the continuous differentiability
of $\Kmsc_s$ as a function of $\zeta$  in (\ref{eq:Kmsc0Zeta}) and
$\zeta$ as a function of $\rho$, we have
\begin{eqnarray*}
\Kmsc_s  &=& D(\Nc(0,1)||\Nc(0,1+\SNR)) - c_2 \zeta +
o(\zeta),\\
&=& D(\Nc(0,1)||\Nc(0,1+\SNR)) - c_2 (c_7 \rho +o(\rho)) +
o(c_7 \rho +o(\rho)),\\
 &=& D(\Nc(0,1)||\Nc(0,1+\SNR)) -
c_2c_7 \rho + o(\rho),
\end{eqnarray*}
for some $c_2, c_7 > 0$. Applying (\ref{eq:proofTheoremdninfty})
to the above equation, we have (\ref{eq:theoremdninfty}). The
proof for the mutual information $\Imsc_s$ is similar.
\hfill{$\blacksquare$}

\vspace{1em} {\em Proof of Theorem
\ref{theo:infinitedensitymodel}}

\noindent Consider a fixed area with size $L \times L$ and a
lattice $\Ic_n$ on it. The sensor spacing $d_n$ for $n$ is given
by
\[
d_n = \frac{L}{n}.
\]
By (\ref{eq:K1flatTop}), we have
\begin{equation}  \label{eq:infinitedensityproof1}
\rho (d_n) = 1+ c_8 \cdot d_n^2 + o(d_n^2)
\end{equation}
for some constant $c_8$.  By the continuous differentiability of
$\Kmsc_s$ (as a function of $\zeta$) and $\zeta$ (as a function of
$\rho$), we have
\[
\zeta = \frac{1}{4} + c_9 \cdot (1-\rho) + o((1-\rho)^2),
\]
and
\[
\Kmsc_s = c_1 \cdot (\zeta - 1/4) + o(\zeta - 1/4),
\]
for some constant $c_9$.  Substituting
(\ref{eq:infinitedensityproof1}) into the above equations gives
\begin{equation}  \label{eq:infinitedensityproof1p1}
\Kmsc_s = c_{10} \cdot d_n^2 + o(d_n^2),
\end{equation}
for some constant $c_{10}$. The node density is given by
\begin{equation} \label{eq:infinitedensityproof2}
\mu_n = \frac{n^2}{L^2} = d_n^{-2}.
\end{equation}
Substituting (\ref{eq:infinitedensityproof2}) into
(\ref{eq:infinitedensityproof1p1}) yields
(\ref{eq:infintedensityKs}).  The total amount of information per
unit area is given by
\begin{equation} \label{eq:infinitedensityproof3}
\mu_n \Kmsc_s =  c_5 +  o(1),
\end{equation}
and it converges to $c_5$ as $n \rightarrow \infty$.

To calculate the energy efficiency, we first calculate the total
communication energy consumed by the minimum hop routing, given by
\begin{eqnarray}
E_t^\prime &=& E_c(d_n) \sum_{i=0}^{n-1}\sum_{j=0}^{n-1}
(|i-\lfloor n/2 \rfloor|+|j-\lfloor n/2 \rfloor|),\nonumber\\
&=&  \Theta(n^3) E_{c}(d_n) = E_0 L^\nu n^{-\nu}\Theta(n^3), \nonumber\\
&=& \Theta( n^{3-\nu}) \label{eq:infinitedensityproof4},
\end{eqnarray}
as $n \rightarrow \infty$ (i.e., $\mu_n \rightarrow \infty$).
Here, $E_t^\prime$ denotes the total energy considering only the
communication energy. The energy efficiency in this case is given
by
\begin{equation}
\eta^\prime =  \frac{\mu_n \Kmsc_s}{E_t^\prime}
~~[\mbox{nats}/J/m^2].
\end{equation}
Applying (\ref{eq:infinitedensityproof3}) and
(\ref{eq:infinitedensityproof4}) to the above equation, we have
the claims. \hfill{$\blacksquare$}

\vspace{1em} {\em Proof of Theorem~\ref{theo:energyasymptotic}}

\noindent Note that
\[
E_t = n^2 E_s + \Theta(n^3)E_c(d_n).
\]
In this  case, $n$ and $d_n$ are fixed, and Theorem
\ref{theo:KLIsvsSNR} is directly applicable. Since the number of
nodes and communication energy are fixed, the sensing energy
increases linearly with the total energy $E_t$. By Assumption {\em
(A.5)}, the measurement SNR increases linearly with the sensing
energy. Applying Theorem \ref{theo:KLIsvsSNR} yields
(\ref{eq:energyAsymptotic1}).
 \hfill{$\blacksquare$}

\section*{Appendix II}

\vspace{1em} \noindent To prove Lemma \ref{lem:4thTerm} (this will
be stated below), we briefly introduce some relevant preliminary
results.

\vspace{1em}
\begin{definition}[Matrix norms \cite{Kent&Mardia:96JSPI,Gohberg&Goldberg&Kaashoek:book}] Let $\Abf$ be an $n
\times n$ matrix with singular value decomposition
\begin{equation}
\Abf = \Ubf \Sbf \Vbf^T = \sum_{i=1}^n s_i \ubf_i \vbf_i^T,
\end{equation}
where $\Ubf$ and $\Vbf$ are unitary matrices with columns $\ubf_i$
and $\vbf_i$, respectively, and $\Sbf = \mbox{diag}(s_1,s_2,
\cdots, s_n)$ with nonnegative elements $s_1 \ge s_2 \ge \cdots
s_n \ge 0$. The {\em operator norm}  of $\|\Abf\|$ is defined as
\begin{equation}
\|\Abf \| = s_1 = \sup_{\xbf \ne {\mathbf 0}} \| \Abf \xbf \|
/\|\xbf\|,
\end{equation}
where $\|\xbf\|$ denotes the 2-norm of $\xbf$. On the other hand,
the {\em trace class norm} of $\Abf$ is defined as
\begin{equation}
\| \Abf \|_1 = \sum_{i} s_i.
\end{equation}
Note that if $\Abf$ is a symmetric matrix with eigenvalues
$\{\lambda_i\}$, then
\begin{equation} \label{eq:matrixNormOp}
\| \Abf \|_1 = \sum_i |\lambda_i|.
\end{equation}
\end{definition}


\vspace{1em} \begin{remark}[The covariance matrix and its
circulant approximation]

Using vector notation, the covariance matrix of the vector
$\ybf_{|\Dc_n|}$ in (\ref{eq:2Dto1DconversionVector}) under $p_1$
is given by
\begin{equation} \label{eq:Sigmabf1usinggamma}
\Sigmabf_{1,|\Dc_n|} = \Ebb_1 \{ \ybf_{|\Dc_n|} \ybf_{|\Dc_n|}^T
\} = [ \sigma_{f_{id}^{-1}(\ibf), f_{id}^{-1}(\jbf)} ],
~~~\sigma_{f_{id}^{-1}(\ibf), f_{id}^{-1}(\jbf)} = \gamma_{\ibf -
\jbf}, ~~~\ibf, \jbf \in \Dc_n,
\end{equation}
where $\gammabf_\hbf$ is defined in (\ref{eq:IDTFTdD}) and
$f_{id}$ is defined in (\ref{eq:indexMapping}).  With slight abuse
of notation, we use $\sigma_{\ibf \jbf}$ for
$\sigma_{f_{id}^{-1}(\ibf), f_{id}^{-1}(\jbf)}$ for the sake of
exposition.

 The circulant approximation $\Cbf_{|\Dc_n|}$ to
$\Sigmabf_{1,|\Dc_n|}$ is obtained by treating $\Dc_n$ as a high
dimensional torus with opposite ends being neighbors, and
$\Cbf_{|\Dc_n|}$ is given by
\begin{equation}
\Cbf_{|\Dc_n|} = [c_{\ibf \jbf}], ~~ c_{\ibf \jbf} =
\gamma_{\pi(\ibf - \jbf)}, ~~~\ibf, \jbf \in \Ic_n,
\end{equation}
where the mapping $\pi: \Zbb^d \rightarrow \Zbb^d$ is defined as
\begin{equation}
\pi(\hbf)=\pi(h_1,h_2,\cdots, h_d) = (h_1^\prime,
h_2^\prime,\cdots, h_d^\prime),
\end{equation}
and
\begin{equation}
h_k^\prime =  h_k I(|h_k| \le n/2) + (n-|h_k|)I(|h_k| > n/2),
~~k=1,\cdots,d. \footnote{The distinction of even and odd $n$ will
not be considered for simplicity, as this is merely a technical
issue. In either case, the asymptotic behavior is the same.}
\end{equation}
Here, $I(\cdot)$ is the indicator function. Note that
$\Sigmabf_{1,|\Dc_n|}$ is a block Toeplitz matrix, while
$\Cbf_{|\Dc_n|}$ is a block circulant matrix.  It is known that
the eigenvalues of the block circulant matrix $\Cbf_{|\Dc_n|}$ are
given by
\begin{equation}
\lambda_\ibf =\sum_{\hbf \in \Dc_n} \gamma_{\pi(\hbf)} e^{\iota
\hbf \cdot \omegabf_\ibf},
\end{equation}
for $\ibf =(i_1, \cdots, i_d) \in \Dc_n$, where
\begin{equation}
\omegabf_\ibf = \left( \omega_{i_1}, \omega_{i_2}, \cdots,
\omega_{i_d} \right) = \left( \frac{2\pi i_1}{n}, \frac{2\pi
i_2}{n},\cdots, \frac{2\pi i_d}{n}  \right).
\end{equation}
Define the periodic approximate spectral density by
\begin{equation}
f_n^c(\omegabf) = (2\pi)^{-d} \sum_{\hbf \in \Dc_n}
\gamma_{\pi(\hbf)} e^{\iota \hbf \cdot \omegabf}.
\end{equation}
Then, the eigenvalues of $\Cbf_{|\Dc_n|}$ are given by
\begin{equation}   \label{eq:eigValSpectrum}
\lambda_\ibf = (2\pi)^d f_n^c(\omegabf_\ibf), ~~~\ibf \in \Dc_n.
\end{equation}
Further, it is shown in \cite[Lemma 4.1.(c)]{Kent&Mardia:96JSPI}
that the periodic approximate spectral density converges uniformly
to the true spectral density $f_1(\omegabf)$, i.e.,
\begin{equation}  \label{eq:eigenCircConvUniform}
\sup_{\omegabf \in [-\pi,\pi)^d} | f_n^c(\omegabf) -
f_1(\omegabf)| \rightarrow 0,
\end{equation}
as $n\rightarrow \infty$.
\end{remark}

\vspace{1em}
\begin{lemma} \label{lem:uniformConvInv}
Under the assumption of Theorem~\ref{theo:KLI}, we have
\begin{itemize}
\item[(a)]  $f_n^c(\omegabf)$ is uniformly continuous for sufficiently large $n$. %
\item[(b)] \begin{equation} \label{eq:eigenCircConvUniform2}
\sup_{\omegabf \in [-\pi,\pi)^d} \left| \frac{1}{f_n^c(\omegabf)}
- \frac{1}{f_1(\omegabf)} \right| \rightarrow 0 ~~\mbox{as}~ n
\rightarrow \infty.
\end{equation}
\item[(c)] $1/f_n^c(\omegabf)$ is uniformly continuous for sufficiently large $n$. %
\end{itemize}
\end{lemma}

\vspace{0.5em} {\em Proof of Lemma~\ref{lem:uniformConvInv}}

\noindent

\begin{itemize}
\item[(a)] By assumption, $f_1(\omegabf)$ is continuous on the
compact domain $[-\pi,\pi]^d$. By the uniform continuity theorem,
$f_1(\omegabf)$ is uniformly continuous. For any $\epsilon >0$,
$||\omegabf - \omegabf^\prime|| < \delta$ imples
\begin{eqnarray*}
\left| f_n^c(\omegabf) - f_n^c(\omegabf^\prime)  \right| &\le&
\left| f_n^c(\omegabf) -f_1(\omegabf) + f_1(\omegabf) -
f_1(\omegabf^\prime) + f_1(\omegabf^\prime)-
f_n^c(\omegabf^\prime) \right|,\\
&\le& | f_n^c(\omegabf) -f_1(\omegabf)| + |f_1(\omegabf) -
f_1(\omegabf^\prime)| + |f_1(\omegabf^\prime)-
f_n^c(\omegabf^\prime)|,\\
&\le& \epsilon/3 + \epsilon/3 + \epsilon/3,
\end{eqnarray*}
for sufficiently large $n$. The convergence of the first and third
terms is by (\ref{eq:eigenCircConvUniform}) and that of the second
term is by the uniform continuity of $f_1(\omegabf)$. %

\item[(b)] Since the spectrum $f_1(\omega)$  has a positive lower
bound by assumption, its inverse $1/f_1(\omegabf)$ is bounded from
 above. In addition, due to (\ref{eq:eigenCircConvUniform}) there
exists $M_1 > 0$ such that
\begin{equation}  \label{eq:uniformConvInv1}
\frac{1}{f_1(\omegabf)} \le M_1 ~~\mbox{and}~~
\frac{1}{f_n^c(\omegabf)} \le M_1,
\end{equation}
for all $\omegabf \in [-\pi,\pi)^d$ and for sufficiently large
$n$. Then, for any $\epsilon >0$
\begin{eqnarray}
\left| \frac{1}{f_n^c(\omegabf)} - \frac{1}{f_1(\omegabf)} \right|
&=& \left| \frac{1}{f_n^c(\omegabf)} \frac{1}{f_1(\omegabf)}
\right| \left| f_n^c(\omegabf) -f_1(\omegabf) \right|,\\
&\le& \epsilon M_1^2
\end{eqnarray}
for all $\omegabf \in [-\pi,\pi)^d$ and for sufficiently large
$n$, by (\ref{eq:eigenCircConvUniform}) and
(\ref{eq:uniformConvInv1}).

\item[(c)] For any $\epsilon >0$, $||\omegabf - \omegabf^\prime||
< \delta$ implies
\begin{eqnarray*}
 \left| \frac{1}{f_n^c(\omegabf)} -
\frac{1}{f_1(\omegabf^\prime)} \right| &\le& \left|
\frac{1}{f_n^c(\omegabf)} -\frac{1}{f_1(\omegabf)} +
\frac{1}{f_1(\omegabf)} - \frac{1}{f_1(\omegabf^\prime)} +
\frac{1}{f_1(\omegabf^\prime)}- \frac{1}{f_n^c(\omegabf^\prime)} \right|,\\
&\le& \left| \frac{1}{f_n^c(\omegabf)} -\frac{1}{f_1(\omegabf)}
\right|+ \left|\frac{1}{f_1(\omegabf)} -
\frac{1}{f_1(\omegabf^\prime)} \right| +\left|
\frac{1}{f_1(\omegabf^\prime)}- \frac{1}{f_n^c(\omegabf^\prime)} \right|,\\
&\le& \epsilon/3 + \epsilon/3 + \epsilon/3,
\end{eqnarray*}
for sufficiently large $n$. The convergence of the first and third
terms is by (\ref{eq:eigenCircConvUniform2}) and that of the
second term is by the uniform continuity of $1/f_1(\omegabf)$.
(The uniform continuity of $1/f_1(\omegabf)$ is obvious due to the
uniform continuity and strict positivity of $f_1(\omegabf)$.)
\hfill{$\blacksquare$}
\end{itemize}

\vspace{1em}
\begin{lemma} \label{lem:4thTerm} Under the conditions of Theorem~\ref{theo:KLI}, we have
\begin{eqnarray*}
\frac{1}{|\Dc_n|} \ybf_{|\Dc_n|}^T \Sigmabf_{1,|\Dc_n|}^{-1}
\ybf_{|\Dc_n|} &\rightarrow& \frac{1}{(2\pi)^d}
\int_{[-\pi,\pi)^d}
\frac{\sigma^2}{(2\pi)^df_1(\omegabf)}d\omegabf,
\end{eqnarray*}
almost surely.
\end{lemma}

\vspace{0.5em} {\em Proof of Lemma \ref{lem:4thTerm}}

\noindent First, it is shown  in  \cite[Lemma
4.1.(a)]{Kent&Mardia:96JSPI} that
\begin{equation}  \label{eq:lemmaImpStepAeq1}
|\Dc_n|^{-1}|| \Sigmabf_{1,|\Dc_n|} - \Cbf_{|\Dc_n|} ||_1 =
O\left( \frac{1}{n} \right),
\end{equation}
as $n \rightarrow \infty$. Let $\{\lambda_{|\Dc_n|}(i),
~i=1,2,\cdots, |\Dc_n|\}$ be the eigenvalues of
$|\Dc_n|^{-1}(\Sigmabf_{1,|\Dc_n|} - \Cbf_{|\Dc_n|})$, where
$|\Dc_n|=n^d$ for $d$-D. Then, by (\ref{eq:matrixNormOp}) and
(\ref{eq:lemmaImpStepAeq1}) we have
\begin{equation}  \label{eq:lemmaImpStepAeq2}
\sum_{i=1}^{n^d} |\lambda_{|\Dc_n|}(i)| = O\left( \frac{1}{n}
\right).
\end{equation}
Since the convergence of the eigenvalues of the block Toeplitz
matrix $\Sigmabf_{1,|\Dc_n|}$ and its block circulant
approximation $\Cbf_{|\Dc_n|}$ is uniform (The eigenvalues of
these matrices are the samples of the corresponding spectra for
sufficiently large $n$; see (\ref{eq:eigValSpectrum}) and
(\ref{eq:eigenCircConvUniform}).), $\min_{i}
|\lambda_{|\Dc_n|}(i)|$ and  $\max_{i} |\lambda_{|\Dc_n|}(i)|$
have the same convergence rate, i.e., there exist $M_2$, $M_3$ and
$r_n$ such that
\begin{equation} \label{eq:lemmaImpStepAeq3}
M_2 r_n \le   \min_{i} |\lambda_{|\Dc_n|}(i)| \le \max_{i}
|\lambda_{|\Dc_n|}(i)| \le M_3 r_n.
\end{equation}
By (\ref{eq:lemmaImpStepAeq2}) and (\ref{eq:lemmaImpStepAeq3}) we
have
\begin{equation}  \label{eq:lemmaImpStepAeq4}
r_n = O \left( \frac{1}{n^{d+1}} \right).
\end{equation}

Since the spectra $f_1(\omega)$ and $f_n^c(\omegabf)$ have
positive lower bounds by assumption, their inverses
$1/f_1(\omegabf)$ and $1/f_n^c(\omegabf)$ are bounded from above.
Hence, the eigenvalues of $\Sigmabf_{1,|\Dc_n|}^{-1}$ and
$\Cbf_{|\Dc_n|}^{-1}$ are bounded from above since the eigenvalues
of these matrices are the samples of the corresponding inverse
spectra for sufficiently large $n$, and thus we have
\begin{equation} \label{eq:lemmaImpStepAeq5}
||\Sigmabf_{1,|\Dc_n|}^{-1}|| < M_1 ~~\mbox{and}~~
||\Cbf_{|\Dc_n|}^{-1}|| < M_1
\end{equation}
for all sufficiently large $n$.

Now consider the error between two quadratic terms.
\begin{eqnarray}
&&\left| |\Dc_n|^{-1} \ybf_{|\Dc_n|}^{T} \Sigmabf_{1,|\Dc_n|}^{-1}
\ybf_{|\Dc_n|} - |\Dc_n|^{-1} \ybf_{|\Dc_n|}^{T}
\Cbf_{|\Dc_n|}^{-1} \ybf_{|\Dc_n|} \right| \nonumber\\
 &=& \left| |\Dc_n|^{-1} \ybf_{|\Dc_n|}^{T} \left(
\Sigmabf_{1,|\Dc_n|}^{-1} -\Cbf_{|\Dc_n|}^{-1} \right) \ybf_{|\Dc_n|} \right|, \nonumber \\
&=& \left| |\Dc_n|^{-1} \ybf_{|\Dc_n|}^{T} \Cbf_{|\Dc_n|}^{-1}
\left( \Cbf_{|\Dc_n|}  - \Sigmabf_{1,|\Dc_n|} \right)
\Sigmabf_{1,|\Dc_n|}^{-1}
 \ybf_{|\Dc_n|} \right|,  \nonumber \\
 &\stackrel{(a)}{\le}& C M_1^2 \sum_{i=1}^{|\Dc_n|} |\lambda_i| y_i^2,  \nonumber\\
 &\stackrel{(b)}{\le}& C M_1^2 M_3 r_n \sum_{i=1}^{|\Dc_n|} y_i^2, \nonumber\\
 &\stackrel{(c)}{\le}& C M_1^2 M_3 O\left( \frac{1}{n} \right) \frac{1}{n^d} \sum_{i=1}^{n^d}
 y_i^2, \nonumber \\
 &\stackrel{(d)}{\rightarrow}& 0 ~~\mbox{a.s.} \label{eq:lemmaImpStepAeq6}
\end{eqnarray}
for some $C > 0$. Here, step (a) is by (\ref{eq:lemmaImpStepAeq5})
and the definition of the trace class norm
(\ref{eq:matrixNormOp}), step (b) is by
(\ref{eq:lemmaImpStepAeq3}), and step (c) is by
(\ref{eq:lemmaImpStepAeq4}). Step (d) is by the strong law of
large numbers (SLLN) on the sample mean of $y_i^2$. Since
$\{y_i\}$ is i.i.d. $\Nc(0,\sigma^2)$ under $p_0$,
 $\frac{1}{n^2} \sum_{i=1}^{n^2} y_i^2 \rightarrow \sigma^2$
almost surely. Thus, the quadratic form using the block circulant
approximation converges almost surely to that based on the true
covariance matrix.

We next consider the asymptotic behavior of $|\Dc_n|^{-1}
\ybf_{|\Dc_n|}^{T} \Cbf_{|\Dc_n|}^{-1} \ybf_{|\Dc_n|}$. Since
$\Cbf_{|\Dc_n|}$ is
 a block circulant matrix, the eigendecomposition is given by
 \cite{Martin:86JAP,Davis:book}
\begin{equation}
\Cbf_{|\Dc_n|} =  \Wbf_{|\Dc_n|} \Lambdabf_{|\Dc_n|}
\Wbf_{|\Dc_n|}^H,
\end{equation}
where $\Wbf_{|\Dc_n|}$ is the $d$-dimensional discrete Fourier
transform (DFT) matrix which is unitary, and
\begin{equation}
\Lambdabf_{|\Dc_n|} =
\mbox{diag}(\lambda_{0,\cdots,0},\cdots,\lambda_{n-1,\cdots,n-1}
).
\end{equation}
The inverse of $\Cbf_{|\Dc_n|}$ is given by
\begin{equation}
 \Cbf_{|\Dc_n|}^{-1} = \Wbf_{|\Dc_n|} \Lambdabf_{|\Dc_n|}^{-1}
 \Wbf_{|\Dc_n|}^H.
\end{equation}
Define
\begin{equation}
\bar{\ybf}_{|\Dc_n|} = \Wbf_{|\Dc_n|}^H \ybf_{|\Dc_n|}.
\end{equation}
Then, $\bar{\ybf}_{|\Dc_n|}$ is a vector of i.i.d. Gaussian random
variables since $\Wbf_{|\Dc_n|}$ is unitary and $\ybf_{|\Dc_n|}$
is a vector with i.i.d. Gaussian elements under $p_0$. Thus,
$|\Dc_n|^{-1} \ybf_{|\Dc_n|}^{T} \Cbf_{|\Ic_n|}^{-1}
\ybf_{|\Dc_n|}$ is given by
\begin{eqnarray}
S_n =|\Dc_n|^{-1} \ybf_{|\Dc_n|}^{T} \Cbf_{|\Dc_n|}^{-1}
\ybf_{|\Dc_n|} &=& |\Dc_n|^{-1} \bar{\ybf}_{|\Dc_n|}^{T}
\Lambdabf_{|\Dc_n|}^{-1}
\bar{\ybf}_{|\Dc_n|},\nonumber\\
&=&  \frac{1}{n^d} \sum_{\ibf \in \Dc_n}
\frac{\bar{Y}_{\ibf}^2}{\lambda_\ibf},\\
&=&  \frac{1}{n^d} \sum_{i_1=0}^{n-1}\cdots \sum_{i_d=0}^{n-1}
\frac{\bar{Y}_{i_1,\cdots, i_d}^2}{\lambda_{i_1,\cdots,i_d}},
\label{eq:lemmaImpStepBeq1}
\end{eqnarray}
where $\{\bar{Y}_{\ibf}, ~\ibf \in \Dc_n\}$ is  i.i.d. zero-mean
Gaussian with variance $\sigma^2$.  For sufficiently large $n$,
fix $K$ ~($0< K < n$) and divide the indices of each dimension
such that
\begin{eqnarray*}
&&\Ic =[0,1,\cdots,n-1] = \Ic(0) \cup \Ic(1) \cup \cdots \Ic(K-1),\\
&&\Ic(i) \cap \Ic(j) = \phi~~~\mbox{if}~ i \ne j, ~~~\mbox{and}\\
&&|\Ic(0)|=\cdots=|\Ic(K-2|=\lfloor n/K  \rfloor, ~~ |\Ic(K-1)| =
n - (K-1) |\Ic(0)|.
\end{eqnarray*}
Then, (\ref{eq:lemmaImpStepBeq1}) is given by
\begin{equation}
S_n =  \frac{1}{K^d} \sum_{j_1=0}^{K-1} \cdots \sum_{j_d=0}^{K-1}
\left( \frac{1}{|\Ic(j_1)|\cdots|\Ic(j_d)|} \sum_{i_1 \in
\Ic(j_1)} \cdots \sum_{i_d \in \Ic(j_d)}
\frac{\bar{Y}_{i_1,\cdots,
i_d}^2}{\lambda_{i_1,\cdots,i_d}}\right).
\label{eq:lemmaImpStepBeq2}
\end{equation}
Now let $i_1,\cdots,i_d(j_1,\cdots,j_d)$ denote the index
representing the center of the $(j_1,\cdots,j_d)^{th}$ hypercube.
Then, by (\ref{eq:eigValSpectrum}) we have
\begin{equation}
\frac{1}{\lambda_{i_1,\cdots,i_d(j_1,\cdots,j_d)}} =
\frac{1}{(2\pi)^d}\frac{1}{f_n^c(\omegabf_\jbf)},
\end{equation}
\begin{equation}
\omegabf_\jbf =(\omega_{j_1},\cdots, \omega_{j_d}) =
\left(\frac{2\pi j_1 }{K},\cdots, \frac{2\pi j_d}{K}\right),
\end{equation}
and
\begin{equation} \label{eq:lemmaImpStepBeq3}
\frac{1}{(2\pi)^d}\frac{1}{f_n^c(\omegabf_\jbf)} - \epsilon^\prime
\le \frac{1}{\lambda_{i_1,\cdots,i_d}} \le
\frac{1}{(2\pi)^d}\frac{1}{f_n^c(\omegabf_\jbf)} + \epsilon^\prime
\end{equation}
for all $(i_1,\cdots,i_d)$ in the $(j_1,\cdots,j_d)^{th}$
hypercube. Here, $\epsilon^\prime$ $(>0)$ is independent of
$(j_1,\cdots,j_d)$ since $1/f_n^c(\omegabf)$ is uniformly
continuous over $\omegabf \in [-\pi,\pi)^d$ by
Lemma~\ref{lem:uniformConvInv} (c). Applying
(\ref{eq:lemmaImpStepBeq3}) to (\ref{eq:lemmaImpStepBeq2}), we
have
\begin{equation}  \label{eq:lemmaImpStepBeq4}
V_n  - \frac{\epsilon^\prime}{n^{d}}\sum_{\ibf\in \Dc_n}
\bar{Y}_\ibf^2 \le S_n \le V_n +
\frac{\epsilon^\prime}{n^{d}}\sum_{\ibf\in \Dc_n} \bar{Y}_\ibf^2,
\end{equation}
where
\begin{equation}
V_n = \frac{1}{K^d} \sum_{j_1=1}^{K} \cdots \sum_{j_d=1}^{K}
\frac{1}{(2\pi)^d} \frac{1}{f_n^c(\omegabf_\jbf)} \left(
\frac{1}{|\Ic(j_1)|\cdots|\Ic(j_d)|} \sum_{i_1 \in \Ic(j_1)}
\cdots \sum_{i_d \in \Ic(j_d)} \bar{Y}_{i_1,\cdots, i_d}^2
\right).
\end{equation}
By the SLLN for the sample mean of $\bar{Y}_\ibf^2$, we have
\begin{equation}
\sigma^2 - \epsilon^{\prime\prime} \le
\frac{1}{|\Ic(j_1)|\cdots|\Ic(j_d)|} \sum_{i_1 \in \Ic(j_1)}
\cdots \sum_{i_d \in \Ic(j_d)} \bar{Y}_{i_1,\cdots, i_d}^2 \le
\sigma^2 + \epsilon^{\prime\prime},
\end{equation}
almost surely for sufficiently large $n$ given $K$.  Thus, $V_n$
is given by
\begin{equation} \label{eq:lemmaImpStepBeq5}
( \sigma^2 - \epsilon^{\prime\prime})  Z_n    \le  V_n  \le (
\sigma^2 + \epsilon^{\prime\prime})  Z_n,
\end{equation}
where
\begin{equation}
Z_n =\frac{1}{K^d} \sum_{j_1=1}^{K} \cdots \sum_{j_d=1}^{K}
\frac{1}{(2\pi)^df_n^c(\omegabf_\jbf)}.
\end{equation}
Now we take $K\rightarrow \infty$, and the Riemann sum $Z_n$
converges to
\begin{equation} \label{eq:lemmaImpStepBeq6}
Z_n  \rightarrow \frac{1}{(2\pi)^d} \int_{-[\pi,\pi)^d}
\frac{1}{(2\pi)^df_1(\omegabf)} d\omegabf
\end{equation}
by Lemma \ref{lem:uniformConvInv} (b) and (c). Since
$\epsilon^\prime$ and $\epsilon^{\prime\prime}$ can be made
arbitrarily small by making $n$ and $K$ large, and
$\frac{1}{(2\pi)^d} \int_{-[\pi,\pi)^d} \frac{1}{(2\pi)^d}
\frac{1}{f_1(\omegabf)} d\omegabf < M_4$ for some $M_4 > 0$ and
$n^{-d}\sum_{\ibf\in \Dc_n} \bar{Y}_\ibf \rightarrow \sigma^2$
a.s., we have by (\ref{eq:lemmaImpStepBeq4}),
(\ref{eq:lemmaImpStepBeq5}) and (\ref{eq:lemmaImpStepBeq6}), that
\begin{equation} \label{eq:lemmaImpStepBeq8}
|\Dc_n|^{-1} \ybf_{|\Dc_n|}^T \Cbf_{|\Dc_n|}^{-1} \ybf_{|\Dc_n|}
\rightarrow (2\pi)^{-d} \int_{\omegabf \in [-\pi,\pi)^2}
\frac{\sigma^2}{(2\pi)^d f_1(\omegabf)} d \omegabf,
\end{equation}
almost surely as $n\rightarrow \infty$. By
(\ref{eq:lemmaImpStepAeq6}) and (\ref{eq:lemmaImpStepBeq8}) we
have
\begin{equation}
|\Dc_n|^{-1} \ybf_{|\Dc_n|}^T \Sigmabf_{1,|\Dc_n|}^{-1}
\ybf_{|\Dc_n|} \rightarrow (2\pi)^{-d} \int_{\omegabf \in
[-\pi,\pi)^2} \frac{\sigma^2}{(2\pi)^d f_1(\omegabf)} d \omegabf,
\end{equation}
almost surely as $n\rightarrow \infty$. This concludes the proof.
\hfill{$\blacksquare$}

\bibliographystyle{ieeetr}



\end{document}